\documentclass[aps,twocolumn,superscriptaddress,groupedaddress]{revtex4}
\usepackage{graphicx}  
\usepackage{dcolumn}   
\usepackage{bm}        
\usepackage{amssymb}   
\begin{document}
\widetext


\title{Measurability of $\phi$, $\omega$ and $\rho$ mesons via di-electron decays in high-temperature states produced in heavy-ion collisions}
\affiliation{Graduate School of Science, Hiroshima University, Kagamiyama, Higashi-Hiroshima, Hiroshima 739-8526, Japan}
\author{Yoshihide Nakamiya} \email{nakamiya@hepl.hiroshima-u.ac.jp}
\author{Kensuke Homma}    \affiliation{Graduate School of Science, Hiroshima University, Kagamiyama, Higashi-Hiroshima, Hiroshima 739-8526, Japan}
\vskip 0.25cm

\begin{abstract}We discuss measurability of $\phi$, $\omega$ and $\rho$ mesons via di-electron decays in high-temperature states produced in heavy-ion collisions, equivalently at different pion multiplicities per heavy-ion collision
$dN_{\pi^{0} + \pi^{\pm}}/dy$ = 1000 and 2700
intended for the most central Au+Au collisions at $\sqrt{s_{NN}}$ = 200 GeV (RHIC) and the most central Pb+Pb collisions at $\sqrt{s_{NN}}$ = 5.5 TeV (LHC), by 
evaluating the signal-to-background ratios and the statistical significance for the idealized detection system in the numerical simulation.
The simulation study provides a guideline to be applicable to a concrete detector design by focusing on only the key experimental issues relevant to the measurement of di-electrons. The results suggest that there are realizable parameter ranges to measure light vector mesons via di-electrons with the reasonable significance level, even in the highest multiplicity case.
\end{abstract}

\maketitle

\section{Introduction}
\label{sect1}
High energy heavy-ion collisions provide a unique opportunity 
to liberate quarks and gluons from nucleons and cause the phase transition 
from nuclear to quark-gluon matter.
Heavy-ion collisions have advantage of studying the QCD phase diagram 
especially in high-temperature and low-baryon density 
domains \cite{virtual_photon1,virtual_photon2, direct_photon, baryon_density1,baryon_density2}.

The mass modification \cite{chiral1,chiral2,chiral3,chiral4,chiral5} of light vector mesons such as $\phi$, $\omega$ and $\rho$ is an important signature of the 
QCD phase transition, because their masses are sensitive to chiral condensate $\langle q\bar{q} \rangle$.
Chiral condensate is one of the most prominent order parameters characterizing
the QCD phase structure. The behavior of the chiral condensates near the critical temperature is studied by several kinds of lattice QCD calculations 
\cite{Lattice1,Lattice2,Lattice3,Lattice4,Lattice5,Lattice6,Lattice7}.  
The light vector mesons can be candles to study the properties of quark-gluon matter produced in heavy-ion collisions.
The mass modification inside quark-gluon matter is potentially 
visible because their lifetimes are supposed to be comparable to duration of the thermal equilibrium state, 
unless the interactions with hadronic matter in the later stage dominate.
In addition, electron-positron pairs, which are referred to as "di-electrons", decaying 
from light vector mesons are considered to be a clear probe, 
since charged leptons carry the original information in the 
early stage without perturbation by hadronic matter via strong coupling in the 
relatively later stage of the system evolution.

In low-temperature and high-baryon density domains, 
the symptoms of the mass modification are intensively discussed \cite{KEK_E325_1,KEK_E325_2,JLab_1,TAPS_1}.
In high-temperature and low-baryon density domains, the enhancement of the 
di-electron yield is observed in the mass range below 0.7 GeV/c$^{2}$ in central Au+Au 
collisions at $\sqrt{s_{NN}}=200$ GeV \cite{virtual_photon1, continum1}, and it indicates a deviation from the di-electron spectrum in p+p collisions \cite{virtual_photon1, continum1, continum2,continum3}. 
Therefore,  quantitative studies of the mass spectra including the light vector mesons grow in importance to understand the phenomena 
in the high temperature domain. 
Furthermore, the measurement of the light vector mesons in higher-temperature states, at the LHC energy, would give a deeper insight into these phenomena.

The detection of di-electrons is, however,  challenging from the
experimental point of view, because a heavy-ion collision event produces 
a huge number of particles. The experiments at RHIC and LHC energies 
report that \textit{O}(100-1000) particles are produced at midrapidity 
in a heavy-ion collision \cite{mul1,mul2,mul3}.
The produced particles are dominated by $\pi^{\pm}$ and $\pi^{0}$ mesons. 
The production cross sections of pions are approximately hundred times larger than that of $\phi$ meson, and ten times larger than that of $\omega/\rho$ 
meson \footnote{
The production ratios between pions and light vector mesons depend on the collision species and energies. 
The exact values of these ratios are calculated one-by-one for the given collision species and energies.
Refer to Table \ref{cross_section} in \ref{app2}.
}.
In addition, the branching ratio (BR) to a di-electron pair is on the order of $10^{-4}$ for $\phi$ meson and $10^{-5}$ for $\omega/\rho$ meson.
Therefore, the measurement of the signal di-electrons requires state-of-the-art technology to identify electrons and positrons among a large amount of background hadrons by combining information
from Ring Imaging Cherenkov counters \cite{RICH}, $dE/dx$ \cite{TPC1,TPC2}, 
Time of Flight \cite{PHENIX_PID}, most notably, the Hadron-Blind Detector \cite{HBD}
and so forth. 

In addition to the small yield of di-electrons from the light vector mesons,
the difficulty of the measurement is severely caused by background 
contaminations from processes other than the light vector meson decays.
Dominant sources of such backgrounds are listed as follows,
\begin{enumerate}
\item Dalitz decays $\pi^{0} \rightarrow \gamma e^{+}e^{-}$ and $\eta \rightarrow \gamma e^{+}e^{-}$,
\item pair creations by decay photons from $\pi^{0}$ and $\eta$ meson,
\item semi-leptonic decays from hadrons,
\item charged hadron contaminations by electron misidentification.
\end{enumerate}
The key issues are in the increase of combinatorial pairs from different sources.
The amount of these pairs has approximately quadratic dependence on multiplicity because combination between all possible electrons and positrons
must be taken, 
while the number of true di-electron pairs from the light vector mesons approximately linearly scales with multiplicity \footnote{
The yields and kinematics of produced particles vary by the collision species and energies. 
They are fully taken into account in the simulation. The details are explained in Section \ref{sect2}.
}.

Detector upgrades are planned in several experiments at RHIC and LHC. 
For instance, the ALICE experiment at LHC plans to upgrade the internal tracking system with the relatively low detector materials and 
enhance low-$p_{T}$ tracking capability \cite{ALICE_upgrade}. 
These upgrades are expected to suppress the backgrounds from the photon-conversion and Dalitz decay processes, and improve the Time-of-Flight measurement 
for the hadron identification.
Moreover, the tracking system will improve the secondary vertex resolution.
It allows to separate the backgrounds from the semi-leptonic decay of open charms.  
The key issues from (1) to (4) above are actually relevant to the items of the upgrade plan.

The measurement of light vector mesons is still important but challenging.
The detector system applied to high-energy heavy ion collisions
is, in general, multipurpose.
Thus it is not necessarily optimized only for the measurement of light vector mesons. 
In this situation, unless the key issues above are simultaneously considered in advance of the other irrelevant design issues, for instance, only installing a state-of-the-art particle identifier might not be sufficient.
In order to prove the performance of an upgraded detector system,
a very detailed detector simulation is required. 
It is usually a time-consuming process to conclude whether a reasonable signal-to-background ratio is 
guaranteed with the upgraded complex detector system or not.
Thus the applicability of such a detailed simulation is rather limited 
to a specific detector system. 
If one knows the key issues,  however, considering only these issues in the idealized case would provide a more general guideline applicable 
to any complex detector system.

In this paper, therefore, we do not focus on details of individual detector designs. Instead, we investigate the 
necessary conditions which must be minimally satisfied in advance 
in order to gain reasonable signal-to-background ratios with idealized 
detector systems for given multiplicities, collision species and energies in heavy-ion collisions.

For this purpose, we first give the estimates on the production yields of
light vector mesons and the other hadrons producing di-electrons
in the final state. They are applied to p+p and A+A collisions at the RHIC and LHC energy regime, in concrete, $dN_{\pi^{0} + \pi^{\pm}}/dy$ = 3, 6, 1000 and 2700 in p+p 200 GeV, p+p 7 TeV, Au+Au 200 GeV and Pb+Pb 5.5 TeV, respectively.
We then implement the key physical background processes in
the numerical simulation including
the Dalitz decay and the photon conversion as well as
the semi-leptonic decay of hadrons under the key experimental conditions:
the amount of the detector material, 
the charged pion contamination, 
the geometrical acceptance, the electron tagging efficiency, 
the measurable momentum cutoff and the smearing effect by momentum resolution.
These experimental conditions are parameterized and implemented into the simulation.  
The flowchart of the simulation is explained in Section \ref{sect2}.
Starting from a set of the idealized baseline parameters,
we explore the expected signal-to-background ratios and the statistical significance
as a function of the individual experimental parameters for given 
pion multiplicities in high-temperature states. 
The results of the simulation study are provided in Section \ref{sect3}.
The non-trivial residual effects, though some of them depend on a specific detection system, are discussed in Section \ref{sect4}. 
In Section \ref{sect5}, we conclude that we can find reasonable parameter ranges to measure the light vector mesons via di-electrons even at $dN_{\pi^{0} + \pi^{\pm}}/dy$ = 2700.

\section{Numerical simulation for idealized detection systems}
\label{sect2}
\begin{figure*}[!htb]
\begin{center}
\begin{tabular}{c}
	\begin{minipage}{0.33\hsize}
	\begin{center}
	\includegraphics[width=6.2cm,height=5cm]{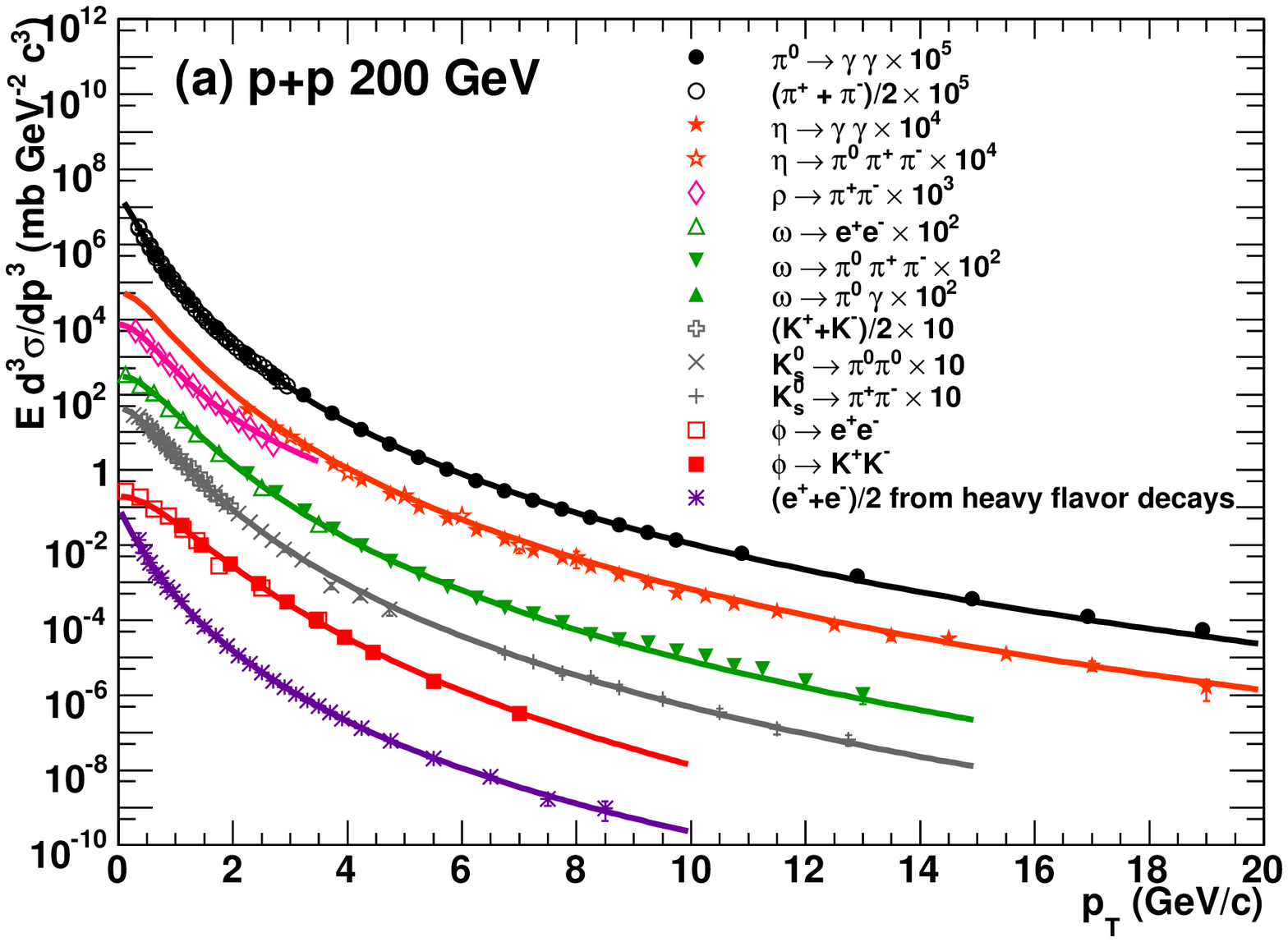}
	\end{center}
	\end{minipage}
	\begin{minipage}{0.33\hsize}
	\begin{center}
	\includegraphics[width=6.2cm,height=5cm]{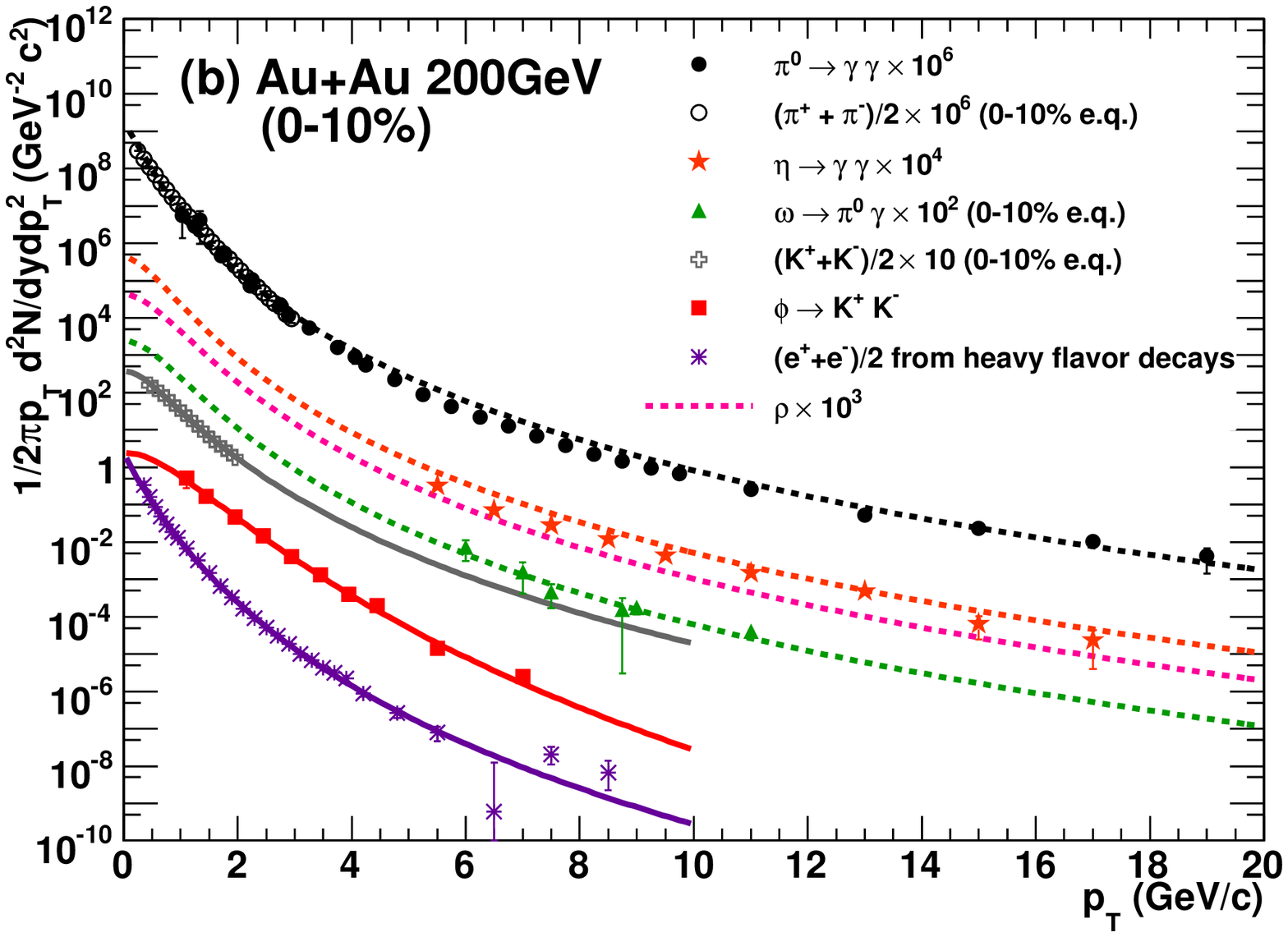}	
	\end{center}
	\end{minipage}
	\begin{minipage}{0.33\hsize}
	\begin{center}
	\includegraphics[width=6.2cm,height=5cm]{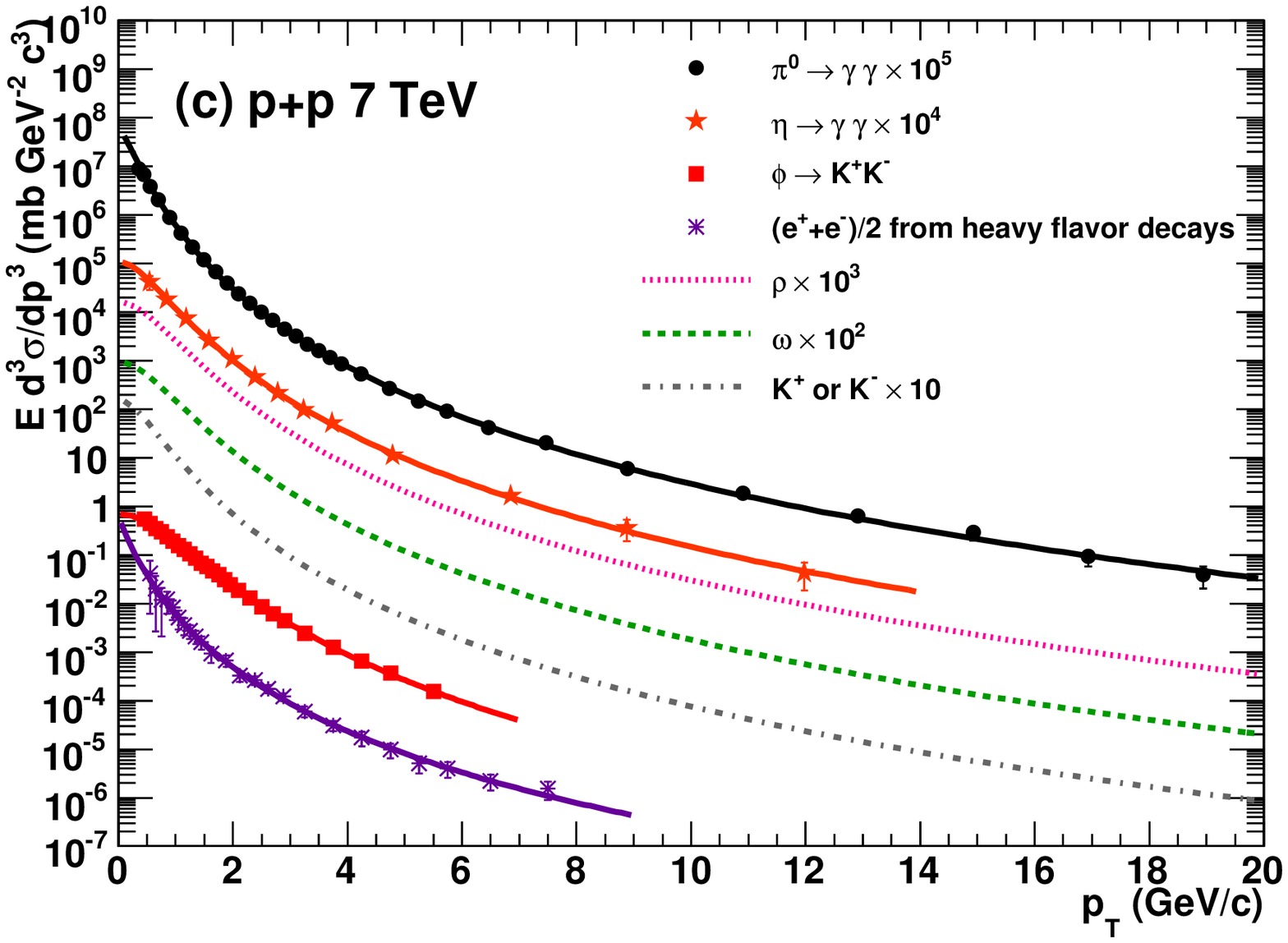}
	\end{center}
	\end{minipage}
\end{tabular}
\caption{\label{Tsallis_input_pt} (color online)
(a) The differential cross sections of different particles in p+p 200 GeV. 
Tsallis fitting curves are depicted as the solid curves on the data points. 
The solid curve on $\pi^{0} \rightarrow \gamma \gamma$ \cite{npi1} and $(\pi^{+} + \pi^{-})/2$ \cite{ch1,ch2} are obtained by the simultaneous fitting.
The curves on the data points of $(K^{+} + K^{-})/2$ \cite{ch1,ch2},  $K^{0}_{s}  \rightarrow \pi^{0}\pi^{0}$ \cite{scale1} and $K^{0}_{s}  \rightarrow \pi^{+}\pi^{-}$ \cite{k0} are also obtained by the simultaneous fitting. 
The star symbols show $\eta \rightarrow \gamma \gamma$ \cite{eta1} and $\eta \rightarrow \pi^{0} \pi^{+} \pi^{-}$ \cite{eta2}.
The open diamonds show $\rho \rightarrow \pi^{+} \pi^{-}$ \cite{rho}.
The triangles show $\omega \rightarrow e^{+}e^{-}, \pi^{0} \pi^{+} \pi^{-}$ and $\pi^{0} \gamma$ \cite{scale1}.
The squares show $\phi \rightarrow e^{+}e^{-}$ and $K^{+}K^{-}$ \cite{scale1}. 
The asterisks show single electrons from heavy flavor decays \cite{charm}.
(b) The invariant $p_{T}$ spectra in Au +Au 200 GeV at the centrality class of 0-10$\%$.  
The dotted curves are scaled by the $N_{part}$ and assumed to be the same spectrum shape as that of p+p 200 GeV.
The scaling curves are consistent with the data points of pions \cite{pi0_auau1,pi0_auau2, cpi_auau1} , $\eta$ \cite{eta_auau1} and $\omega$ \cite{ome_auau1}, 
respectively.
The solid curves are the fitting results to the data points of $K^{\pm}$ \cite{cpi_auau1}, $\phi$ \cite{phi_auau1} and single electrons from heavy flavor decays \cite{charm_auau1}. For $K^{\pm}$, the Tsallis parameter $q$ is fixed since there is no data point in the high $p_{T}$ region.
(c) The differential cross section in p+p 7 TeV. The data points of $\pi^{0}/\eta$ \cite{pi0eta7TeV}, $\phi$ \cite{phi7TeV} and single electrons from heavy flavor 
decays \cite{charm7TeV} are shown in this figure. Tsallis fitting curves are depicted as the solid curves. 
The dotted curves of $\rho$, $\omega$ and $K^{\pm}$ are obtained by assuming the same spectrum shape of $\pi^{0}$ and normalizing the individual production ratios with respect to pions in p+p 200 GeV. 
}
\end{center}
\end{figure*}

The numerical simulation is developed to estimate the signal-to-background ratios and the statistical significance of light vector mesons 
via di-electron decays in heavy-ion collisions.
Instead of directly simulating the multi-particle production with detailed dynamics in heavy-ion collisions, 
high multiplicity states are first represented by the form of the total pion multiplicity $dN_{\pi^{0} + \pi^{\pm}}/dy$.
The production of the relevant particles other than pions is determined based on the individual production cross sections relative to that of pions 
as a function of the transverse momentum $p_{T}$.
They are evaluated by the measured data points, or the extrapolation via the proper scaling for missing data points.
In addition, the key experimental parameters for the di-electron measurement are set as the inputs.
As the idealized baseline parameters on the experimental conditions, we choose the following set of parameters: 
\begin{enumerate}
\item photon conversion probability $P_{cnv}$: 1 $\%$,
\item rejection factor of charged pions $R_{\pi^{\pm}}$, which is defined by the inverse of the probability 
that charged pions are identified as electrons: 500,
\item geometrical acceptance $\epsilon_{acc}$: 100 $\%$,
\item electron tagging efficiency $\epsilon_{tag}$: 100 $\%$,
\item transverse momentum threshold $p_{T}^{th}$: 0.1 GeV/c,
\item transverse momentum resolution $\sigma^{ref}_{p_{T}}$ (we quote the ALICE-TPC resolution \cite{momreso_alice}):
$\sqrt{ \left(0.01 \cdot p_{T} \right)^{2} + \left( 0.0056 \right)^{2}}$ GeV/c.
\end{enumerate}
Some of experimental parameters are correlated in fact, but these correlations are neglected in this simulation.
\begin{figure}[!htb]
\begin{center}
\includegraphics[height=5.7cm,width=8.05cm]{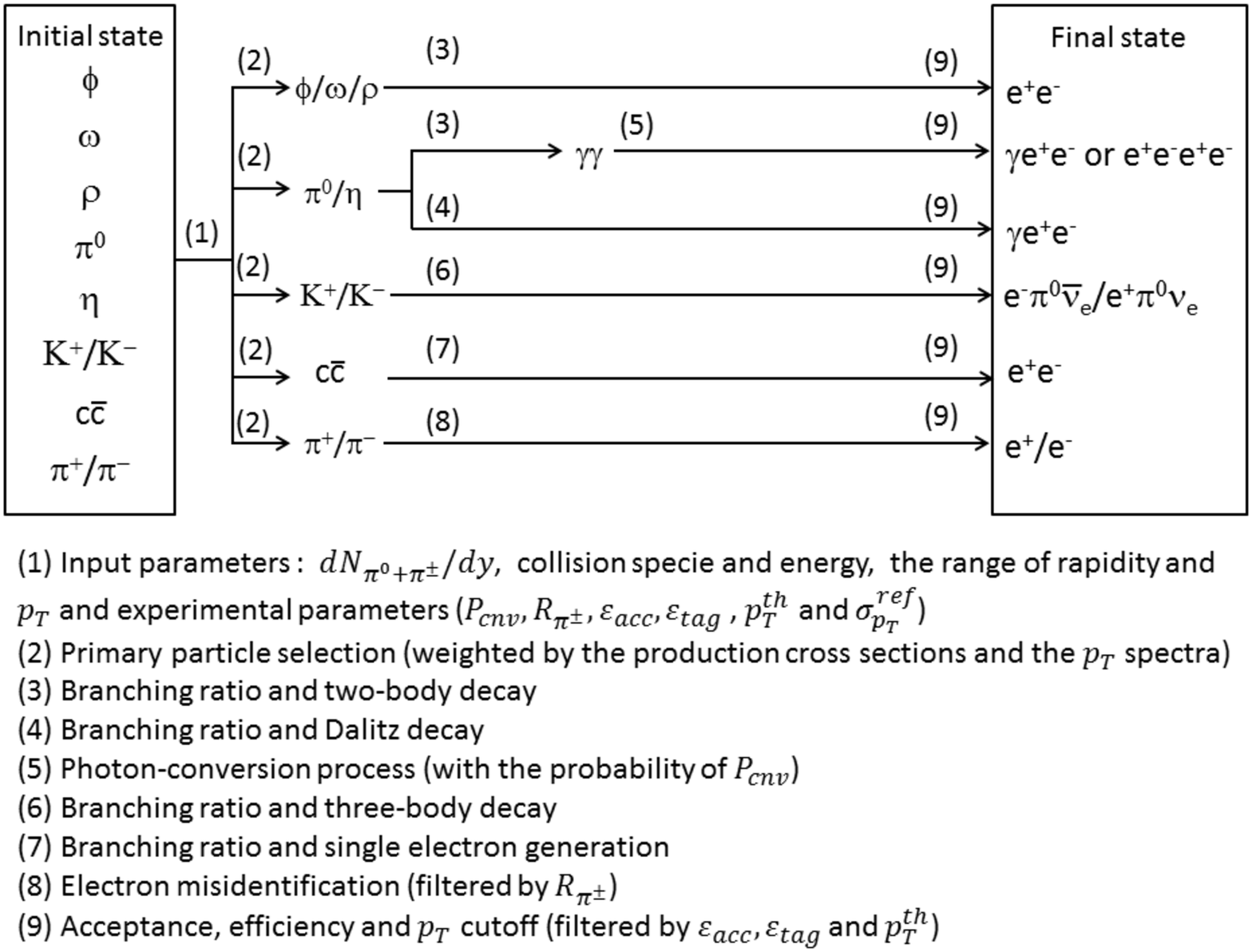}
\caption{\label{flowchart} 
The flowchart of  the numerical simulation. 
}
\end{center}
\end{figure}
Figure \ref{flowchart} shows the flowchart of the numerical simulation. 
The step (1) sets the input parameters above. 
In the step (2), primary particles are generated with the weights of the invariant $p_{T}$ spectra.
The $p_{T}$ spectra are provided by the experimental data and the proper scaling for missing data.
The details of the input $p_{T}$ spectra for each are explained later.
Rapidity $y$ of a particle is uniformly generated in $|y| \leq 0.5$ \cite{mul1, rapidity1}. 
Primary particles branch into subsequent decay processes according to their branching ratios. 
The branching ratios are summarized in Table \ref{pdg_info} of \ref{app3}.
$\phi$, $\omega$ and $\rho$ mesons decay into di-electrons through the two-body decay process in the step (3).
The phase space of di-electrons from the light vector mesons is determined by the Gounaris-Sakurai model \cite{GSmodel}.
$\pi^{0}$ and $\eta$ mesons branch into 2$\gamma$ or the Dalitz decay process ($\gamma e^{+} e^{-}$) in the step (3) or (4).
The decaying $\gamma$'s are subsequently converted into di-electrons with the given photon-conversion probability in the step (5). 
Kinematics of di-electron in the photon-conversion process, that is, energy and scattering angle, are simulated by the well-established GEANT algorithm
\cite{geant4, lepton_pair1, lepton_pair2}. All photon-conversion points are fixed to the primary vertex points \footnote{The contribution to the di-electron background shape depends on where the photon conversion takes place,  in other words, 
depends on the arrangement of the detector materials. 
They should be considered in association with the track reconstruction algorithms. 
The biases from photon-conversion electrons at the off-axis point are discussed in Section \ref{sect4}.
}. The phase space of Dalitz decaying di-electrons is determined by the Kroll-Wada formula \cite{Kroll_Wada1, Kroll_Wada2}. 
The detailed formula is expressed in \ref{app4}. Charged kaons decay into electrons through the three-body decay process in the step (6).
In the step (7), electrons and positrons from open charms are directly generated to be consistent with the input $p_{T}$ spectra of single electrons \footnote{
The $p_{T}$ spectrum of single electrons originates from not only charm quarks but also beauty quarks.
The contributions from them are calculated by the fixed-order-plus-next-to-leading-log perturbative
QCD calculation (FONLL) \cite{FONLL} and its calculation is compared to the measurements \cite{bottom1, bottom2, bottom3}.
The results suggest that $N_{b \rightarrow e} / \left( N_{c \rightarrow e} + N_{b \rightarrow e} \right)$ is smaller than 0.2 at $p_{T} \leq$ 2.0 GeV/c in p+p 200 GeV 
and $N_{b \rightarrow e} / N_{c \rightarrow e}$ are smaller than 0.3 at $p_{T} \leq$ 2.0 GeV/c in p+p 7 TeV, where $N_{b \rightarrow e}$ and 
$N_{c \rightarrow e}$ is the number of electrons from beauty quarks and charm quarks, respectively. 
These ratios tend to drop rapidly as the $p_{T}$ reduces.
Therefore we neglect the contributions from beauty quarks in the calculation of particle production and decay kinematics in this simulation.  
}
as shown in Fig.\ref{Tsallis_input_pt}. 
They are randomly generated in azimuth with the branching ratio of 9.5 $\%$ \cite{charm}. 
The correlation of a di-electron pair originating from the open charm production is neglected in this simulation. 
The effect on the correlation is discussed in Section \ref{sect4}. 
Charged pions are identified as electrons with the given probability corresponding to the rejection factor of charge pions in the step (8).
At the final stage of the simulation, final-state electrons are filtered by the geometrical acceptance, the electron tagging efficiency
and $p_{T}$ threshold in the step (9). The algorithms of decaying processes mentioned above are developed based on the EXODUS simulator \cite{EXODUS}.

The main targets of this simulation are central Au+Au collisions  at $\sqrt{s_{NN}}$ = 200 GeV and central Pb+Pb collisions at $\sqrt{s_{NN}}$ = 5.5 TeV.
The results of p+p 200 GeV and p+p 7 TeV simulations are used as the references.
The $dN_{\pi^{0} + \pi^{\pm}}/dy$ and the invariant $p_{T}$ spectra are applied to the simulation taking the given collision species and energies into account. 
The input $dN_{\pi^{0} + \pi^{\pm}}/dy$ is estimated by the measured $dN_{ch}/dy$ \cite{mul1,mul2,mul3}.
$dN_{\pi^{0} + \pi^{\pm}}/dy$ = 3, 6 and 1000 are set for the simulation of p+p 200 GeV, p+p 7 TeV and central Au+Au 200 GeV collisions, respectively.
For the simulation of central Pb+Pb 5.5 TeV collisions, the $dN_{\pi^{0} + \pi^{\pm}}/dy$ is estimated by the extrapolation of the scaling curve as a function of 
collision energy \cite{mul3}. The extrapolated $dN_{\pi^{0} + \pi^{\pm}}/dy$ corresponds to 2700.

The input $p_{T}$ spectra for the p+p 200 GeV simulation are determined by the measured data points with the fits in the panel (a) of Fig.\ref{Tsallis_input_pt}. 
The Tsallis function, whose properties are explained in \ref{app1}, is used as the fitting function.

The invariant $p_{T}$ spectra in central Au+Au collisions are well known to be suppressed and scaled by the number of participant nucleons $N_{part}$ in the high $p_{T}$ region. 
The scaling parameter $N_{part}$ is calculated by the Monte Carlo simulation based on the Glauber model \cite{GlauberMC, GlauberModel}.
The $N_{part}$ scaling helps to extend the $p_{T}$ spectra in p+p collisions to those in heavy-ion collisions, even if there is lack of data points.
Therefore we can determine the input $p_{T}$ spectra in the wide range by combining the existing data points with the scaling properties.
The panel (b) of Fig.\ref{Tsallis_input_pt} shows the data points and the scaling curves in Au+Au 200 GeV at the centrality class of 0-10 $\%$ \footnote{
We use the published data points of $\pi^{\pm}$ and $K^{\pm}$ at the centrality class of 0-5 $\%$ 
and those of $\omega$ at the centrality class of 0-20 $\%$.
The mismatch of the centrality class is corrected by the weights with the $N_{part}$. 
The corrected data points, which are equivalent to the data at the centrality class of 0-10 $\%$, are used for the comparison to the scaling curves 
in the panel (b) of Fig.\ref{Tsallis_input_pt}.
}.
The dotted curves are obtained by the $N_{part}$ scaling. The spectrum shape is assumed to be the same as that of p+p 200 GeV.
These curves are reasonably consistent with the data points.
The solid curves on the data points of $\phi$ mesons, charged kaons and single electrons from heavy flavor decays are obtained by directly fitting with the Tsallis function. For charged kaons, the Tsallis parameter $q$ is fixed because of the missing data in the high $p_{T}$ region.

The panel (c) of Fig.\ref{Tsallis_input_pt} shows the invariant $p_{T}$ spectra in p+p 7 TeV.
The $p_{T}$ spectra of $\pi^{0}$, $\eta$, $\phi$ and single electrons are determined by the data points and the fits.
The dotted curves show the $p_{T}$ spectra of the other hadrons.
The spectrum shape is estimated by the $m_{T}$ scaling based on the $\pi^{0}$ data points, where $m_{T} = \sqrt{p_T^{2} + m_{0}^{2}}$ and $m_{0}$ is the rest mass of a particle. 
Their absolute production cross sections are estimated by the inclusive ratios between pions and the other hadrons in p+p 200 GeV. 
The invariant $p_{T}$ spectra in p+p 7 TeV are commonly used for the simulations of p+p 7 TeV and Pb+Pb 5.5 TeV, 
since the relative production cross sections between pions and the other hadrons are expected to be common for both collision systems, as long as the particle production between 7 TeV and 5.5 TeV has little dependence on the collision energy.

Table \ref{cross_section} in \ref{app2} summarizes these production cross sections and inclusive yields over all $p_{T}$ ranges in p+p 200 GeV, p+p 7 TeV and Au+Au 200 GeV at the centrality class of 0-10 $\%$.

\begin{figure*}[!ht]
\begin{center}
\begin{tabular}{c}
	\begin{minipage}{0.5\hsize}
	\begin{center}
	\includegraphics[width=8cm,height=4.5cm]{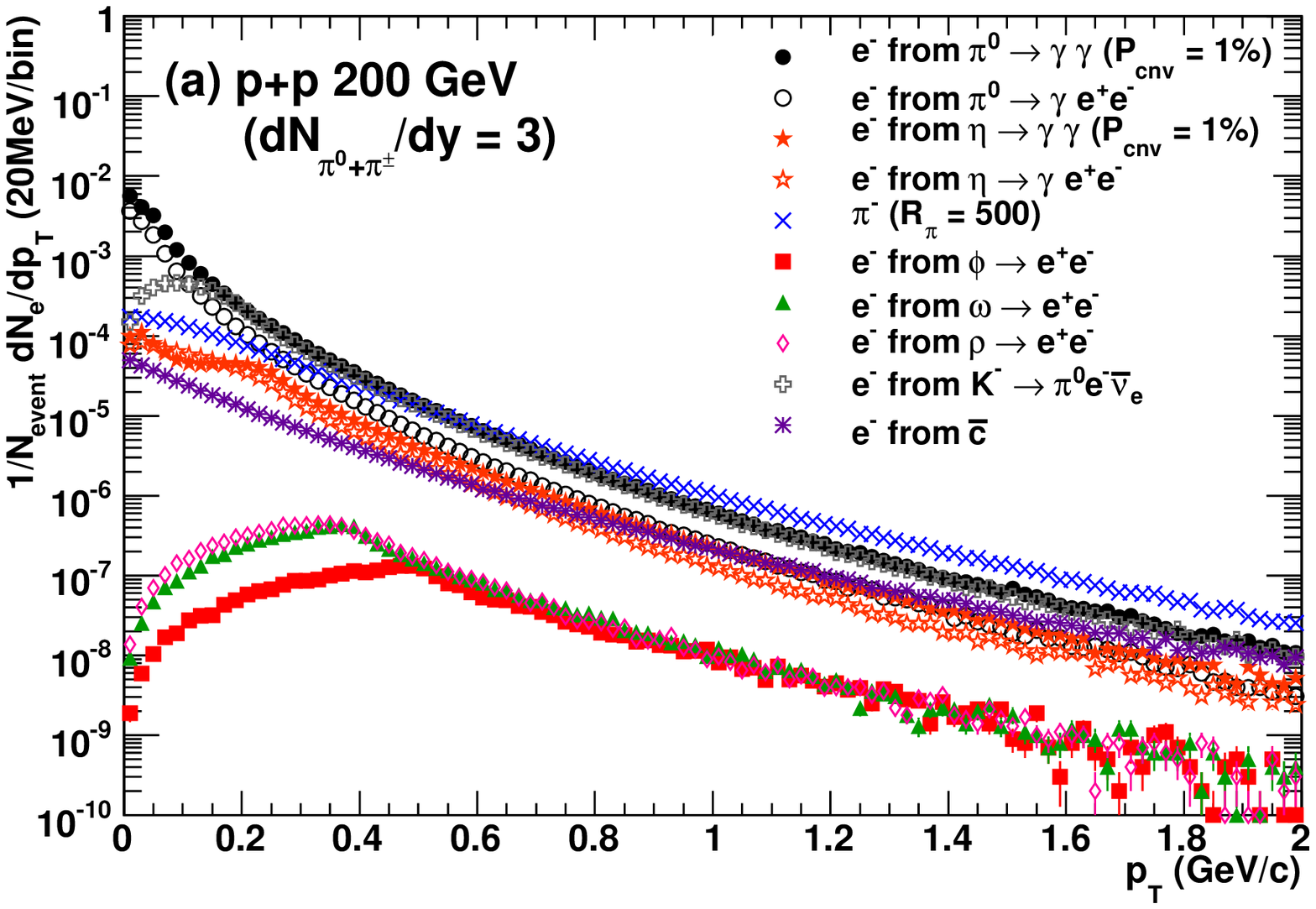}
	\end{center}
	\end{minipage}
	\begin{minipage}{0.5\hsize}
	\begin{center}
	\includegraphics[width=8cm,height=4.5cm]{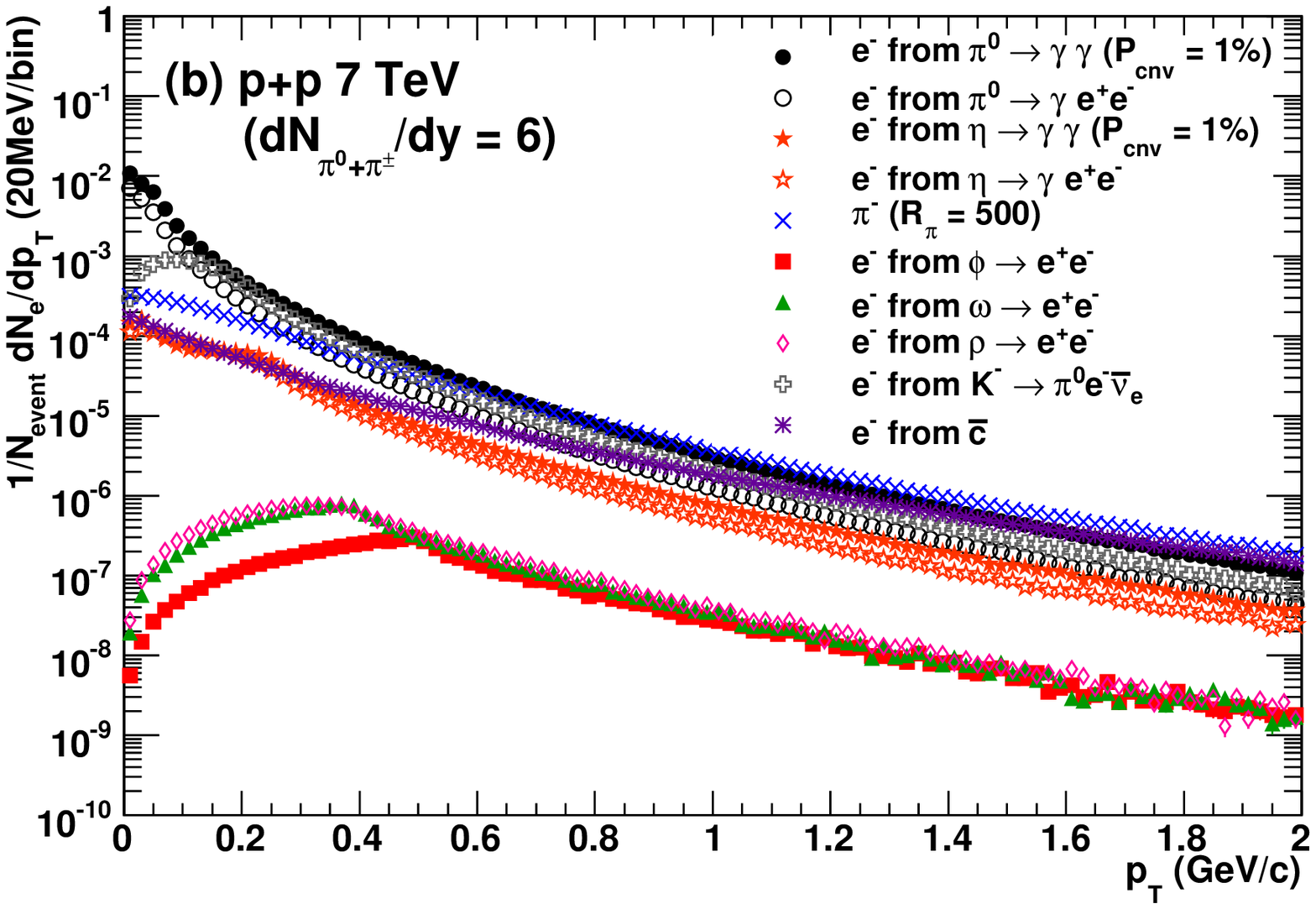}	
	\end{center}
	\end{minipage}
\end{tabular}
\begin{tabular}{c}
	\begin{minipage}{0.5\hsize}
	\begin{center}
\includegraphics[width=8cm,height=4.5cm]{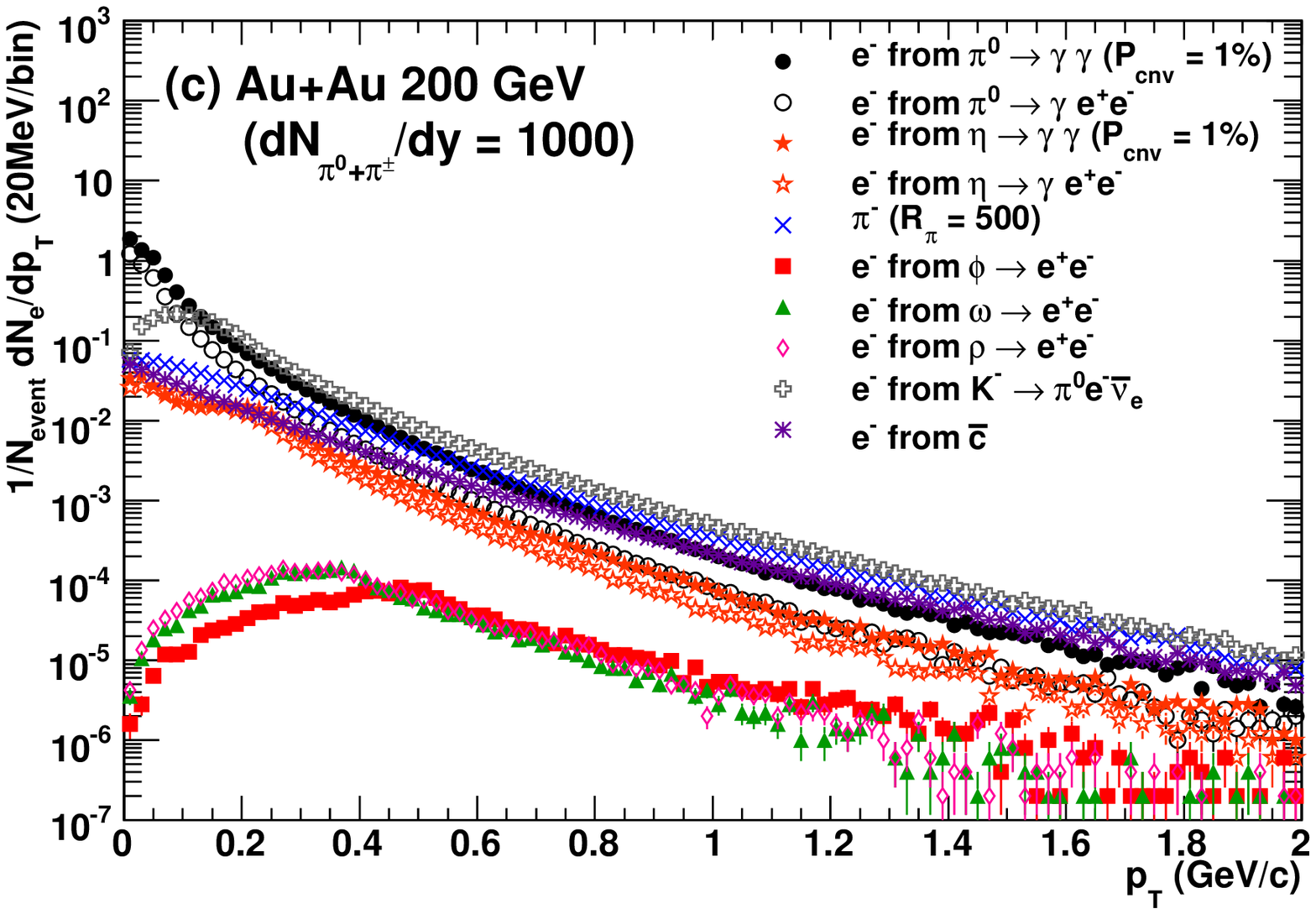}
	\end{center}
	\end{minipage}
	\begin{minipage}{0.5\hsize}
	\begin{center}
\includegraphics[width=8cm,height=4.5cm]{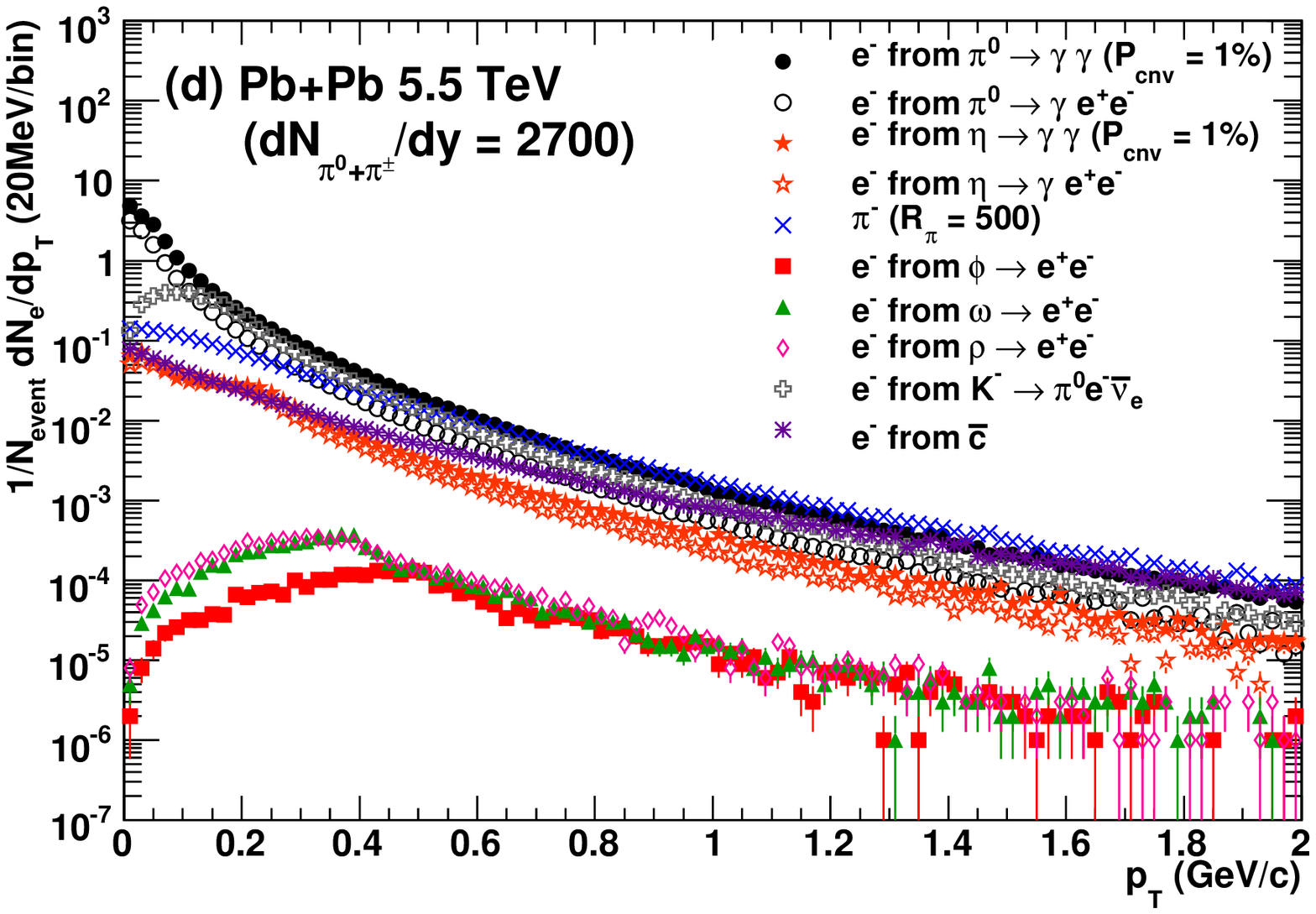}
	\end{center}
	\end{minipage}
\end{tabular}
\caption{\label{electron_pt}
(color online) The transverse momentum spectra of final-state electrons and misidentified charged pions from individual sources for the given $dN_{\pi^{0} + \pi^{\pm}}/dy$ with $P_{cnv}$ = 1 $\%$, $R_{\pi^{\pm}}$ = 500, $\epsilon_{acc}$ = 100 $\%$, $\epsilon_{tag}$ = 100 $\%$, $\sigma^{ref}_{p_{T}} = \sqrt{ \left(0.01 \cdot p_{T} \right)^{2} + \left( 0.0056 \right)^{2}}$ GeV/c and without $p_{T}$ cutoff.
}
\end{center}
\end{figure*}

Figure \ref{electron_pt} shows the simulated results of the $p_{T}$ spectra 
for the final-state electrons from individual sources with the baseline parameter set in p+p collisions 
at $\sqrt{s}$ = 200 GeV, p+p collisions at $\sqrt{s}$ = 7 TeV, central Au+Au collisions at  $\sqrt{s_{NN}}$ = 200 GeV
 and central Pb+Pb collisions at $\sqrt{s_{NN}}$ = 5.5 TeV, separately.
The parents of electrons are all indicated with different symbols specified inside the plot. 

\begin{figure}[!h]
\includegraphics[scale=0.4]{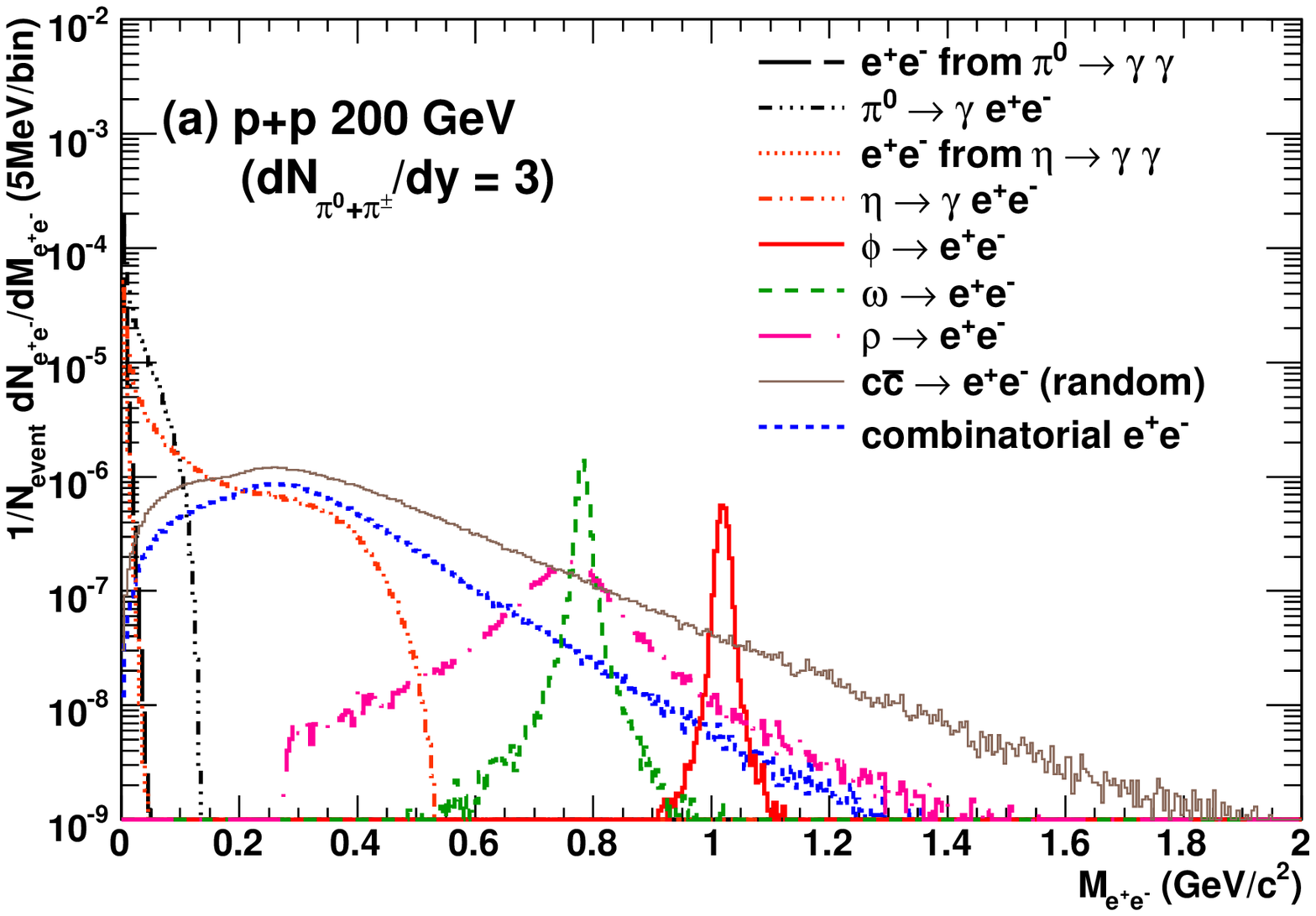}
\includegraphics[scale=0.4]{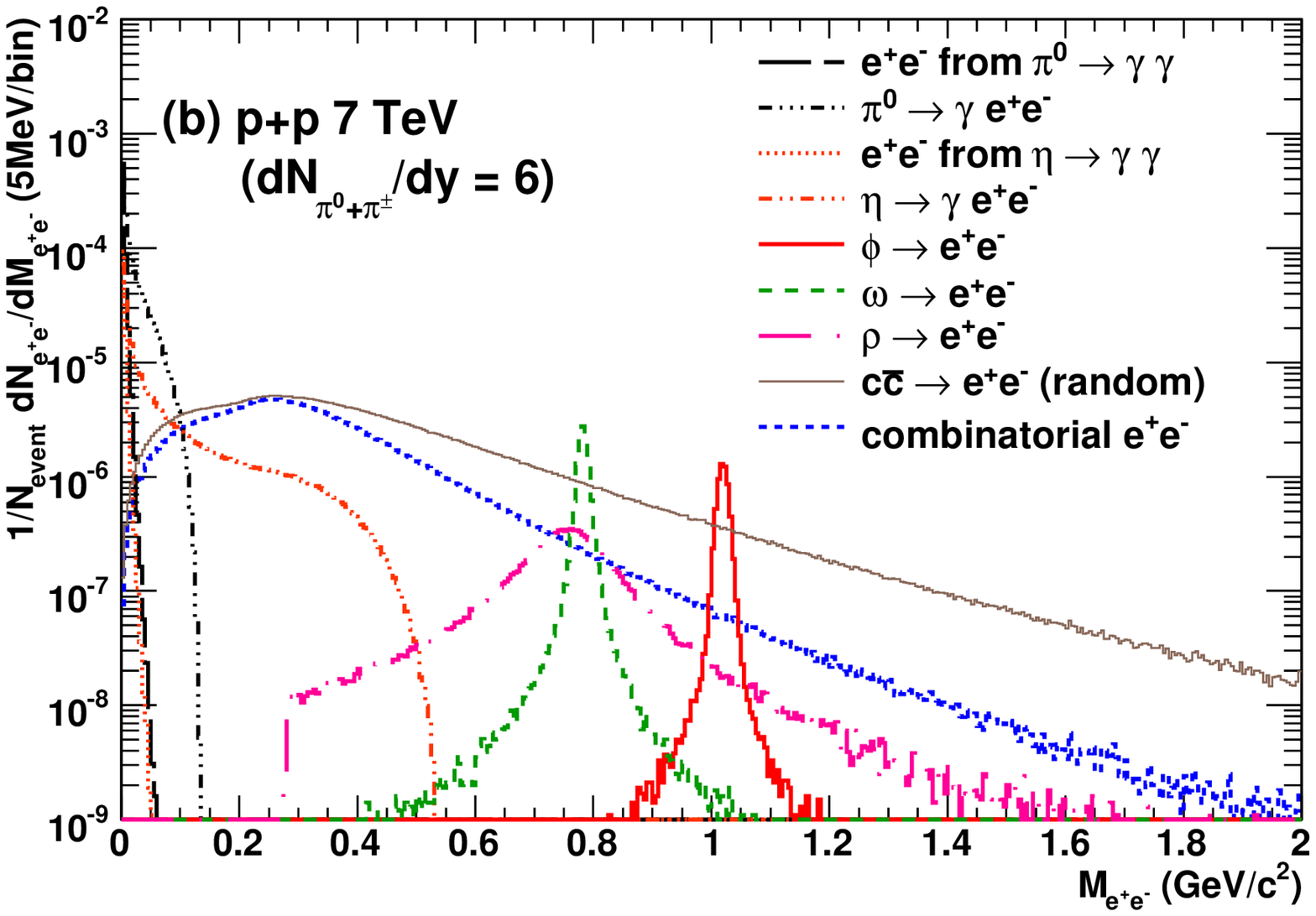}
\includegraphics[scale=0.4]{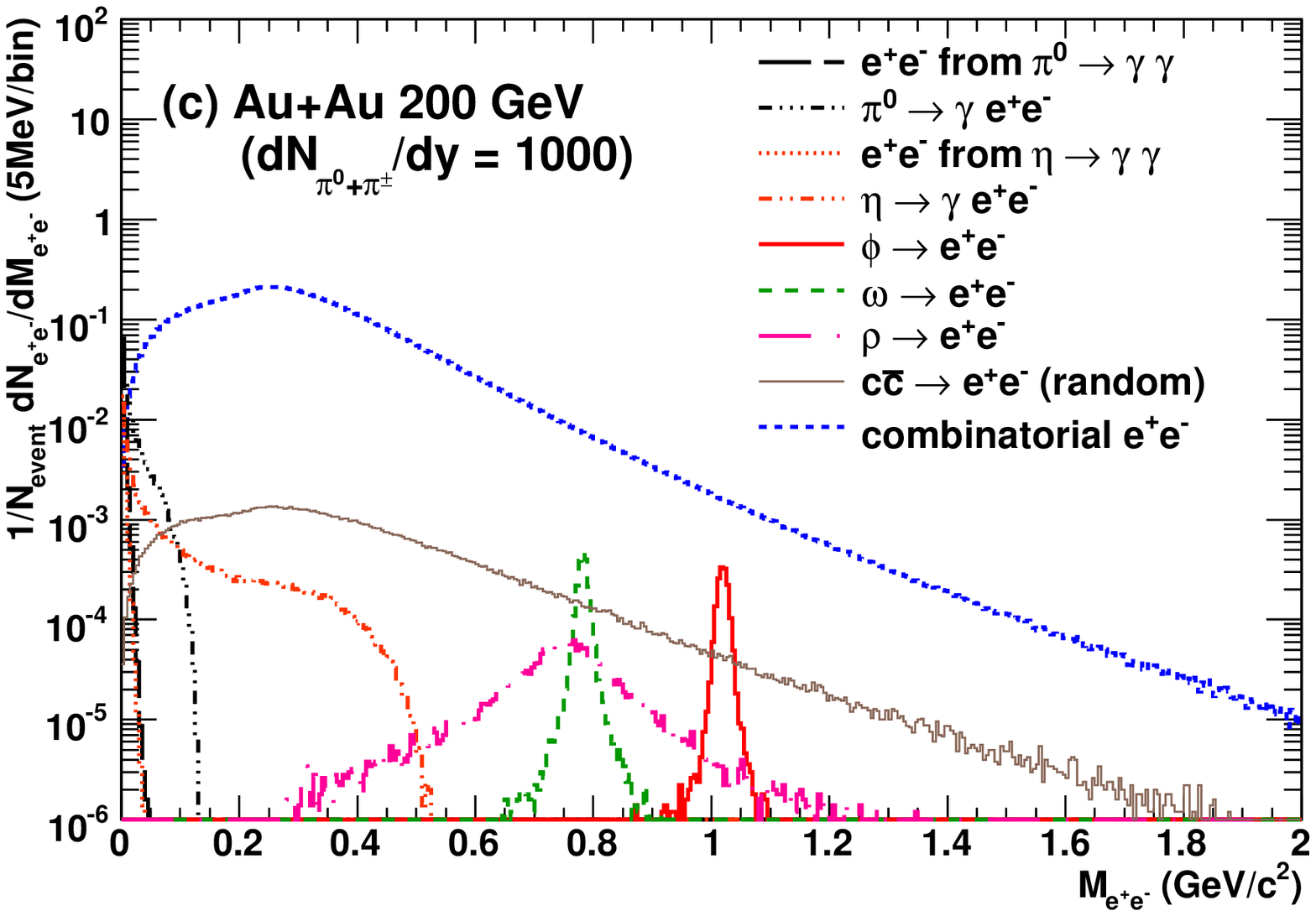}
\includegraphics[scale=0.4]{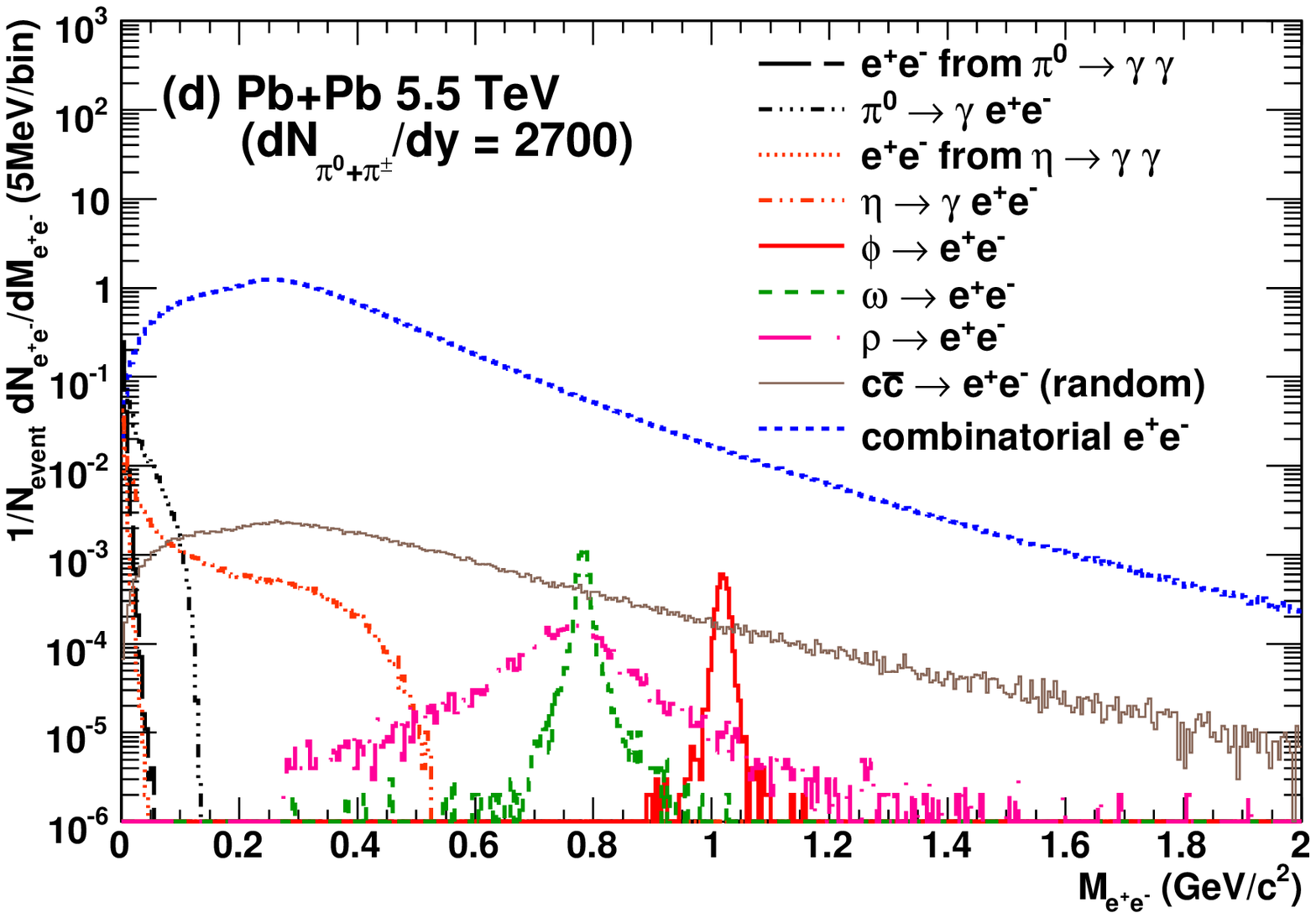}
\caption{\label{mass_cocktail} 
(color online) 
The invariant mass spectra of di-electrons from individual sources for the given $dN_{\pi^{0} + \pi^{\pm}}/dy$ 
with $P_{cnv}$ = 1 $\%$, $R_{\pi^{\pm}}$ = 500, $\epsilon_{acc}$ = 100 $\%$, $\epsilon_{tag}$ = 100 $\%$, $p_{T}^{th}$ = 0.1 GeV/c 
and $\sigma^{ref}_{p_{T}} = \sqrt{ \left(0.01 \cdot p_{T} \right)^{2} + \left( 0.0056 \right)^{2}}$ GeV/c.
The mass spectra from individual origins are shown with different curves specified inside the plot. 
The curves of the combinatorial pairs are reconstructed by all combinations between electrons and positrons
but only true combinations are excluded.
}
\end{figure}

\begin{figure}[!hbt]
	\includegraphics[scale=0.4]{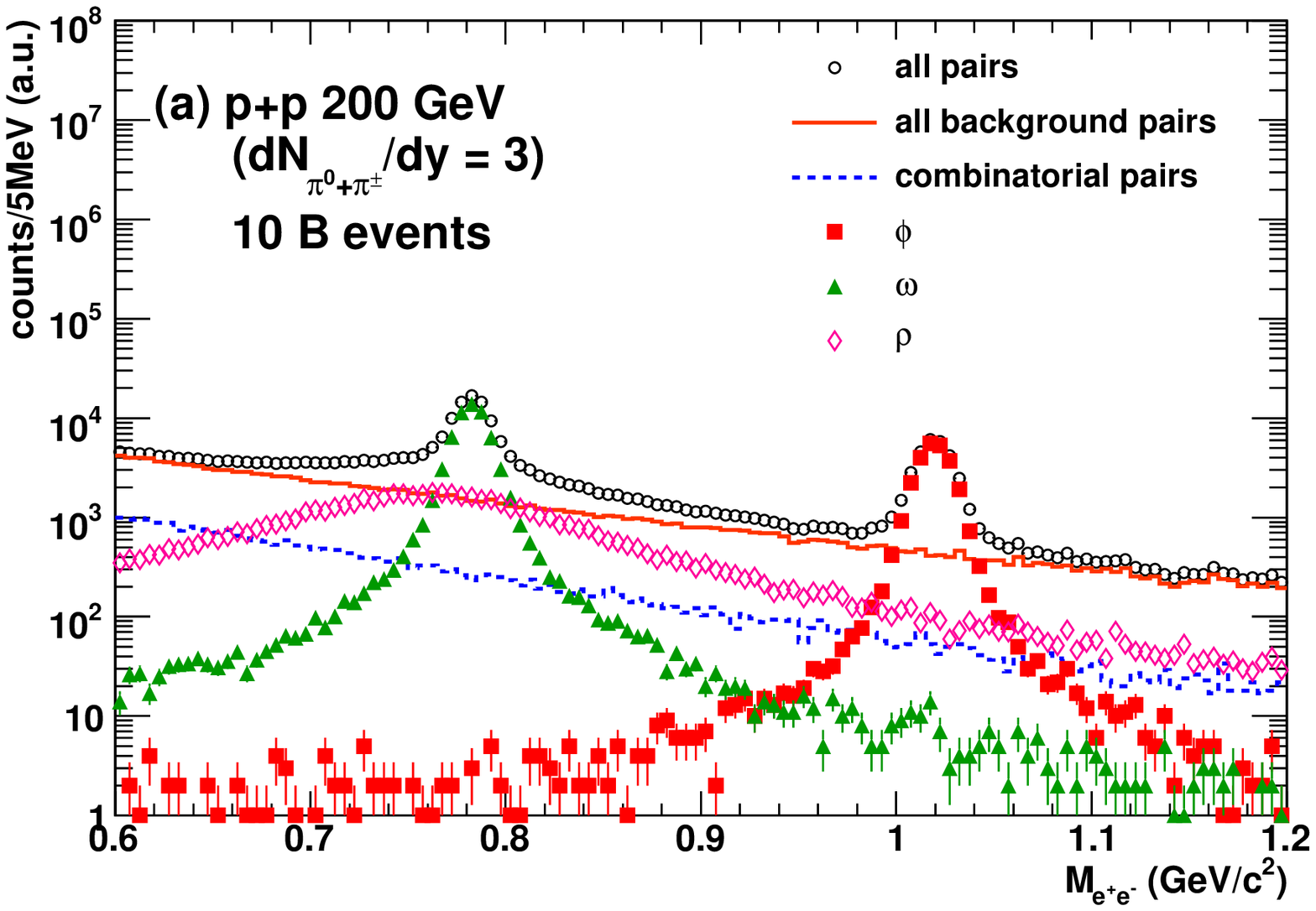}
	\includegraphics[scale=0.4]{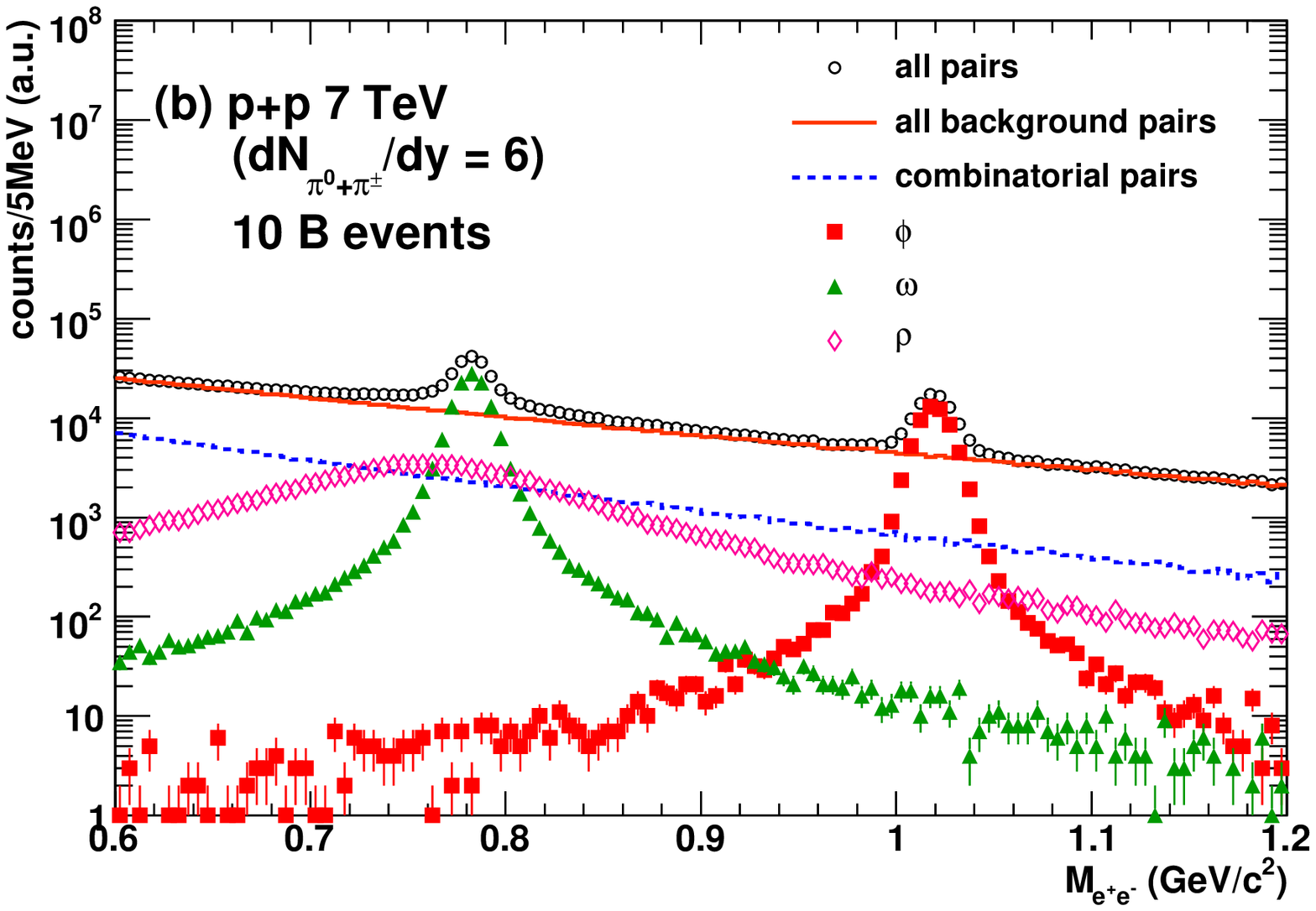}
	\includegraphics[scale=0.4]{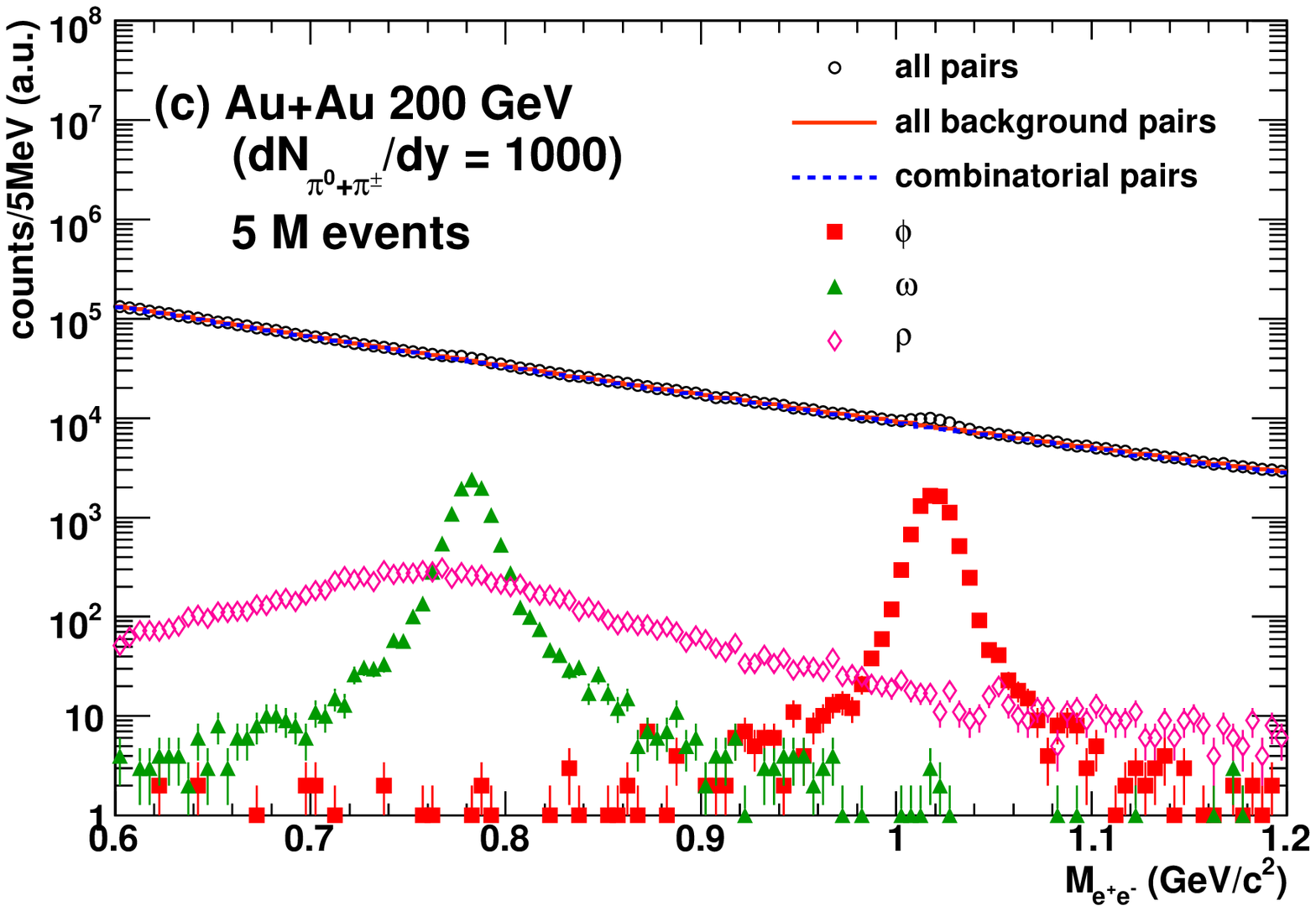}
	\includegraphics[scale=0.4]{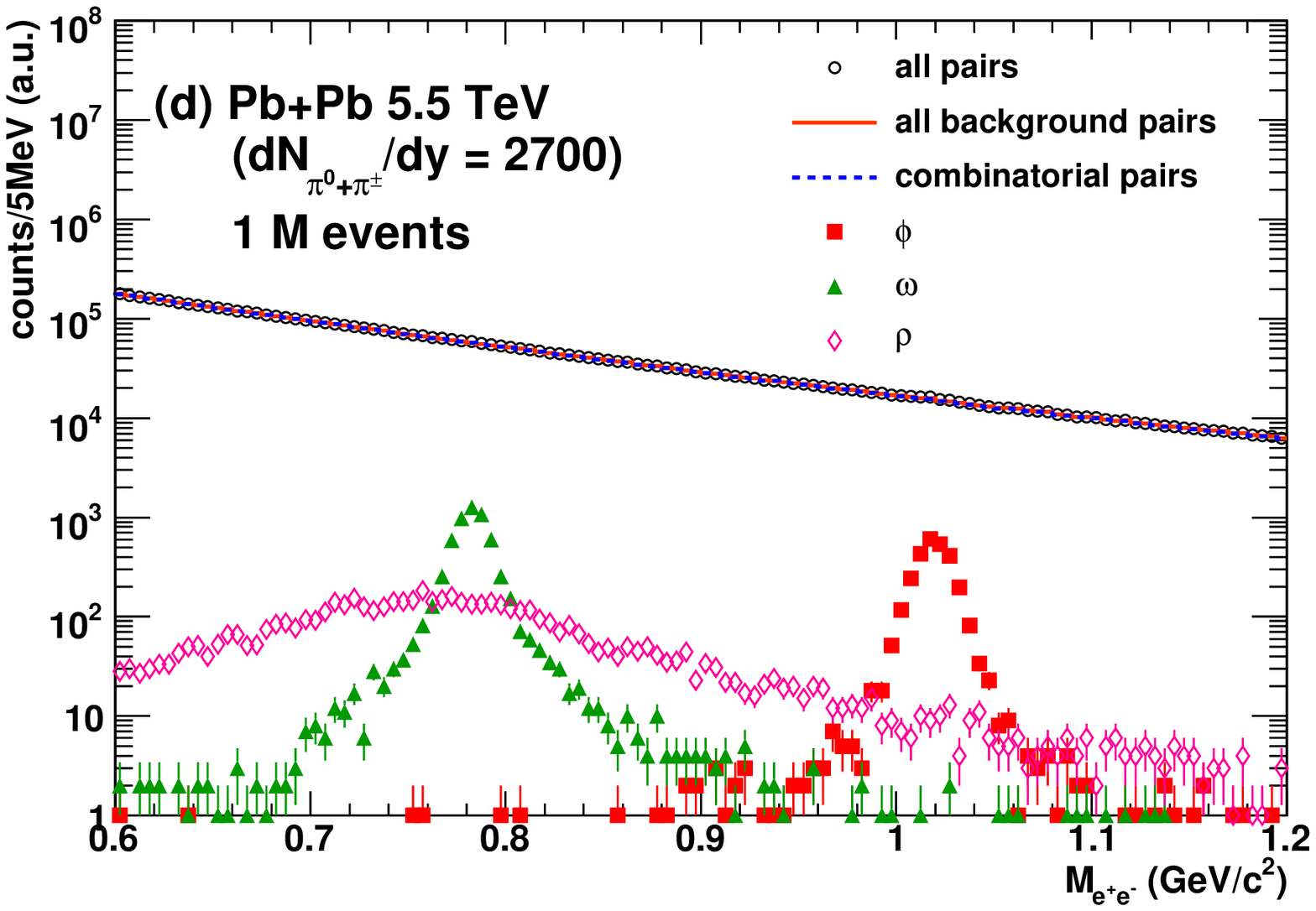}
\caption{\label{mass_signal}
(color online) The inclusive mass spectra compared to the components of signal pairs, all background pairs and combinatorial background pairs 
for the given $dN_{\pi^{0} + \pi^{\pm}}/dy$ with $P_{cnv}$ = 1 $\%$, $R_{\pi^{\pm}}$ = 500, $\epsilon_{acc}$ = 100 $\%$, $\epsilon_{tag}$ = 100 $\%$, $p_{T}^{th}$ = 0.1 GeV/c and $\sigma^{ref}_{p_{T}} = \sqrt{ \left(0.01 \cdot p_{T} \right)^{2} + \left( 0.0056 \right)^{2}}$ GeV/c.
The simulated number of events for each collision system is shown inside the plot.
}
\end{figure}

Figure \ref{mass_cocktail} shows the invariant mass distributions of di-electron pairs 
with $dN_{\pi^{0} + \pi^{\pm}}/dy$ = 3, 6, 1000, and 2700, respectively.
The components from individual di-electron sources
are indicated with different types of curves in the figure.
The mass shapes of $\phi$, $\omega$ and $\rho$ characterize their short lifetimes and show Breit-Wigner resonance peaks.
The invariant mass spectrum of photon-conversion pairs obeys dynamics of the pair-creation process in materials.
The invariant mass spectrum of Dalitz decaying pairs has a character whose leading edge is the summation of masses of decay products 
and the distribution continues up to their parent masses. 
The mass spectrum of $c\bar{c} \rightarrow e^{+}e^{-}$ is reconstructed by randomly pairing di-electrons in azimuth. \\
\quad The inclusive invariant mass spectra are shown in Fig.\ref{mass_signal}.
The component of signal pairs, combinatorial background pairs and all background pairs is superimposed in the figures.
The fraction of individual components depends on the given multiplicities, collision species and energies.
The peaks of the light vector mesons are clearly seen at the multiplicities in p+p collisions, but hardly seen above $dN_{\pi^{0} + \pi^{\pm}}/dy$ = 1000, 
though the statistical significance is not necessarily small.
The quantitative evaluations of the signal-to-background ratios and the statistical significance are discussed in the next section.  	
\section{signal-to-background ratios and the statistical significance}
\label{sect3}
The feasibility to measure $\phi/\omega/\rho \rightarrow e^{+}e^{-}$ is 
evaluated by the signal-to-background ratios and the statistical significance 
in the signal mass region.
The signal mass region for each meson is defined as the invariant mass range of $M_{\phi, \omega,\rho} \pm 3 \times \sqrt{\Gamma_{\phi, \omega, \rho}^{2} + \sigma_{\phi, \omega, \rho}^{2}} $, 
where $M_{\phi, \omega, \rho}$ is the mass center and $\Gamma_{\phi, \omega, \rho}$ is the decay width. 
$M_{\phi, \omega, \rho}$ and $\Gamma_{\phi, \omega, \rho}$ are cited from the particle data group \cite{PDG2010}. 
The mass resolutions $\sigma_{\phi, \omega, \rho}$ are calculated by the single particle simulation \footnote{
The mass resolutions are calculated as follows. 
$\phi,$ $\omega$ and $\rho$ mesons are singly generated under the condition that their mass widths are set at zero, respectively.
The mass distributions fluctuate around individual mass centers due to only the transverse momentum resolution $\sigma^{ref}_{p_{T}}$.
The mass resolutions $\sigma_{\phi, \omega, \rho}$ are estimated by the fits with the Gauss function.  
} 
and result in 7.6, 5.7 and 5.6 MeV/$c^{2}$ for $\phi$, $\omega$ and $\rho$ mesons, respectively.

\begin{figure}[!hbt]
	\includegraphics[scale=0.45]{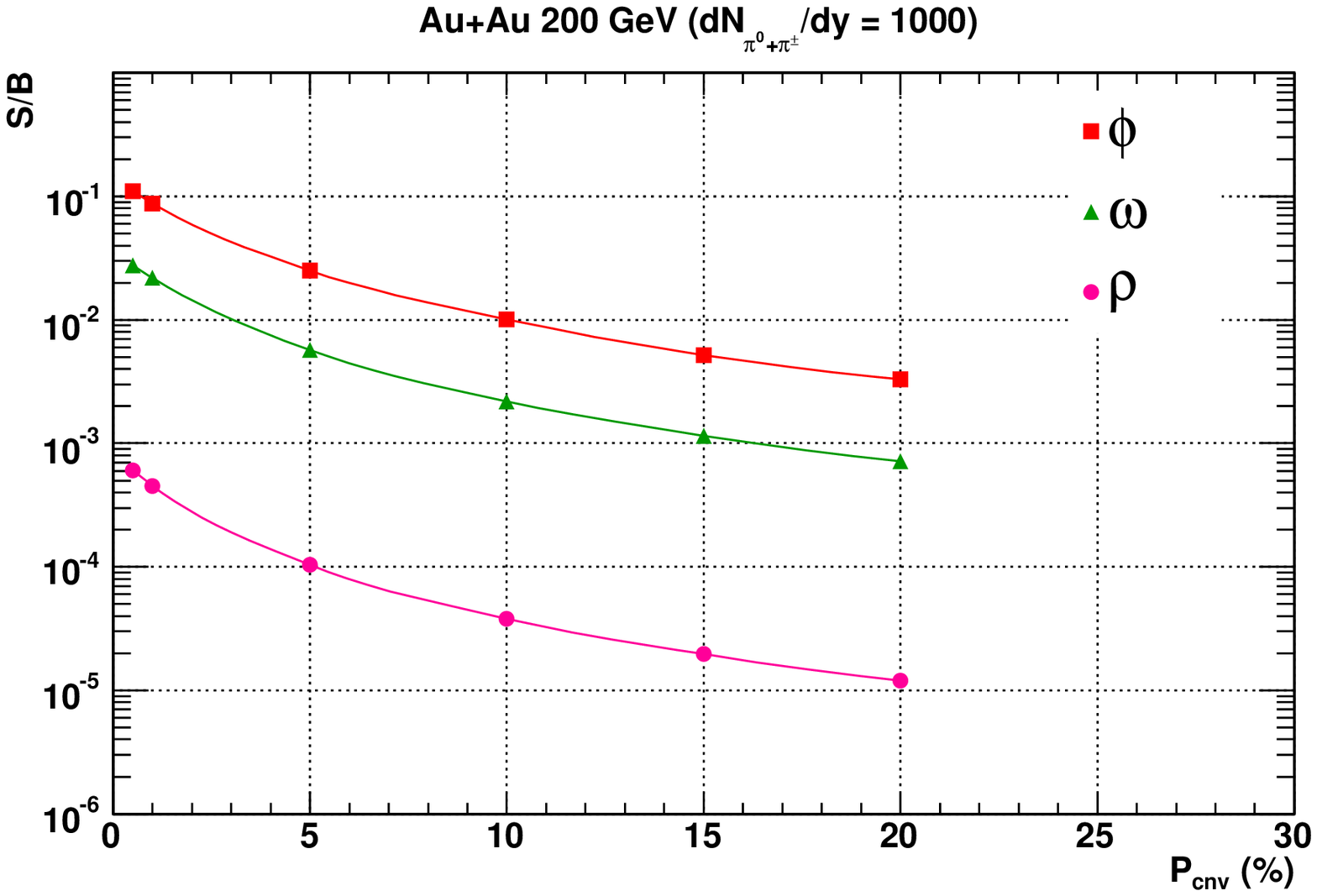}
	\includegraphics[scale=0.45]{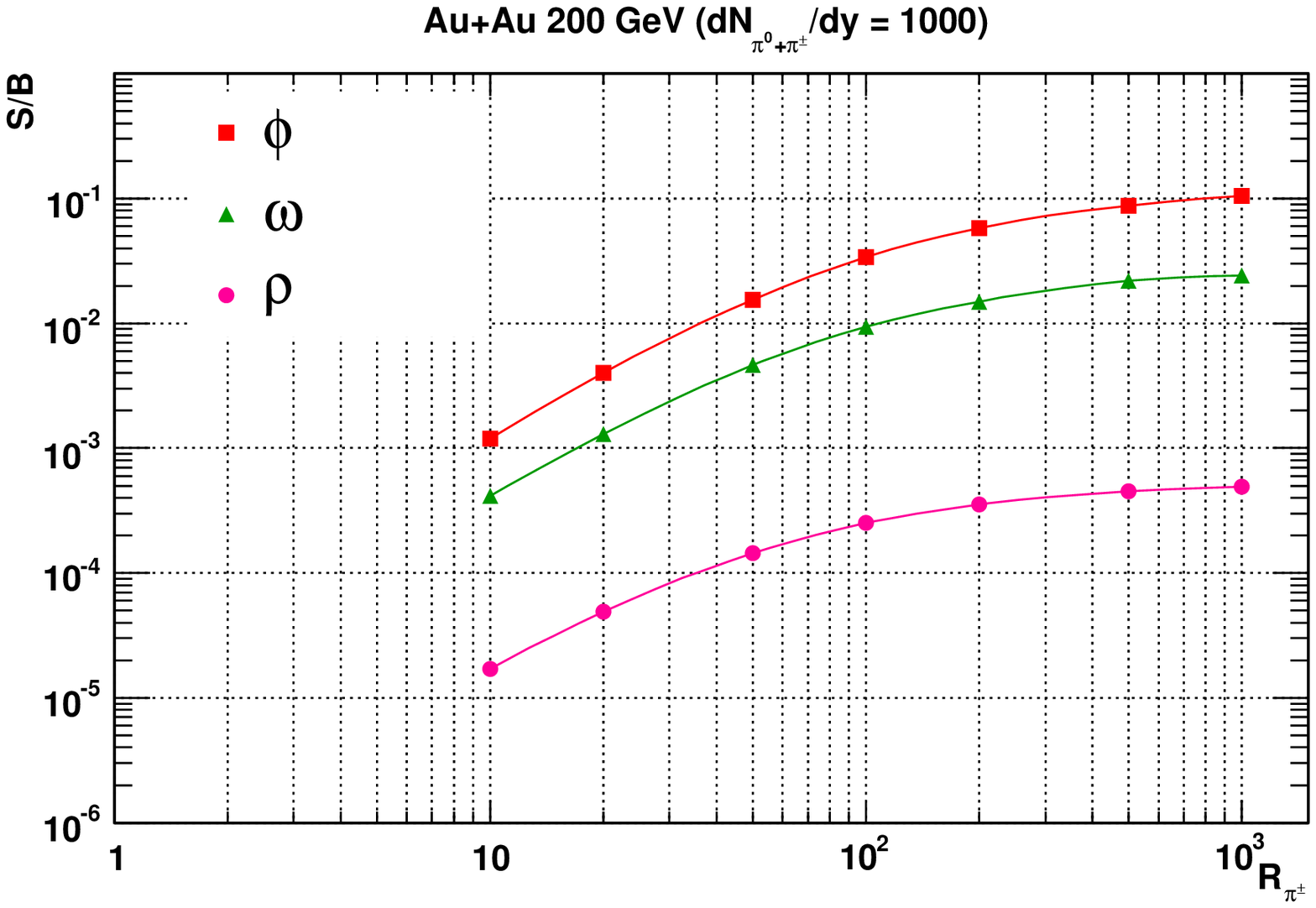}
	\includegraphics[scale=0.45]{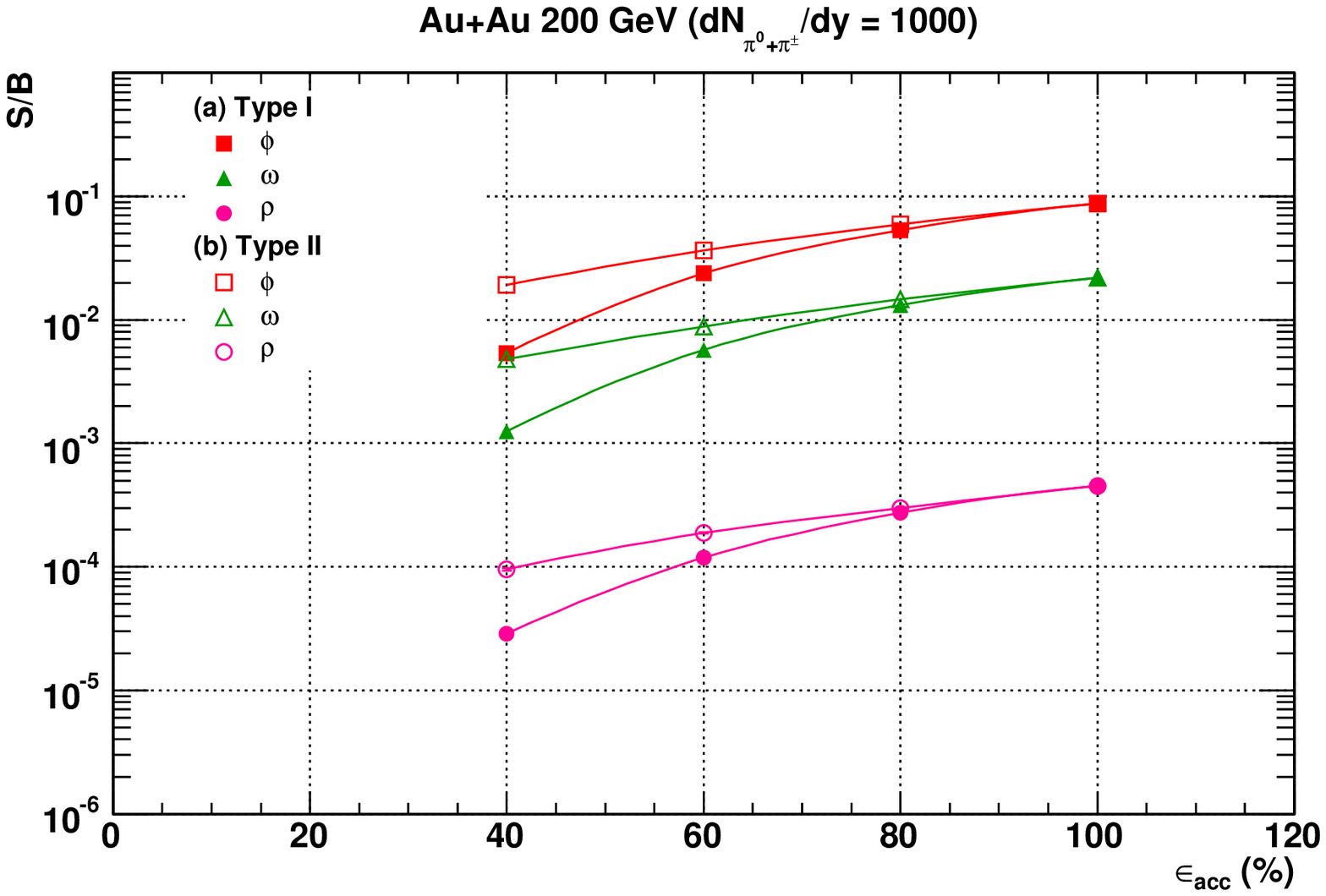}
\caption{\label{sn_mul1000} (color online) 
The signal-to-background ratio $S/B$ of $\phi$, $\omega$ and $\rho$ meson as a function of the experimental parameters $P_{cnv}$, $R_{\pi^{\pm}}$ and $\epsilon_{acc}$ in central Au+Au collisions at $\sqrt{s_{NN}}$ = 200 GeV ($dN_{\pi^{0} + \pi^{\pm}}/dy$ = 1000).
Only one parameter is changed by fixing the other parameters at the baseline values for each plot.
Type I on the bottom figure shows the azimuthal coverage of $0 \leq \phi \leq \phi_{1}$.
Type  I\hspace{-.1em}I shows two separated coverages of  $0 \leq \phi \leq \frac{\phi_{1}}{2}$ and $\pi \leq \phi  \leq \pi + \frac{\phi_{1}}{2}$.
}
\end{figure}

\begin{figure}[hbt]
	\includegraphics[scale=0.45]{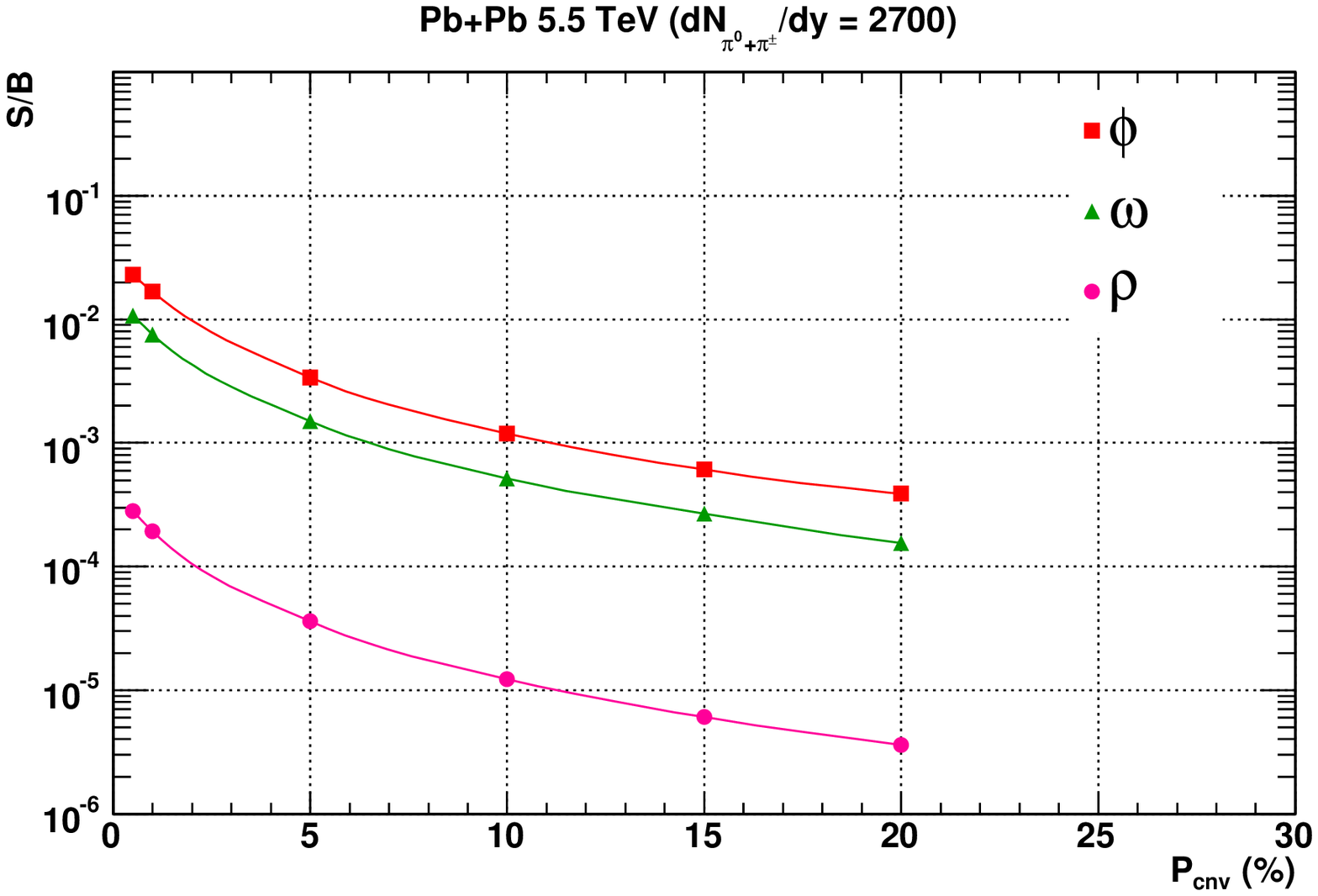}
	\includegraphics[scale=0.45]{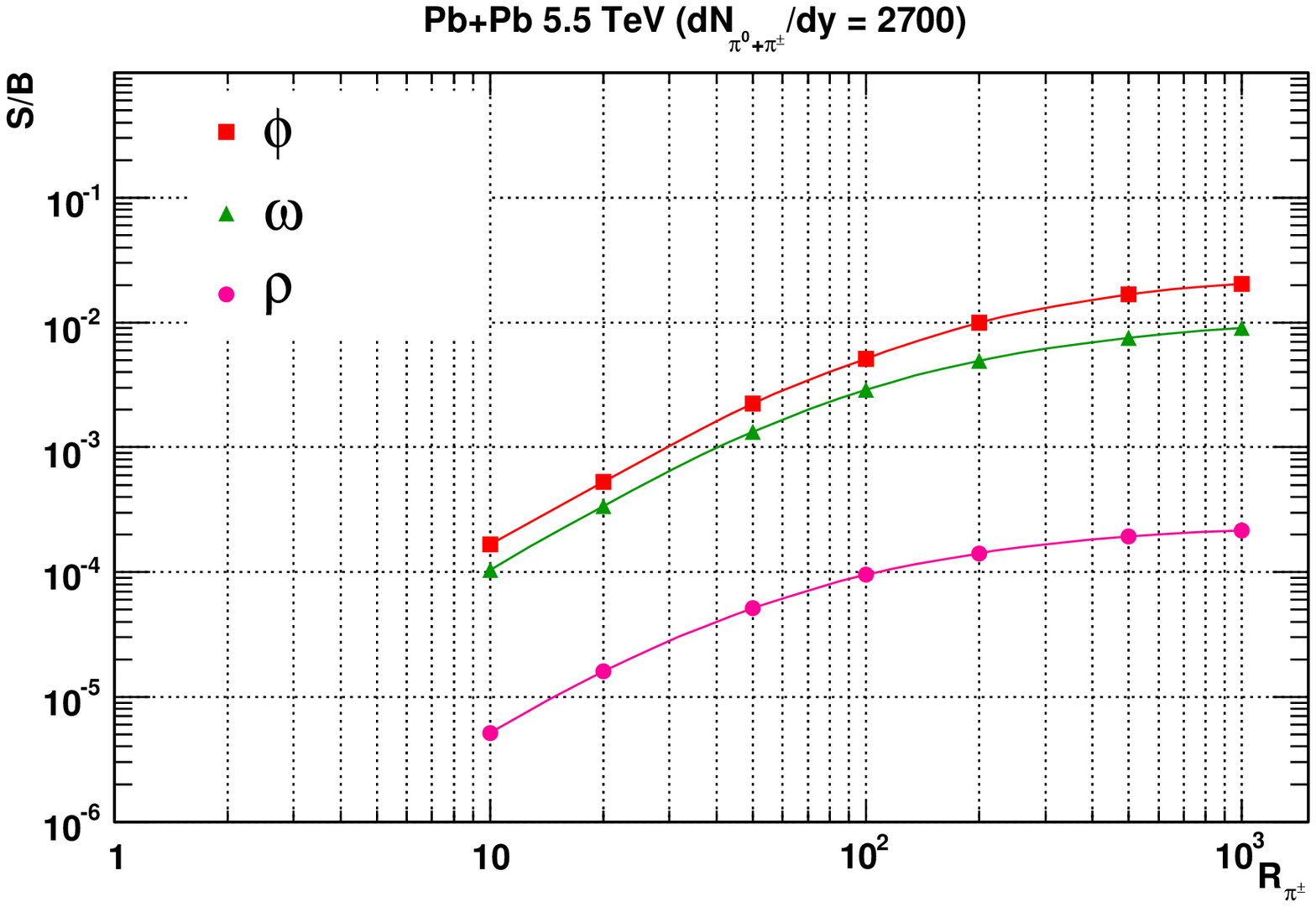}
	\includegraphics[scale=0.45]{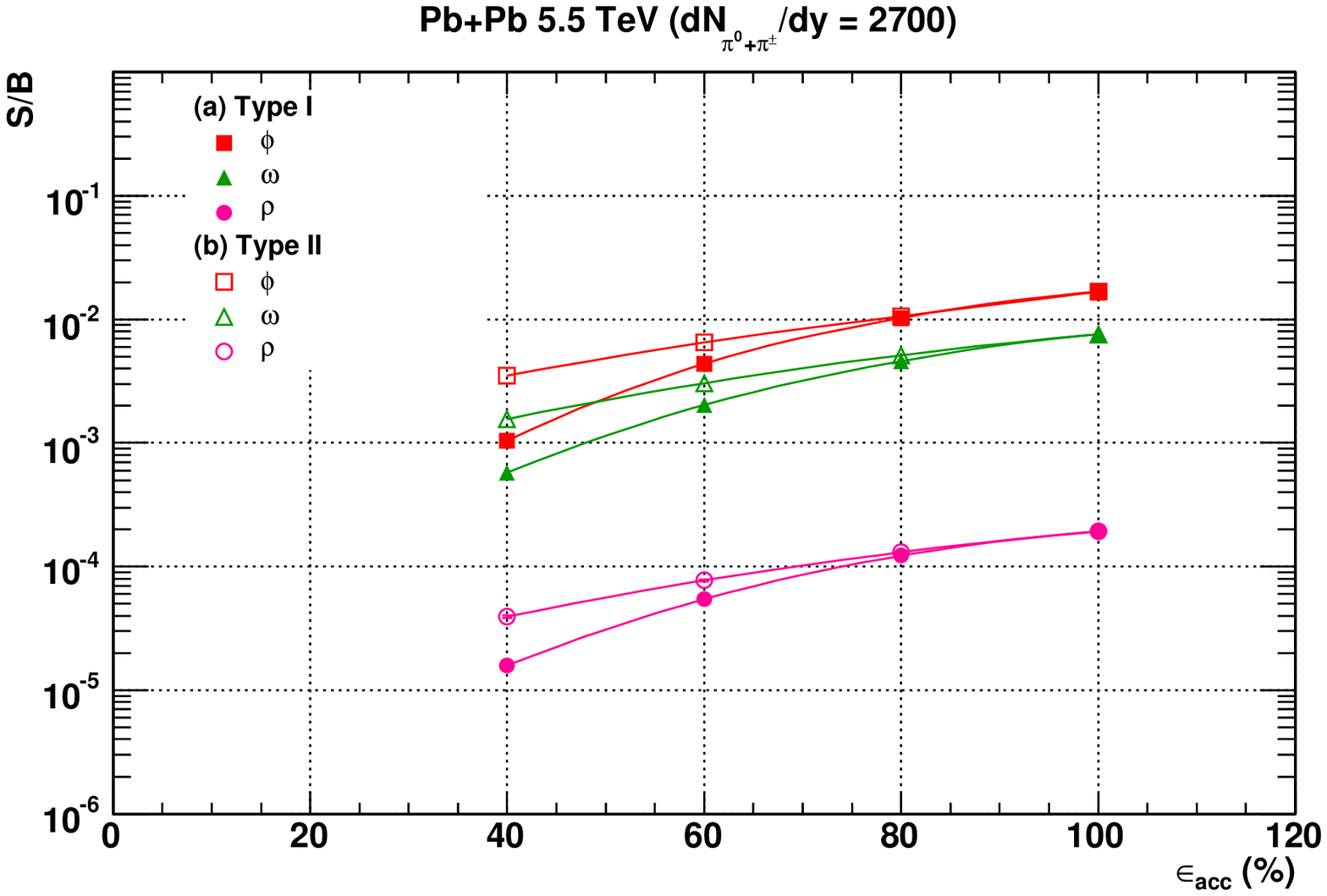}
\caption{\label{sn_mul2700} (color online) 
The signal-to-background ratio $S/B$ of $\phi$, $\omega$ and $\rho$ meson as a function of the experimental parameters 
$P_{cnv}$, $R_{\pi^{\pm}}$ and $\epsilon_{acc}$ in central Pb+Pb collisions at $\sqrt{s_{NN}}$ = 5.5 TeV ($dN_{\pi^{0} + \pi^{\pm}}/dy$ = 2700).
Only one parameter is changed by fixing the other parameters at the baseline values for each plot.
Type I of the bottom figure shows the azimuthal coverage of $0 \leq \phi \leq \phi_{1}$.
Type  I\hspace{-.1em}I shows two separated coverages of  $0 \leq \phi \leq \frac{\phi_{1}}{2}$ and $\pi \leq \phi  \leq \pi + \frac{\phi_{1}}{2}$.
}
\end{figure}

Figure \ref{sn_mul1000} and \ref{sn_mul2700} show the signal-to-background ratios $S/B$ as a function of the experimental parameters 
in central Au+Au collisions at $\sqrt{s_{NN}}$ = 200 GeV ($dN_{\pi^{0} + \pi^{\pm}}/dy$ = 1000) and 
central Pb+Pb collisions at $\sqrt{s_{NN}}$ = 5.5 TeV ($dN_{\pi^{0} + \pi^{\pm}}/dy$ = 2700), respectively.
Only one parameter is changed by fixing the other parameters at the baseline values:  $P_{cnv}$ = 1 $\%$, $R_{\pi^{\pm}}$ = 500, $\epsilon_{acc}$ = 100 $\%$, 
$\epsilon_{tag}$ = 100 $\%$, $p_{T}^{th}$ = 0.1 GeV/c and $\sigma^{ref}_{p_{T}} = \sqrt{ \left(0.01 \cdot p_{T} \right)^{2} + \left( 0.0056 \right)^{2}}$ GeV/c. 
The top-left figure shows the $S/B$ as a function of photon-conversion probability $P_{cnv}$.
The minimum amount of detector materials typically corresponds to $P_{cnv}$ = 1-2 $\%$, because photon conversions from the beam pipe 
and the first layer of the innermost detector are unavoidable in any detector system, even though electron trajectories coming from the off-axis point are rejected by tracking algorithm.
Thus the tendency below $P_{cnv}$ = 10 $\%$ is important for the detector system with typical amount of the materials.

The dependence on the rejection factor of charged pions $R_{\pi^{\pm}}$ is shown in the top-right plot of Fig.\ref{sn_mul1000} and \ref{sn_mul2700}.
Typical devices for the electron identification have the rejection factor of a few hundreds in the stand-alone operation 
\cite{RICH, continum2, ALICE_detector, ATLAS_detector}, although it varies by the principle of detection. 
Therefore, the information in the range of $R_{\pi^{\pm}}$ = 100-1000 are useful. 
The $S/B$ can be changed by a factor of 3-5 for $\phi/\omega$ meson in this range.

The bottom-left figure shows the $S/B$ as a function of the azimuthal acceptance $\epsilon_{acc}$.
The $S/B$ depends on decay kinematics of the signal particles and the backgrounds.
Therefore the geometrical configuration in azimuthal coverage as well as the absolute acceptance in azimuth should be taken into account.
Two types of geometrical configurations are considered in this simulation. 
Type I simply covers the azimuthal range of $0 \leq \phi \leq \phi_{1}$.
Type  I\hspace{-.1em}I covers two separated domains which are symmetrically arranged in azimuth with respect to the collision point, that is, 
the coverage is set to $0 \leq \phi \leq \frac{\phi_{1}}{2}$ and $\pi \leq \phi  \leq \pi + \frac{\phi_{1}}{2}$.
Both of them have the same total acceptance in azimuth with different geometry.
The difference between the two geometrical configurations increases in the case of the imperfect coverage.
If $\epsilon_{acc}$ is 40 $\%$, for instance, the $S/B$ differs by a factor of 3-4 for $\phi/\omega$ meson depending on the detector geometry. 
  
\begin{figure*}[!h]
\begin{center}
\begin{tabular}{c}
	\begin{minipage}{0.5\hsize}
	\begin{center}
	\includegraphics[scale=0.39]{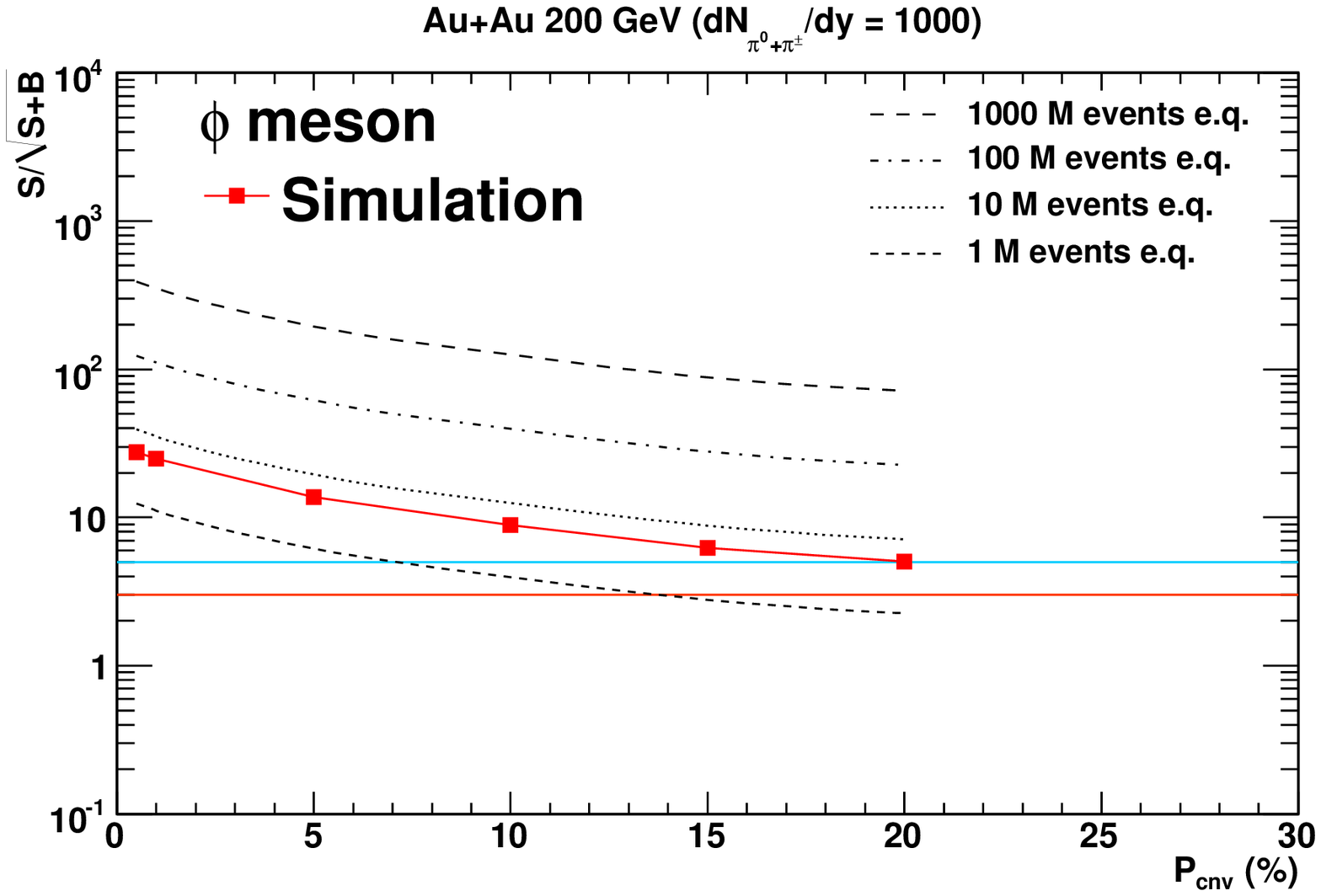}
	\end{center}
	\end{minipage}
	\begin{minipage}{0.5\hsize}
	\begin{center}
	\includegraphics[scale=0.39]{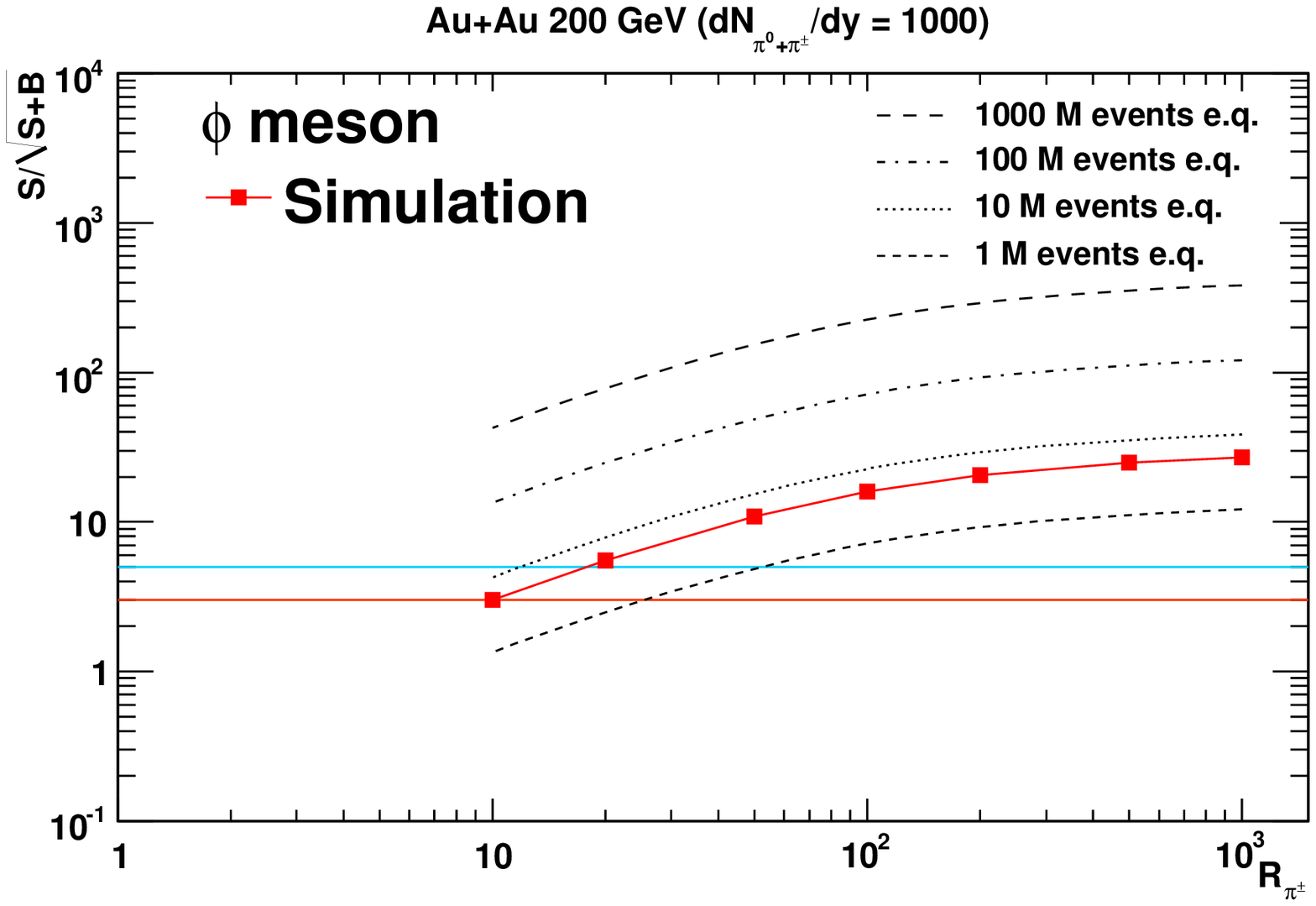}
	\end{center}
	\end{minipage}
\end{tabular}
\begin{tabular}{c}
	\begin{minipage}{0.5\hsize}
	\begin{center}
	\includegraphics[scale=0.39]{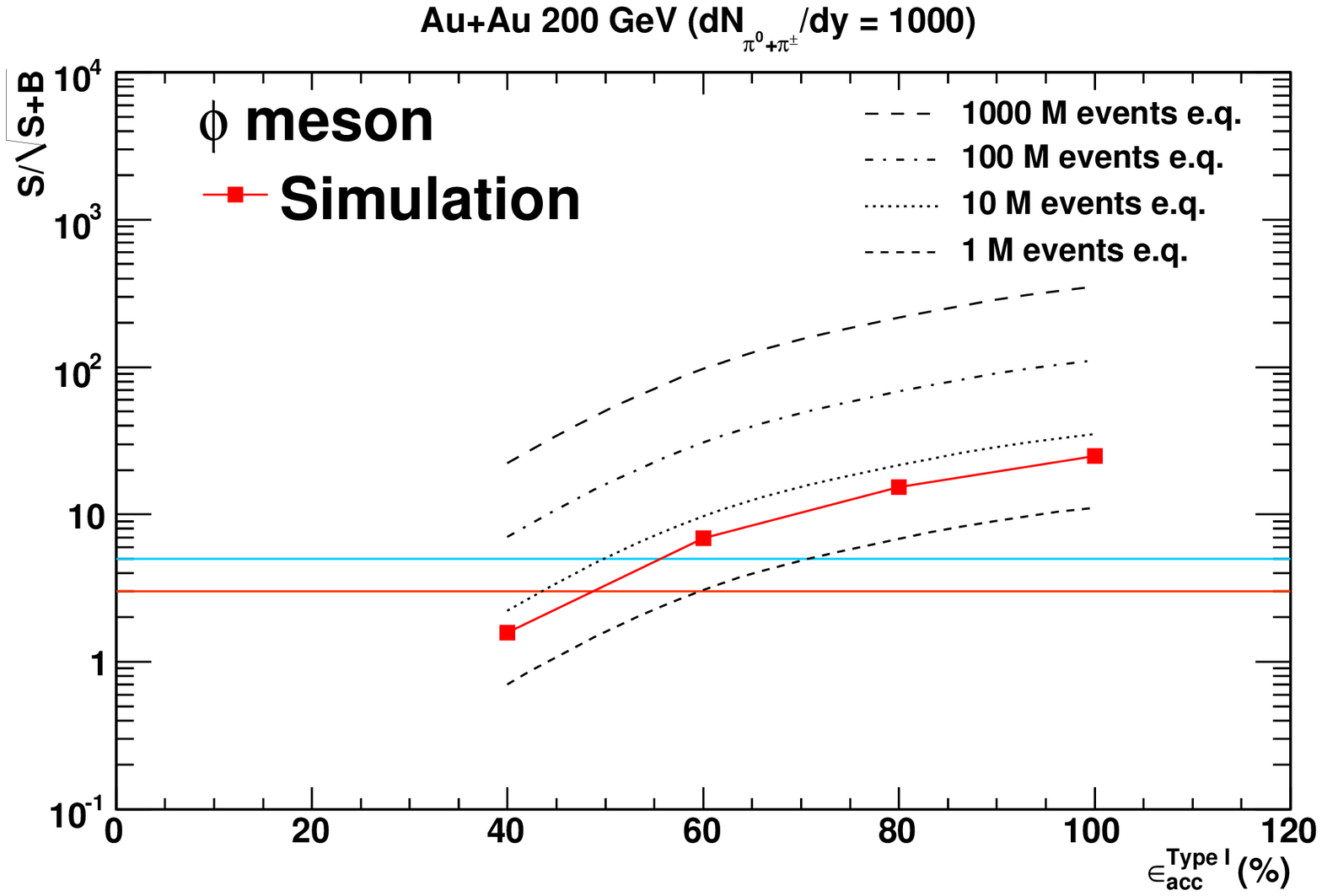}
	\end{center}
	\end{minipage}
	\begin{minipage}{0.5\hsize}
	\begin{center}
	\includegraphics[scale=0.39]{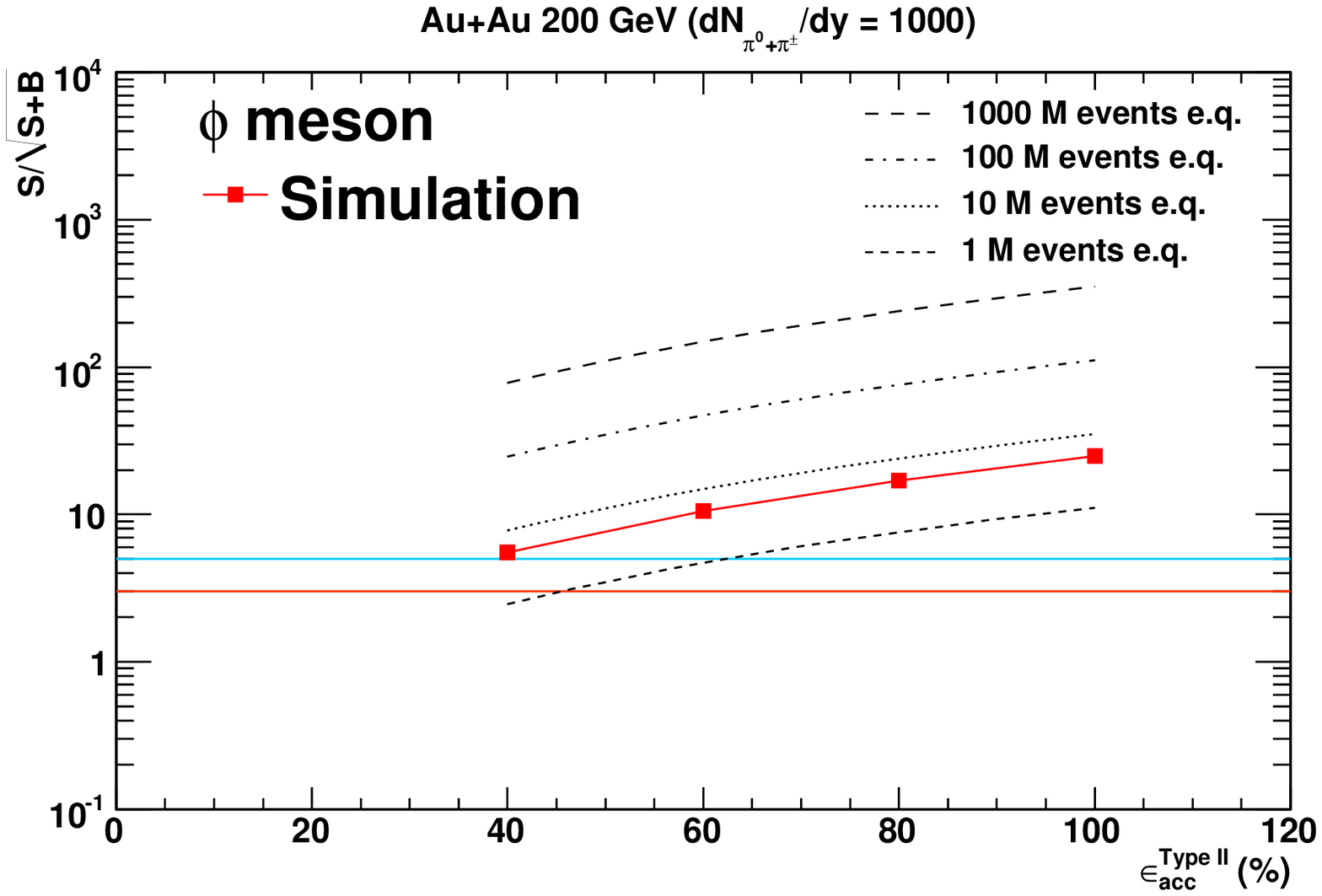}
	\end{center}
	\end{minipage}
\end{tabular}
\begin{tabular}{c}
	\begin{minipage}{0.5\hsize}
	\begin{center}
	\includegraphics[scale=0.39]{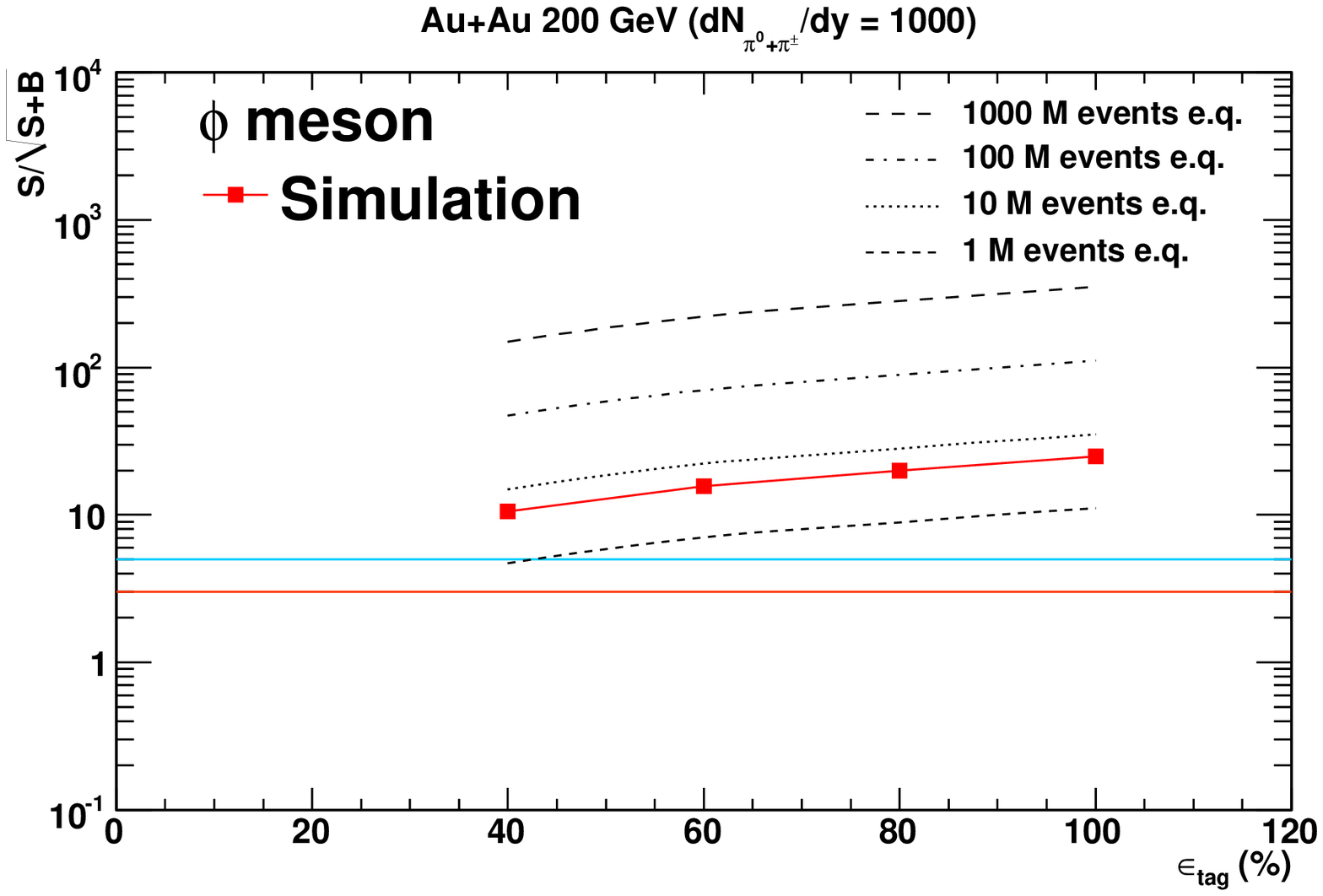}
	\end{center}
	\end{minipage}
	\begin{minipage}{0.5\hsize}
	\begin{center}
	\end{center}
	\end{minipage}
\end{tabular}
\caption{\label{phi_sg_mul1000} (color online)
The statistical significance $S/\sqrt{S+B}$ of $\phi$ mesons as a function of the experimental parameters $P_{cnv}$, $R_{\pi^{\pm}}$, $\epsilon_{acc}$ and $\epsilon_{tag}$ in central Au+Au collisions at $\sqrt{s_{NN}}$ = 200 GeV ($dN_{\pi^{0} + \pi^{\pm}}/dy$ = 1000).
Only one parameter is changed by fixing the other parameters at the baseline values for each plot.
The results of the simulation are shown as the symbols and the empirical curves are superimposed on the data points as the solid curves.
The other dotted curves are the scaled curves with the square root of the expected number of events found in the highest centrality class.
Two horizontal lines indicate $S/\sqrt{S+B}$ = 3 and 5.
 }
\end{center}
\end{figure*}

\begin{figure*}[!h]
\begin{center}
\begin{tabular}{c}
	\begin{minipage}{0.5\hsize}
	\begin{center}
	\includegraphics[scale=0.39]{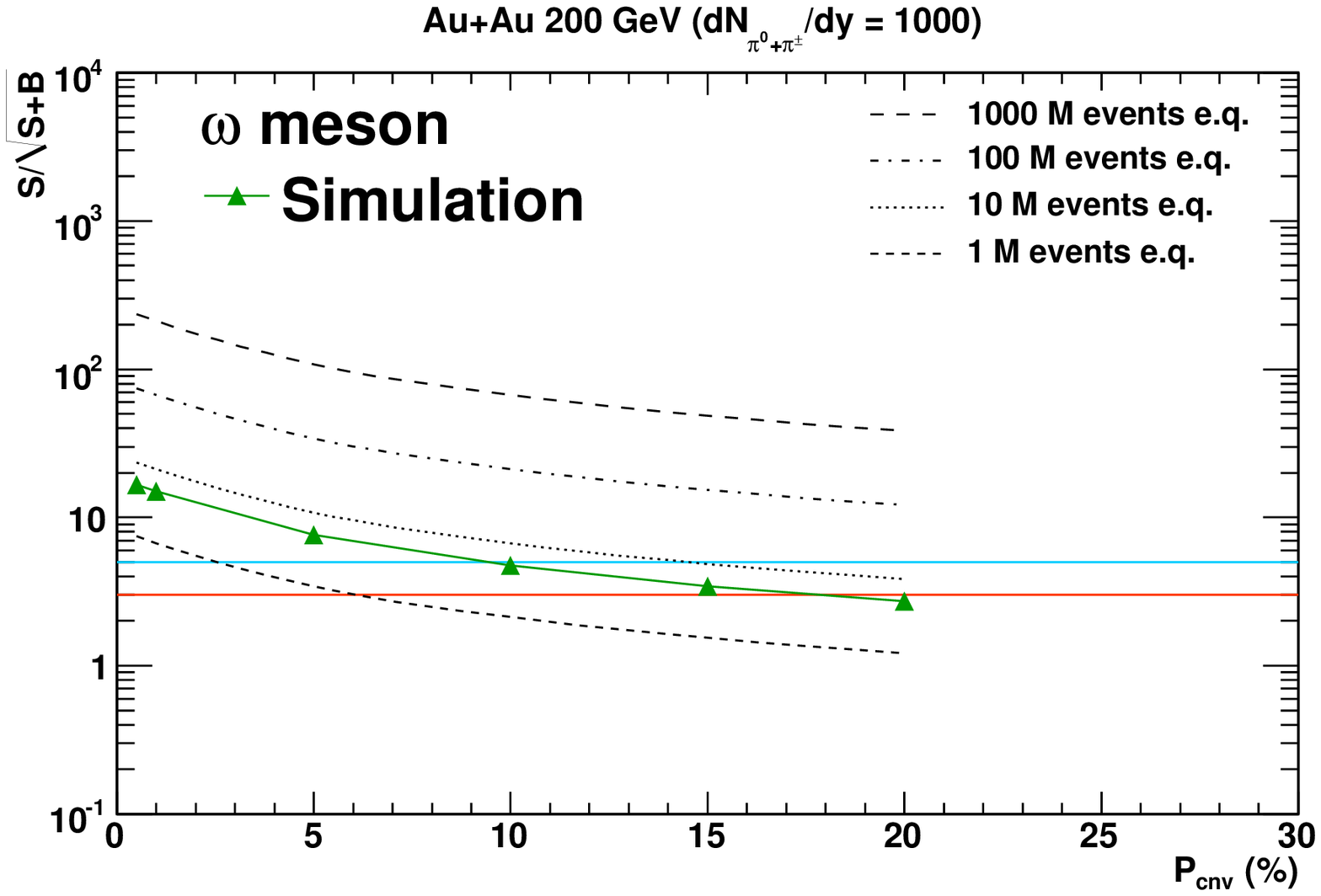}
	\end{center}
	\end{minipage}
	\begin{minipage}{0.5\hsize}
	\begin{center}
	\includegraphics[scale=0.39]{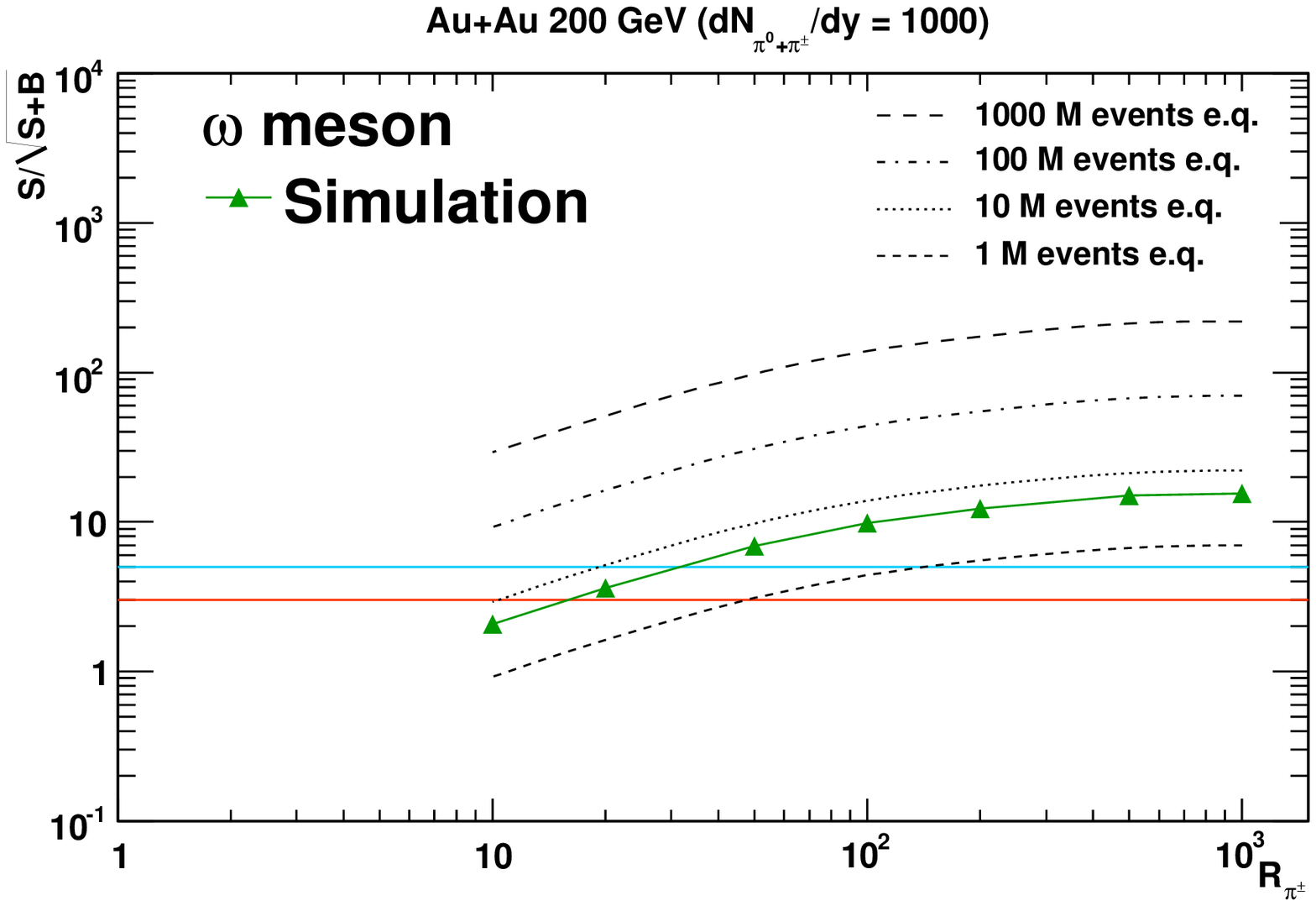}
	\end{center}
	\end{minipage}
\end{tabular}
\begin{tabular}{c}
	\begin{minipage}{0.5\hsize}
	\begin{center}
	\includegraphics[scale=0.39]{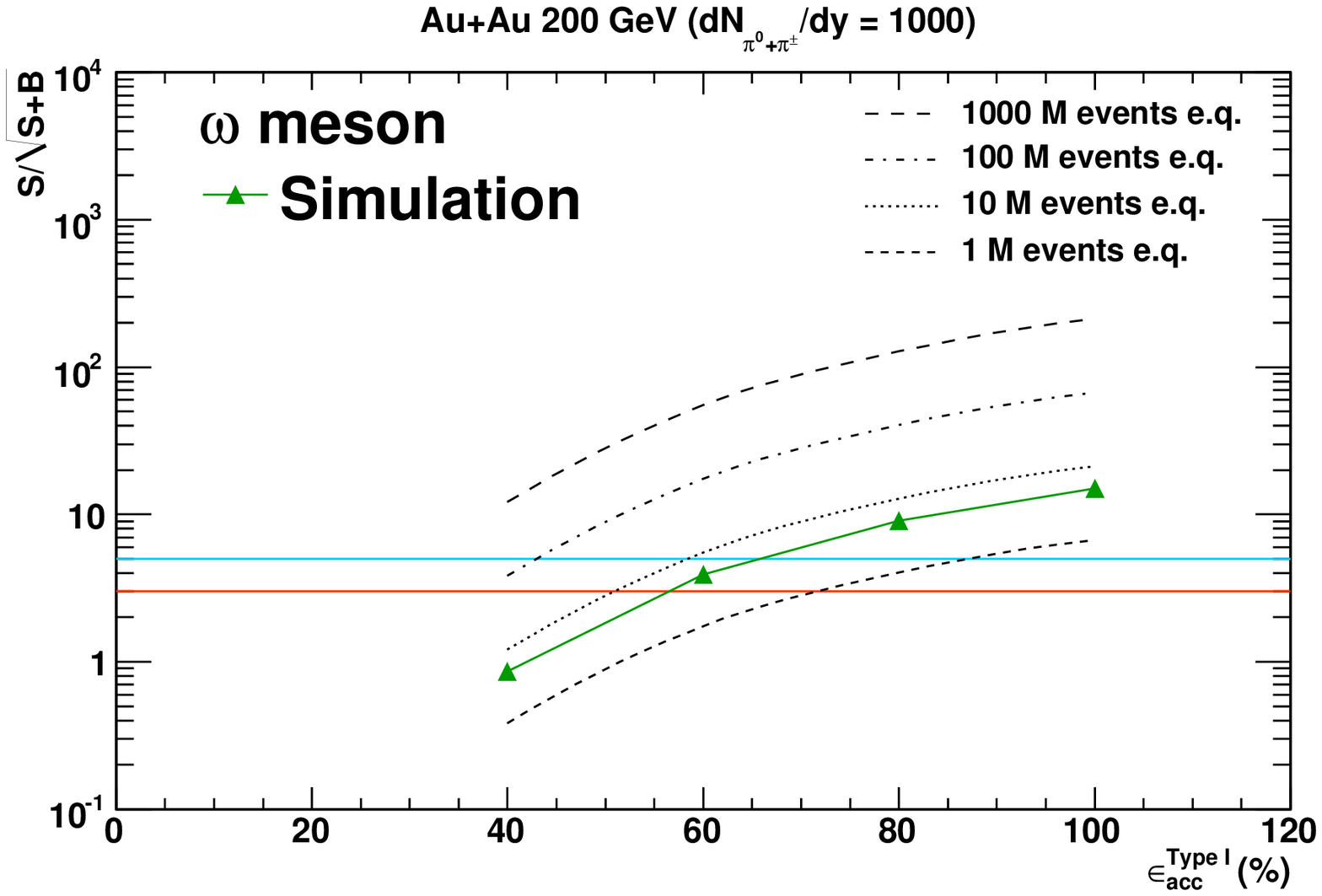}
	\end{center}
	\end{minipage}
	\begin{minipage}{0.5\hsize}
	\begin{center}
	\includegraphics[scale=0.39]{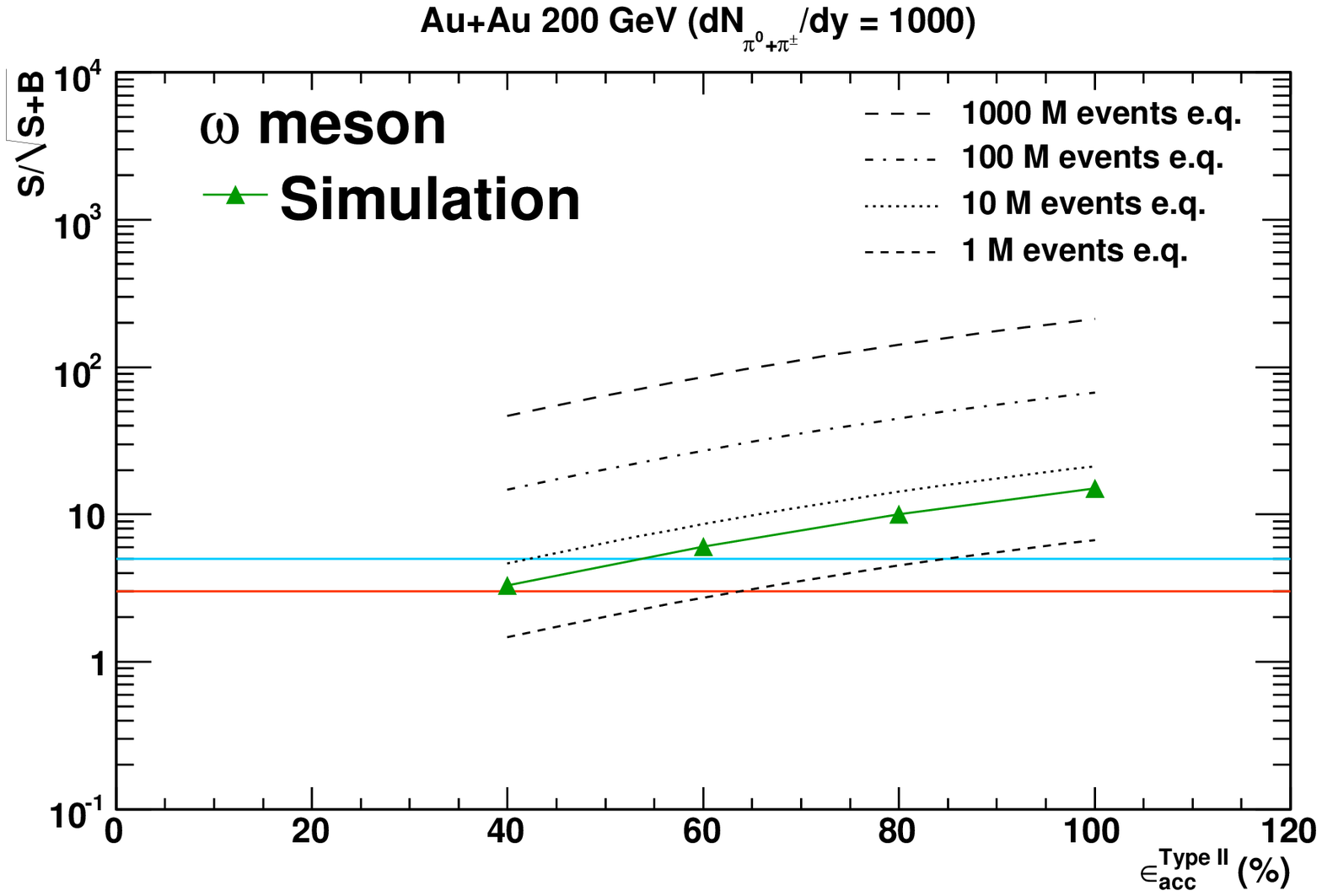}
	\end{center}
	\end{minipage}
\end{tabular}
\begin{tabular}{c}
	\begin{minipage}{0.5\hsize}
	\begin{center}
	\includegraphics[scale=0.39]{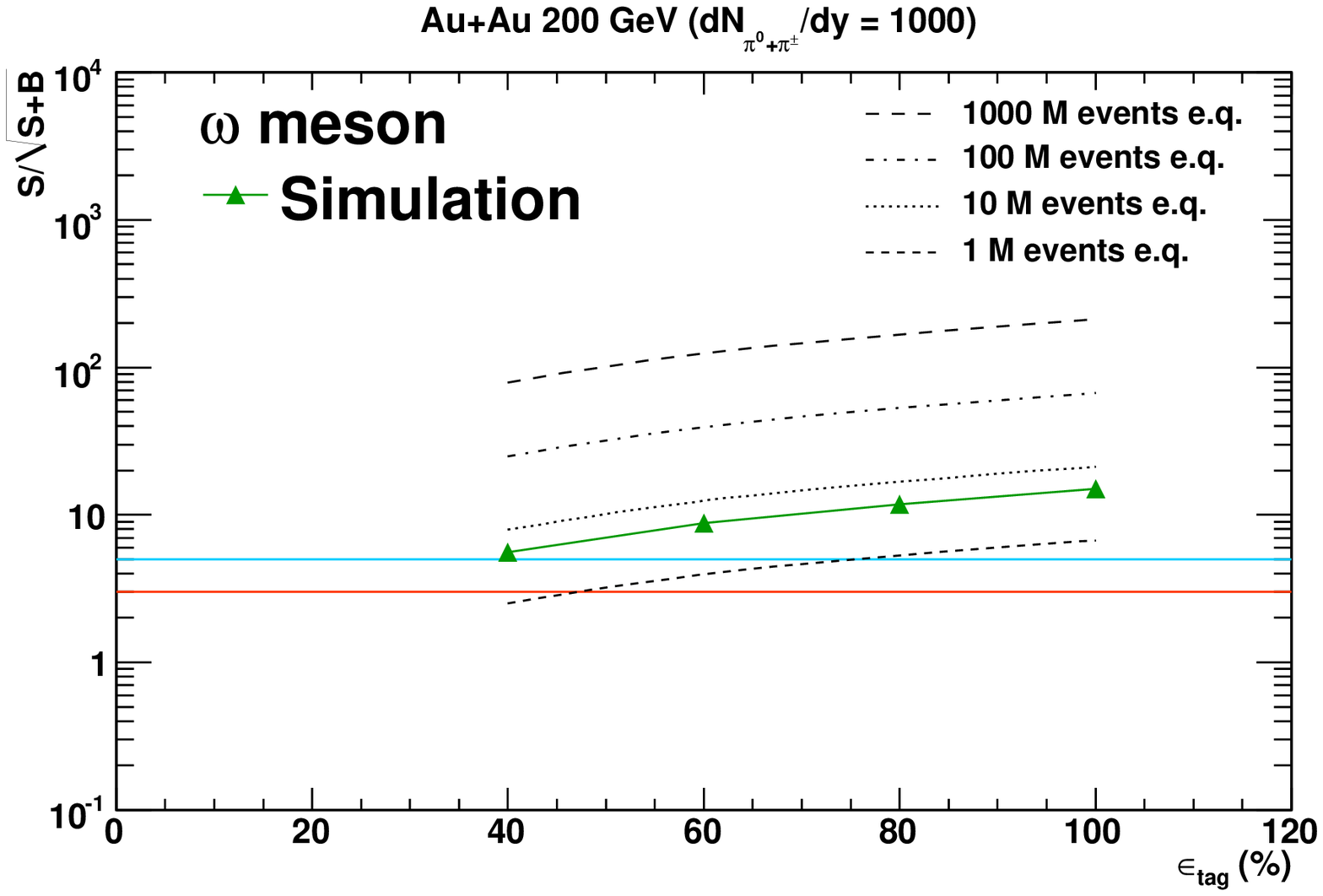}
	\end{center}
	\end{minipage}
	\begin{minipage}{0.5\hsize}
	\begin{center}
	\end{center}
	\end{minipage}
\end{tabular}
\caption{\label{ome_sg_mul1000} (color online)
The statistical significance $S/\sqrt{S+B}$ of $\omega$ mesons as a function of the experimental parameters $P_{cnv}$, $R_{\pi^{\pm}}$, $\epsilon_{acc}$ and $\epsilon_{tag}$ in central Au+Au collisions at $\sqrt{s_{NN}}$ = 200 GeV ($dN_{\pi^{0} + \pi^{\pm}}/dy$ = 1000).
Only one parameter is changed by fixing the other parameters at the baseline values for each plot.
The results of the simulation are shown as the symbols and the empirical curves are superimposed on the data points as the solid curves.
The other dotted curves are the scaled curves with the square root of the expected number of events found in the highest centrality class.
Two horizontal lines indicate $S/\sqrt{S+B}$ = 3 and 5.
}
\end{center}
\end{figure*}

\begin{figure*}[!h]
\begin{center}
\begin{tabular}{c}
	\begin{minipage}{0.5\hsize}
	\begin{center}
	\includegraphics[scale=0.39]{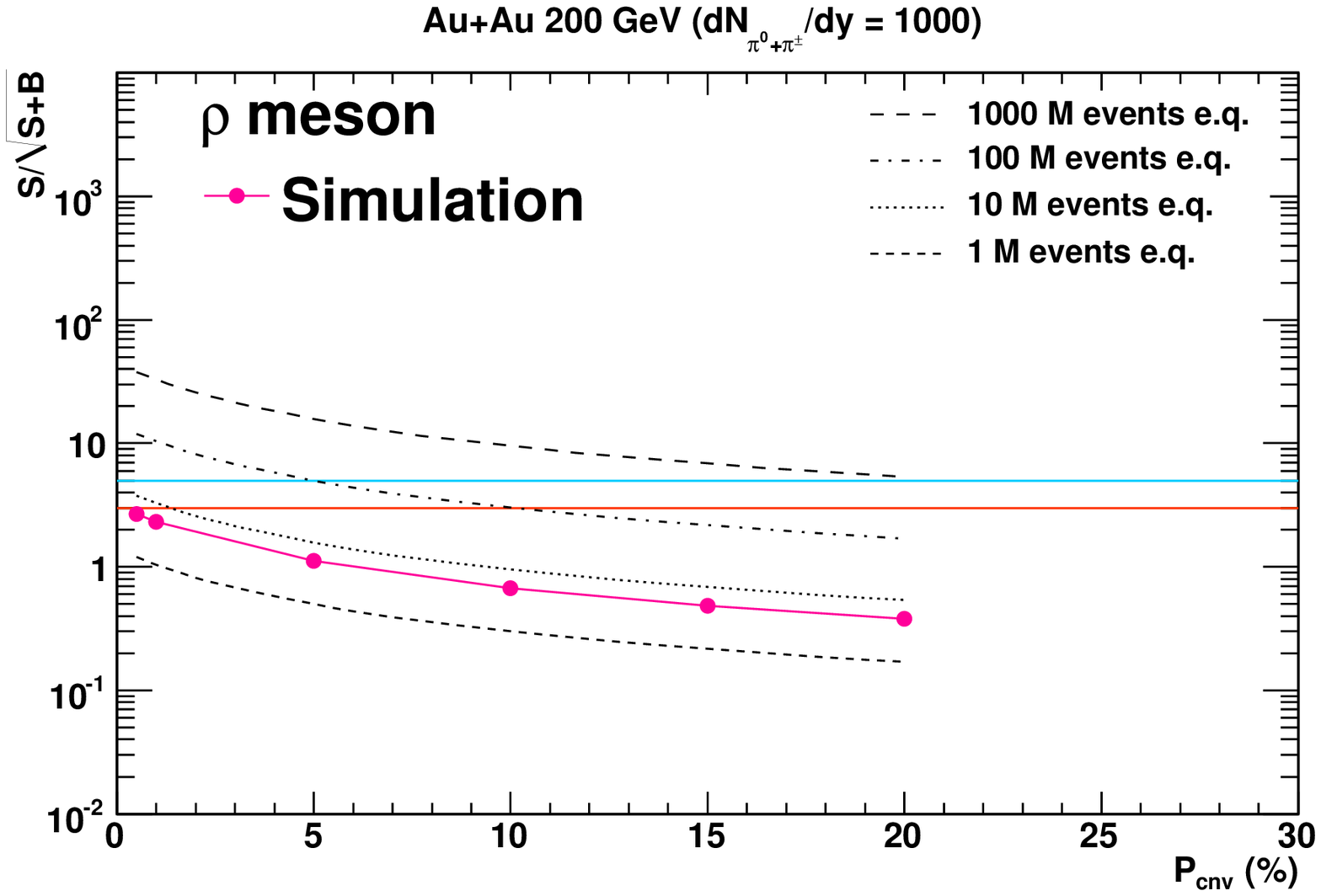}
	\end{center}
	\end{minipage}
	\begin{minipage}{0.5\hsize}
	\begin{center}
	\includegraphics[scale=0.39]{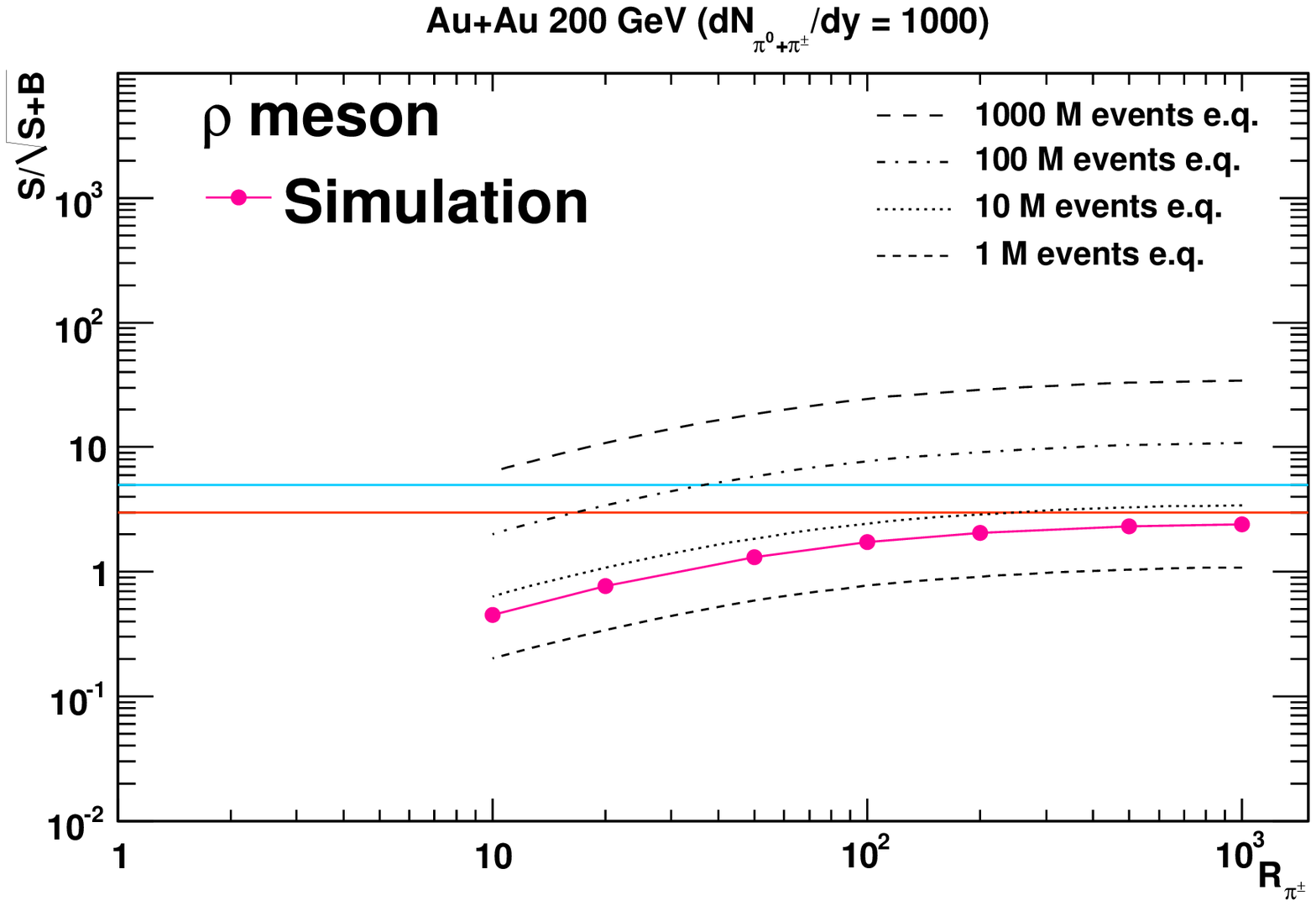}
	\end{center}
	\end{minipage}
\end{tabular}
\begin{tabular}{c}
	\begin{minipage}{0.5\hsize}
	\begin{center}
	\includegraphics[scale=0.39]{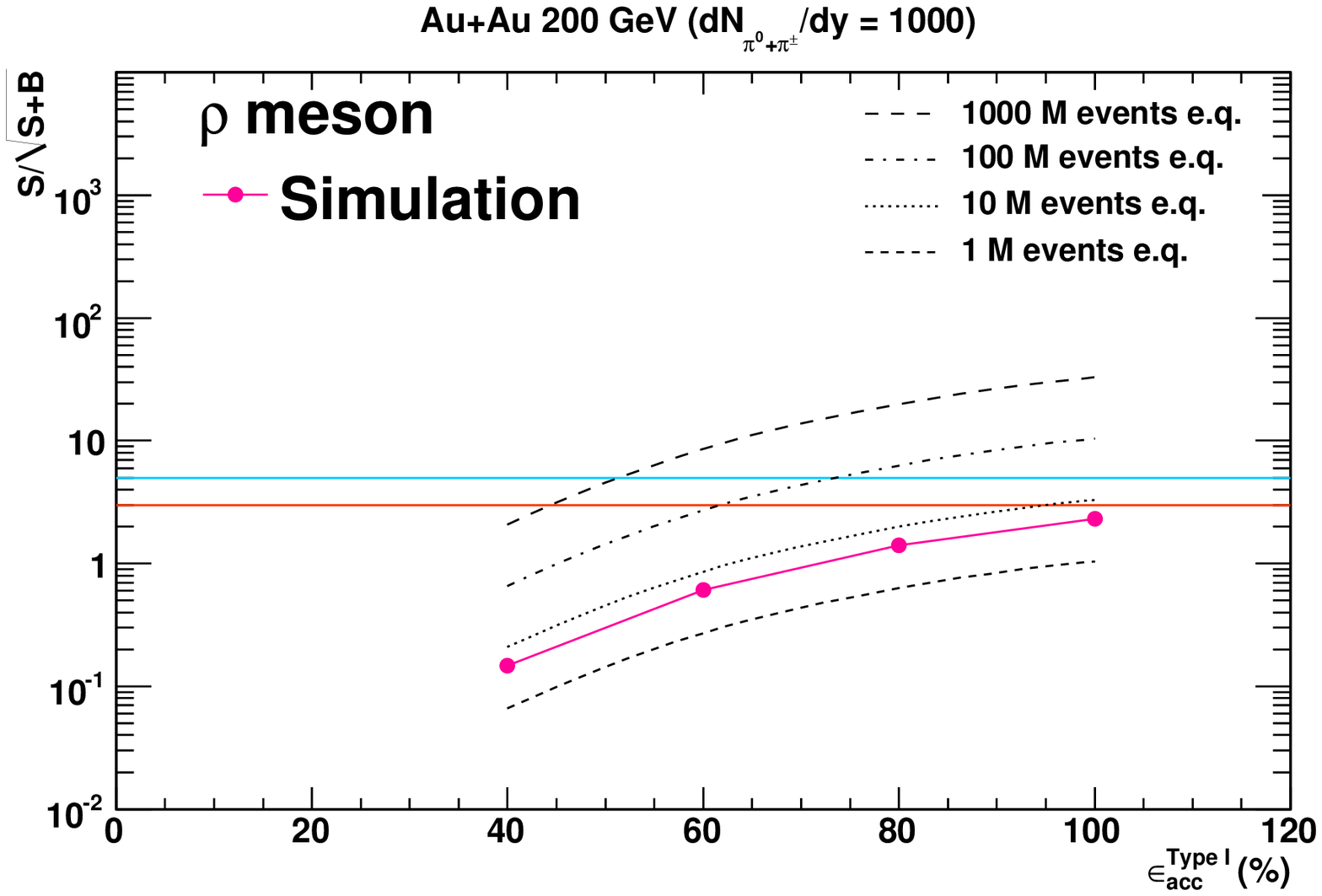}
	\end{center}
	\end{minipage}
	\begin{minipage}{0.5\hsize}
	\begin{center}
	\includegraphics[scale=0.39]{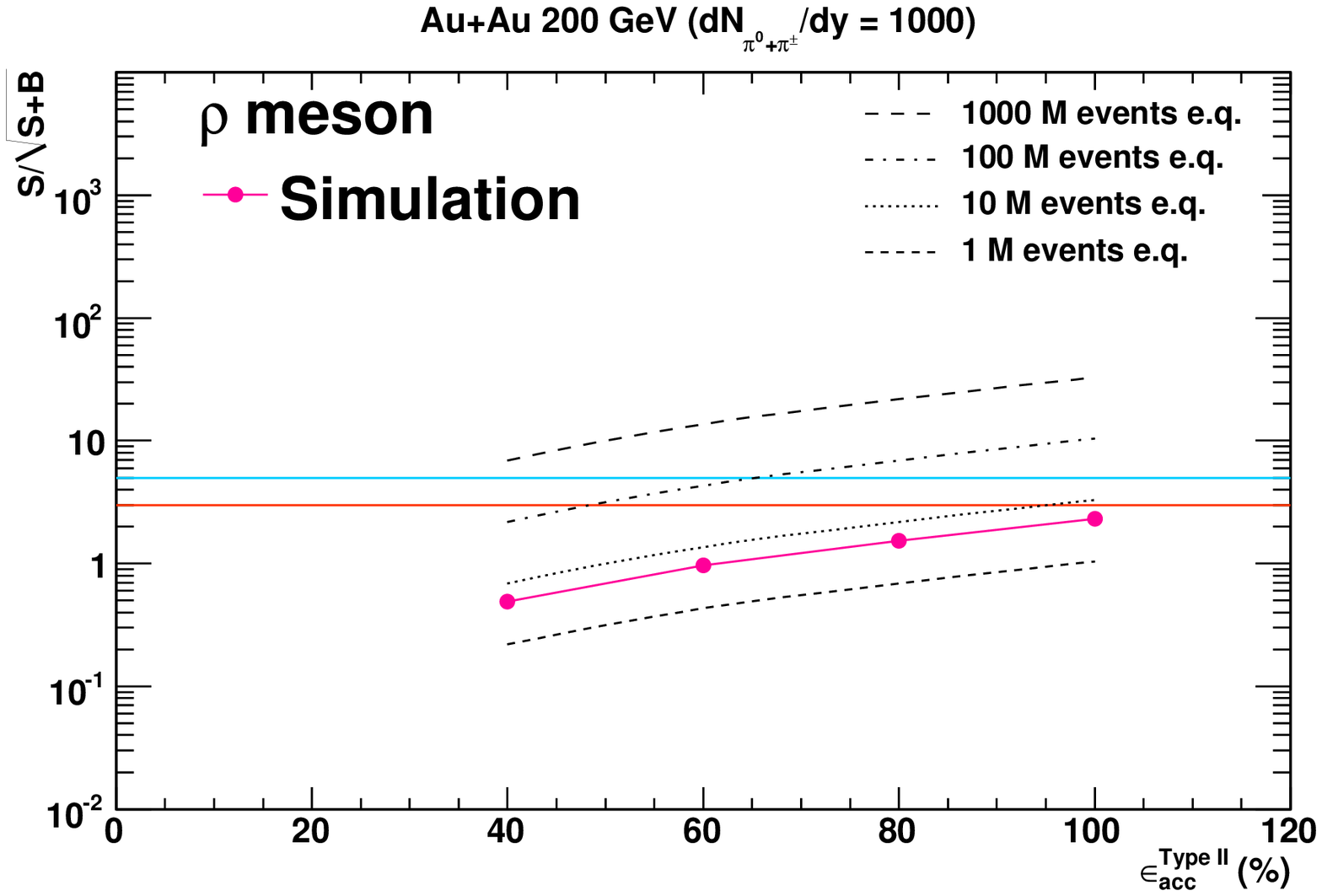}
	\end{center}
	\end{minipage}
\end{tabular}
\begin{tabular}{c}
	\begin{minipage}{0.5\hsize}
	\begin{center}
	\includegraphics[scale=0.39]{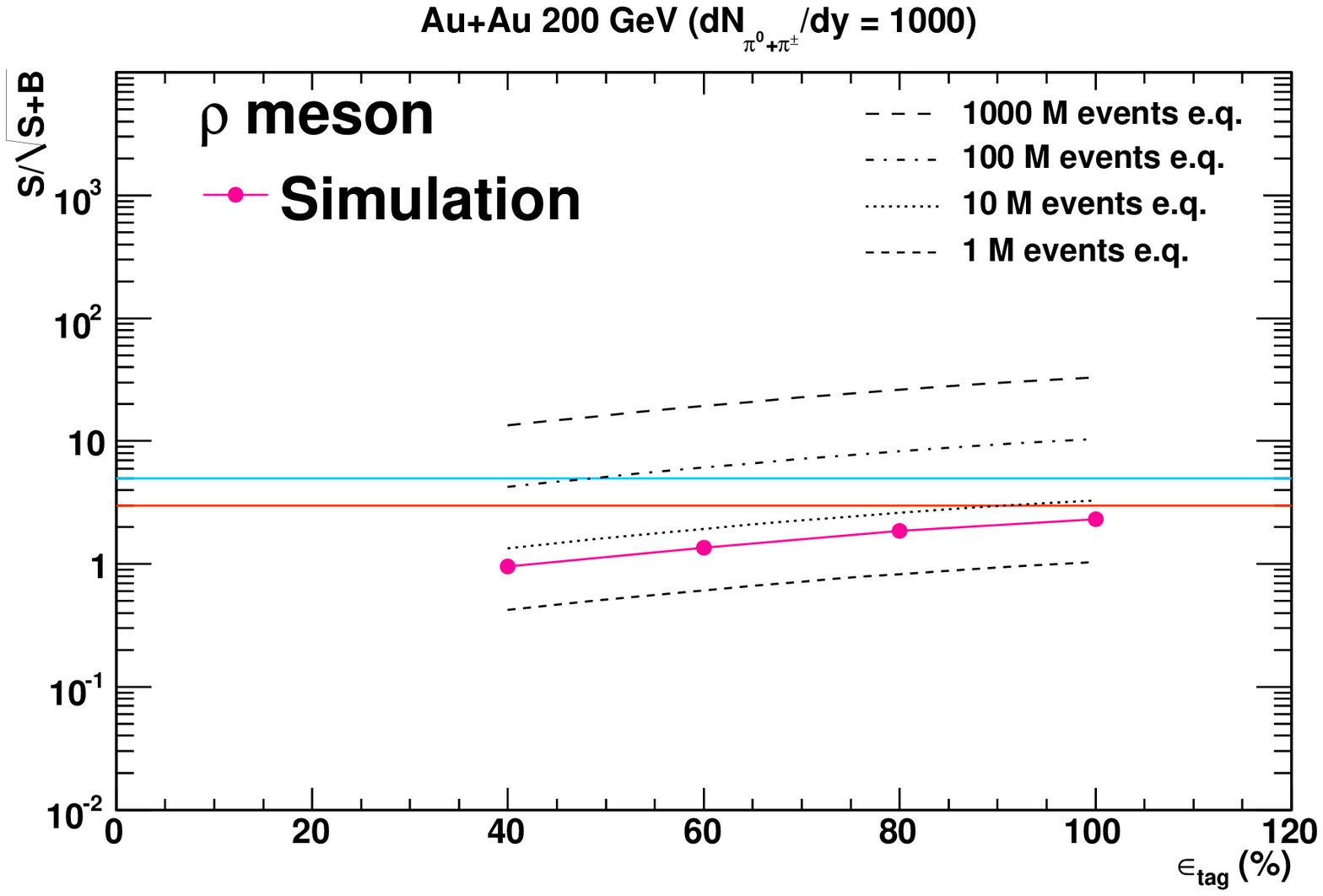}
	\end{center}
	\end{minipage}
	\begin{minipage}{0.5\hsize}
	\begin{center}
	\end{center}
	\end{minipage}
\end{tabular}
\caption{\label{rho_sg_mul1000} (color online)
The statistical significance $S/\sqrt{S+B}$ of $\rho$ mesons as a function of the experimental parameters $P_{cnv}$, $R_{\pi^{\pm}}$, $\epsilon_{acc}$ and $\epsilon_{tag}$ in central Au+Au collisions at $\sqrt{s_{NN}}$ = 200 GeV ($dN_{\pi^{0} + \pi^{\pm}}/dy$ = 1000).
Only one parameter is changed by fixing the other parameters at the baseline values for each plot.
The results of the simulation are shown as the symbols and the empirical curves are superimposed on the data points as the solid curves.
The other dotted curves are the scaled curves with the square root of the expected number of events found in the highest centrality class.
Two horizontal lines indicate $S/\sqrt{S+B}$ = 3 and 5.
}
\end{center}
\end{figure*}

\begin{figure*}[!h]
\begin{center}
\begin{tabular}{c}
	\begin{minipage}{0.5\hsize}
	\begin{center}
	\includegraphics[scale=0.39]{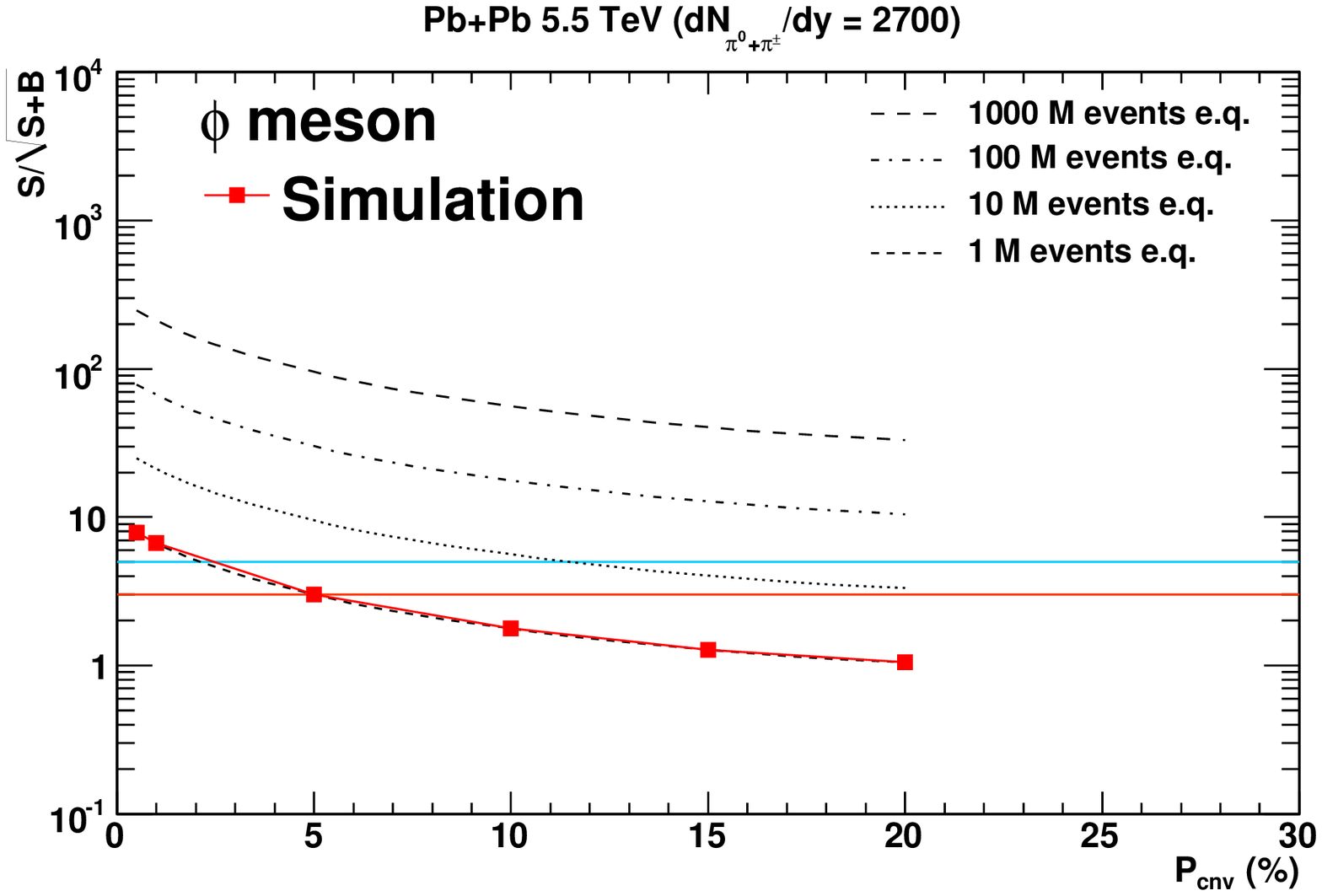}
	\end{center}
	\end{minipage}
	\begin{minipage}{0.5\hsize}
	\begin{center}
	\includegraphics[scale=0.39]{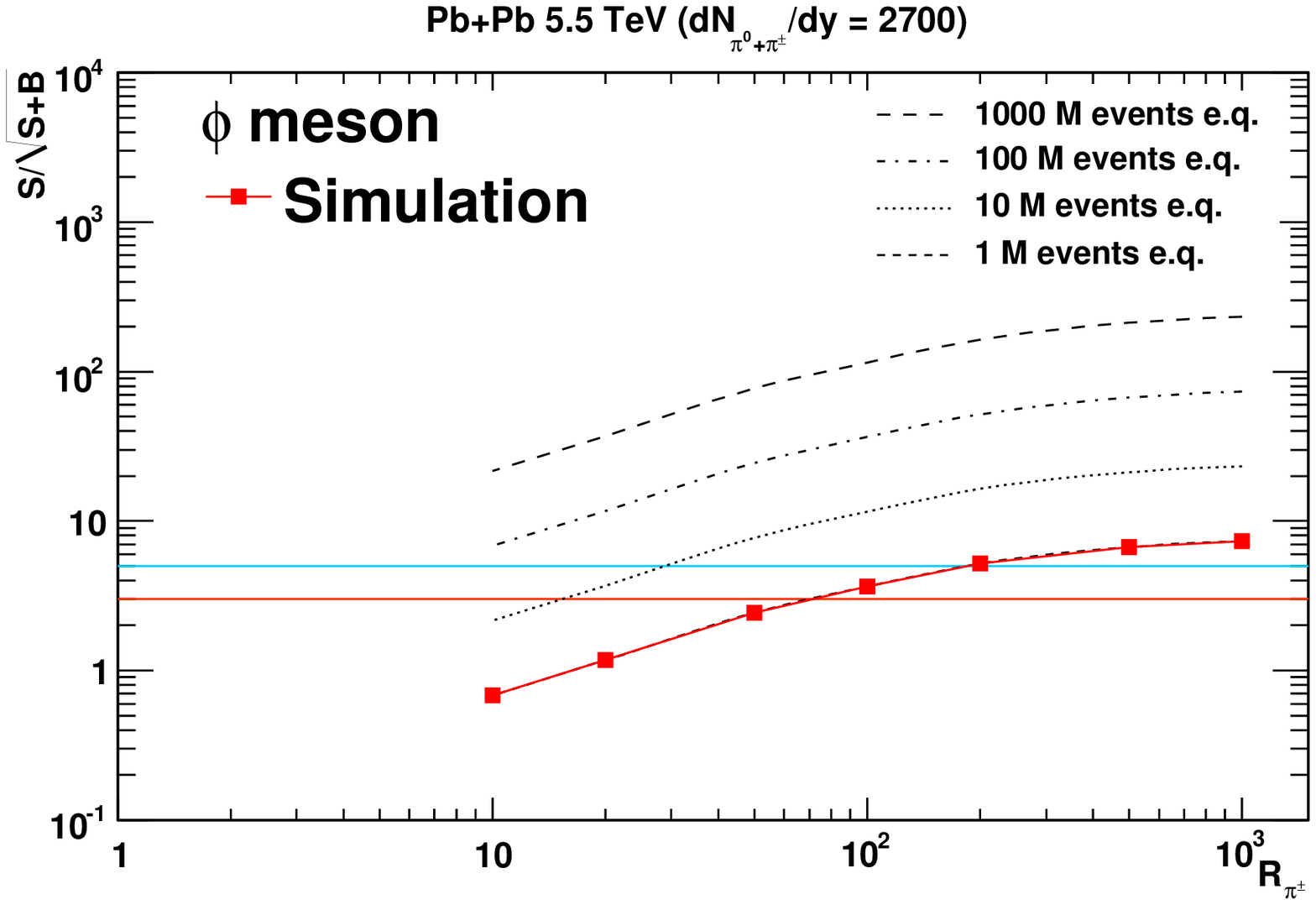}
	\end{center}
	\end{minipage}
\end{tabular}
\begin{tabular}{c}
	\begin{minipage}{0.5\hsize}
	\begin{center}
	\includegraphics[scale=0.39]{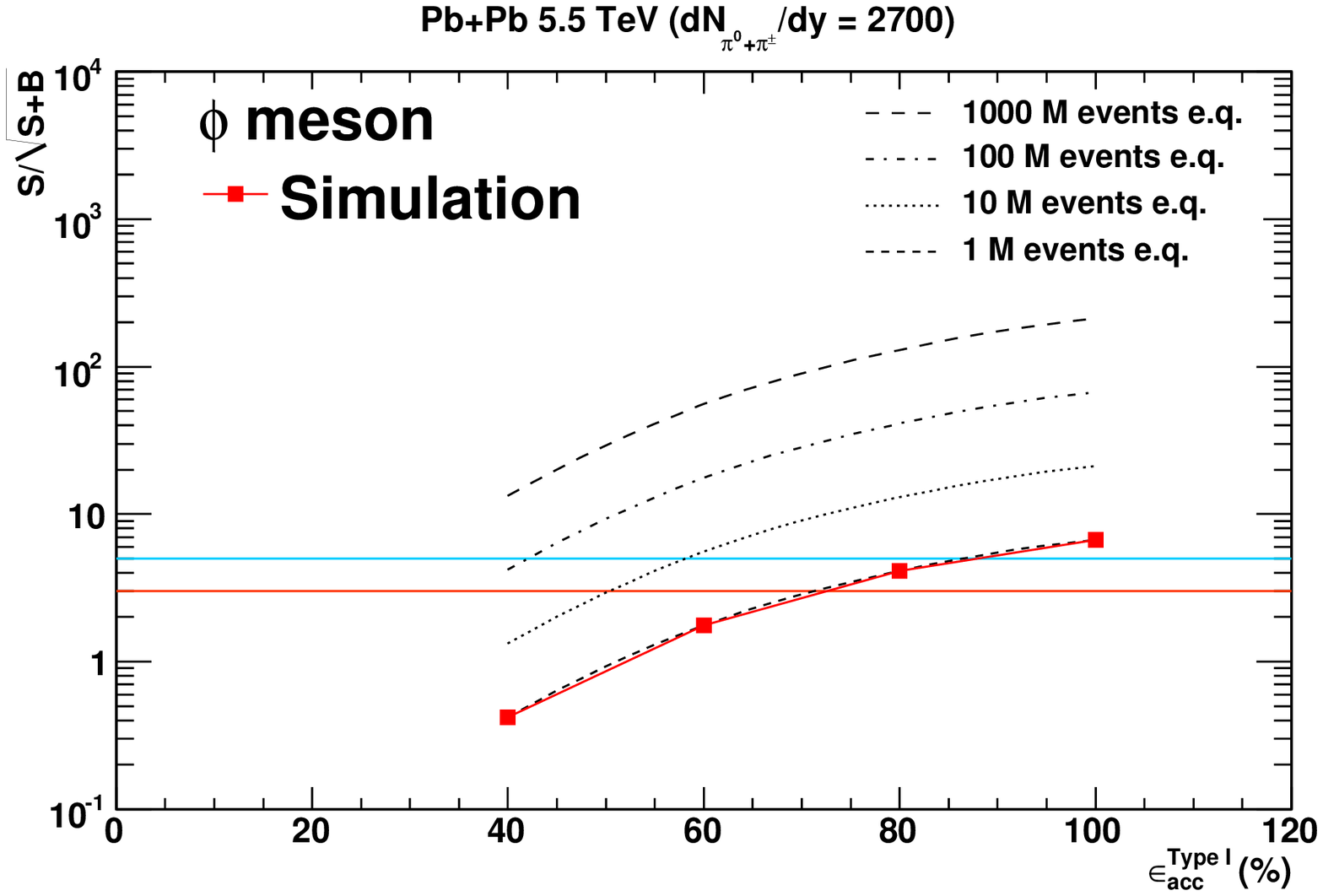}
	\end{center}
	\end{minipage}
	\begin{minipage}{0.5\hsize}
	\begin{center}
	\includegraphics[scale=0.39]{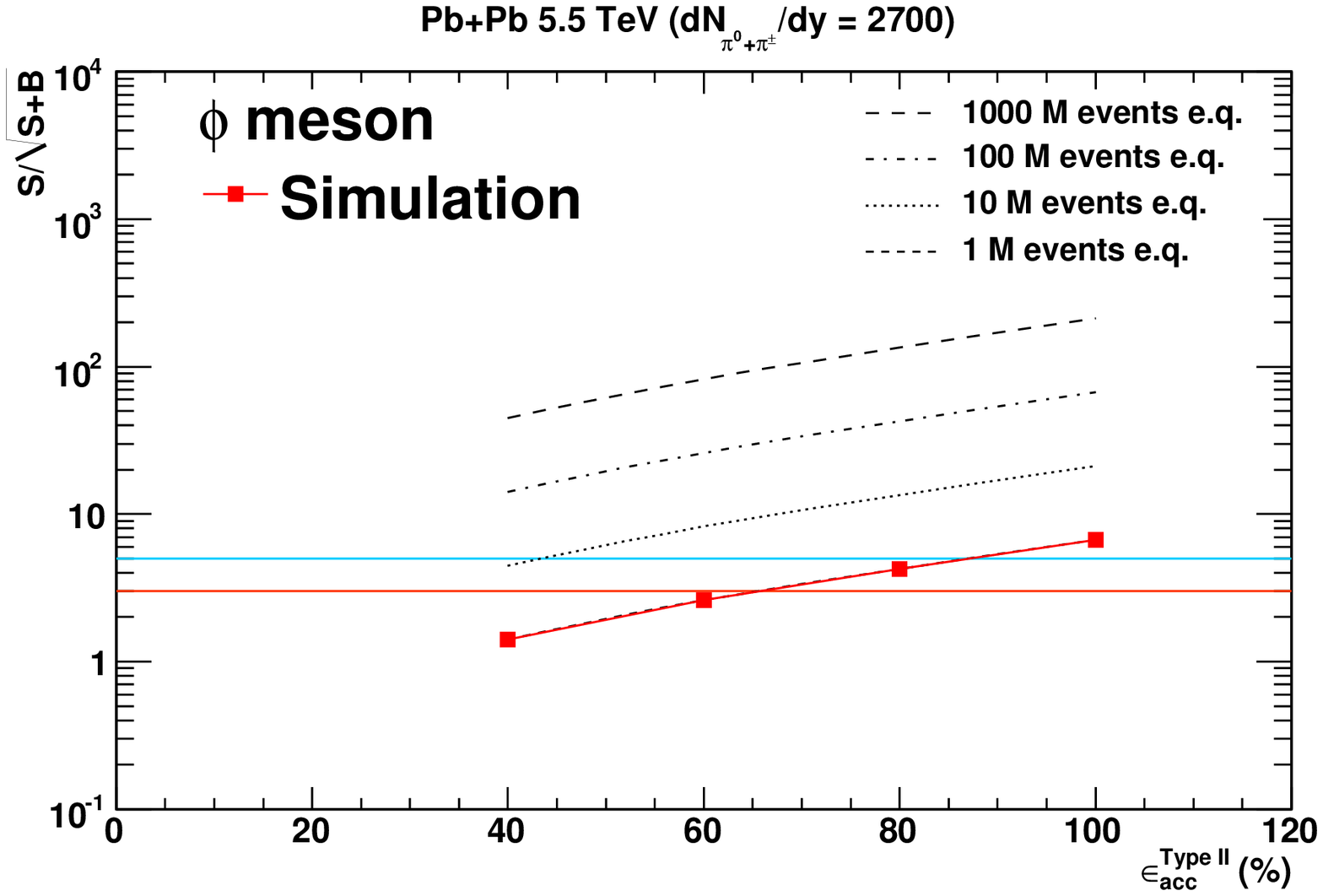}
	\end{center}
	\end{minipage}
\end{tabular}
\begin{tabular}{c}
	\begin{minipage}{0.5\hsize}
	\begin{center}
	\includegraphics[scale=0.39]{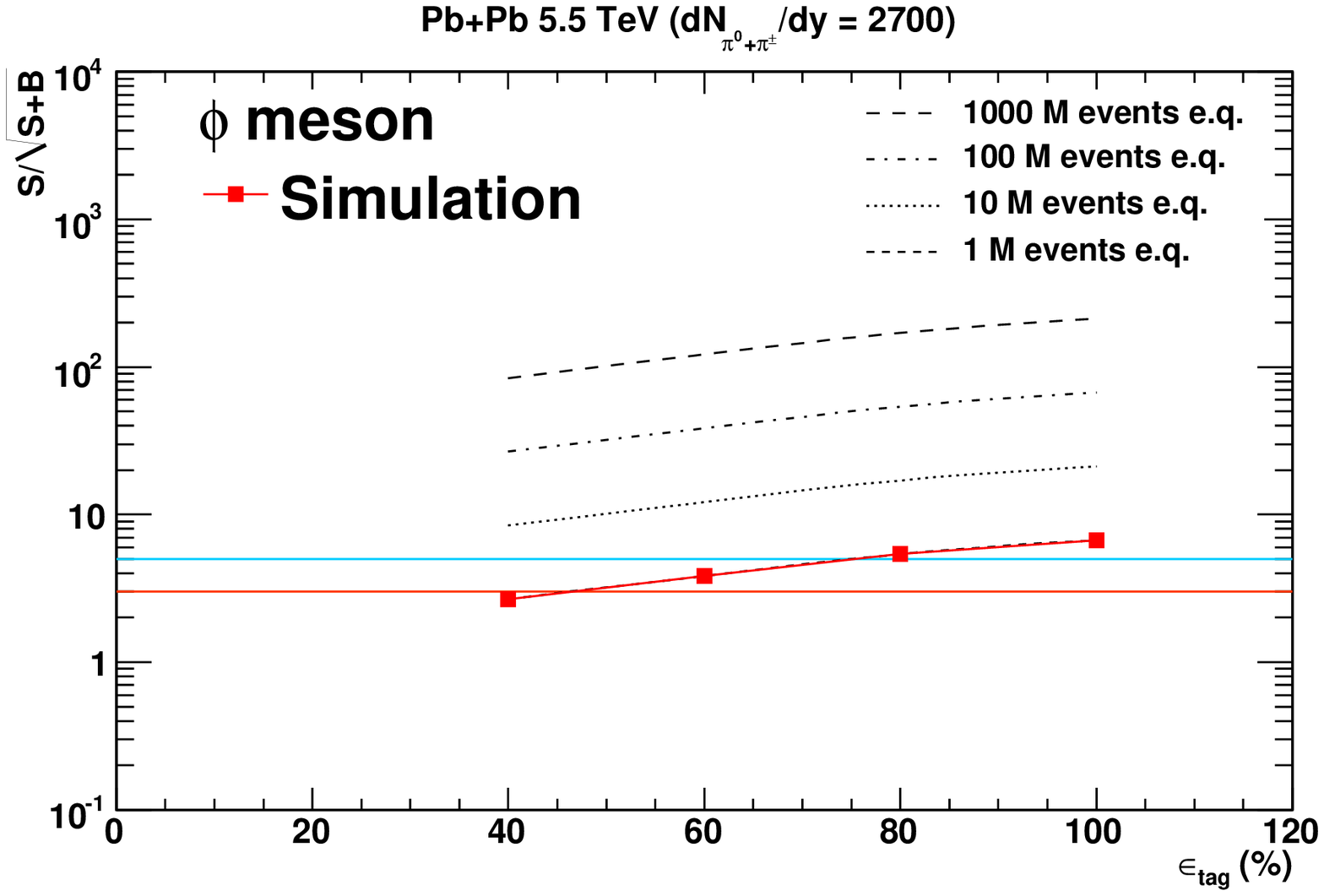}
	\end{center}
	\end{minipage}
	\begin{minipage}{0.5\hsize}
	\begin{center}
	\end{center}
	\end{minipage}
\end{tabular}
\caption{\label{phi_sg_mul2700} (color online) 
The statistical significance $S/\sqrt{S+B}$ of $\phi$ mesons as a function of the experimental parameters $P_{cnv}$, $R_{\pi^{\pm}}$, $\epsilon_{acc}$ and $\epsilon_{tag}$ in central Pb+Pb collisions at $\sqrt{s_{NN}}$ = 5.5 TeV ($dN_{\pi^{0} + \pi^{\pm}}/dy$ = 2700).
Only one parameter is changed by fixing the other parameters at the baseline values for each plot.
The results of the simulation are shown as the symbols and the empirical curves are superimposed on the data points as the solid curves.
The other dotted curves are the scaled curves with the square root of the expected number of events found in the highest centrality class.
Two horizontal lines indicate $S/\sqrt{S+B}$ = 3 and 5.
}
\end{center}
\end{figure*}

\begin{figure*}[!h]
\begin{center}
\begin{tabular}{c}
	\begin{minipage}{0.5\hsize}
	\begin{center}
	\includegraphics[scale=0.39]{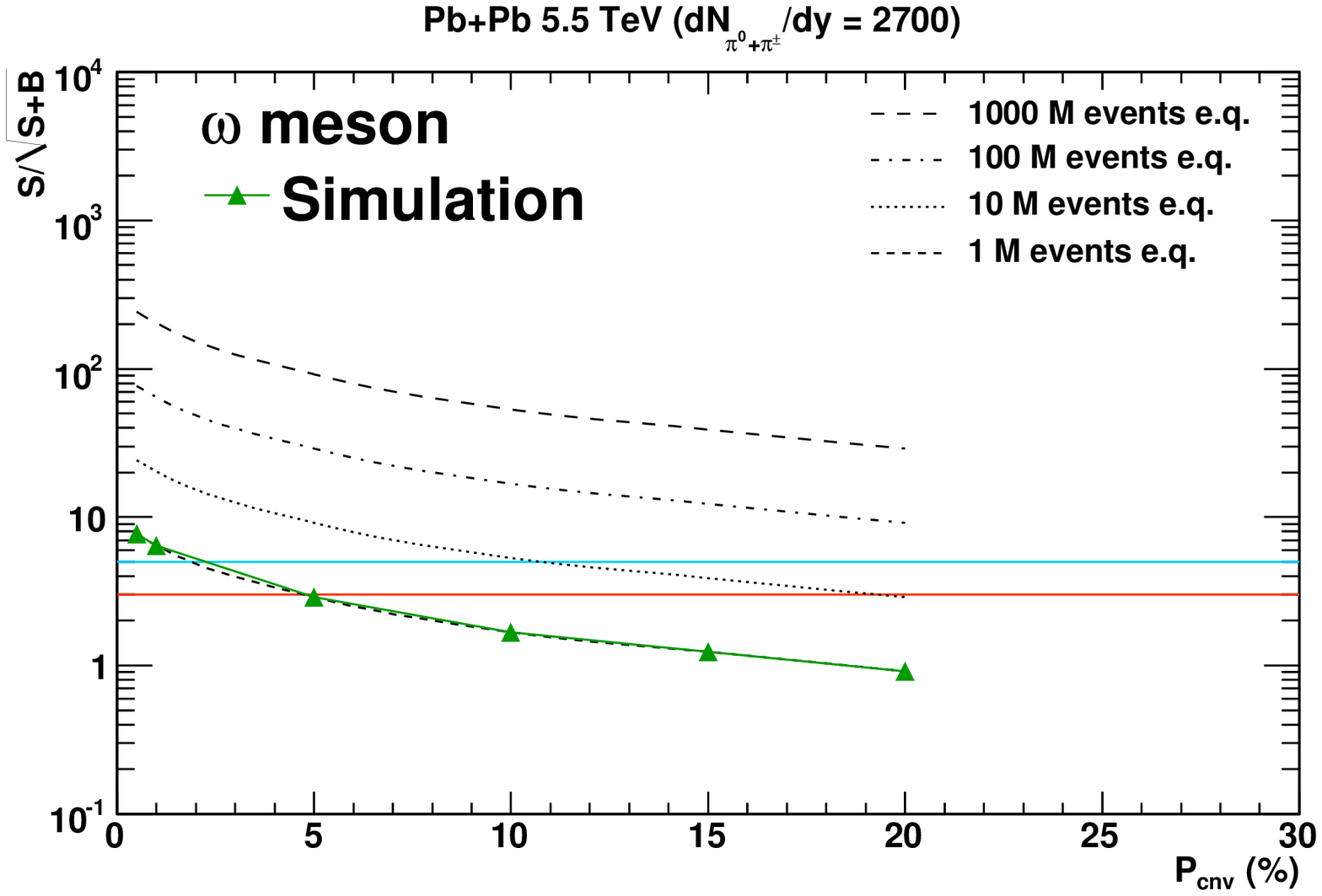}
	\end{center}
	\end{minipage}
	\begin{minipage}{0.5\hsize}
	\begin{center}
	\includegraphics[scale=0.39]{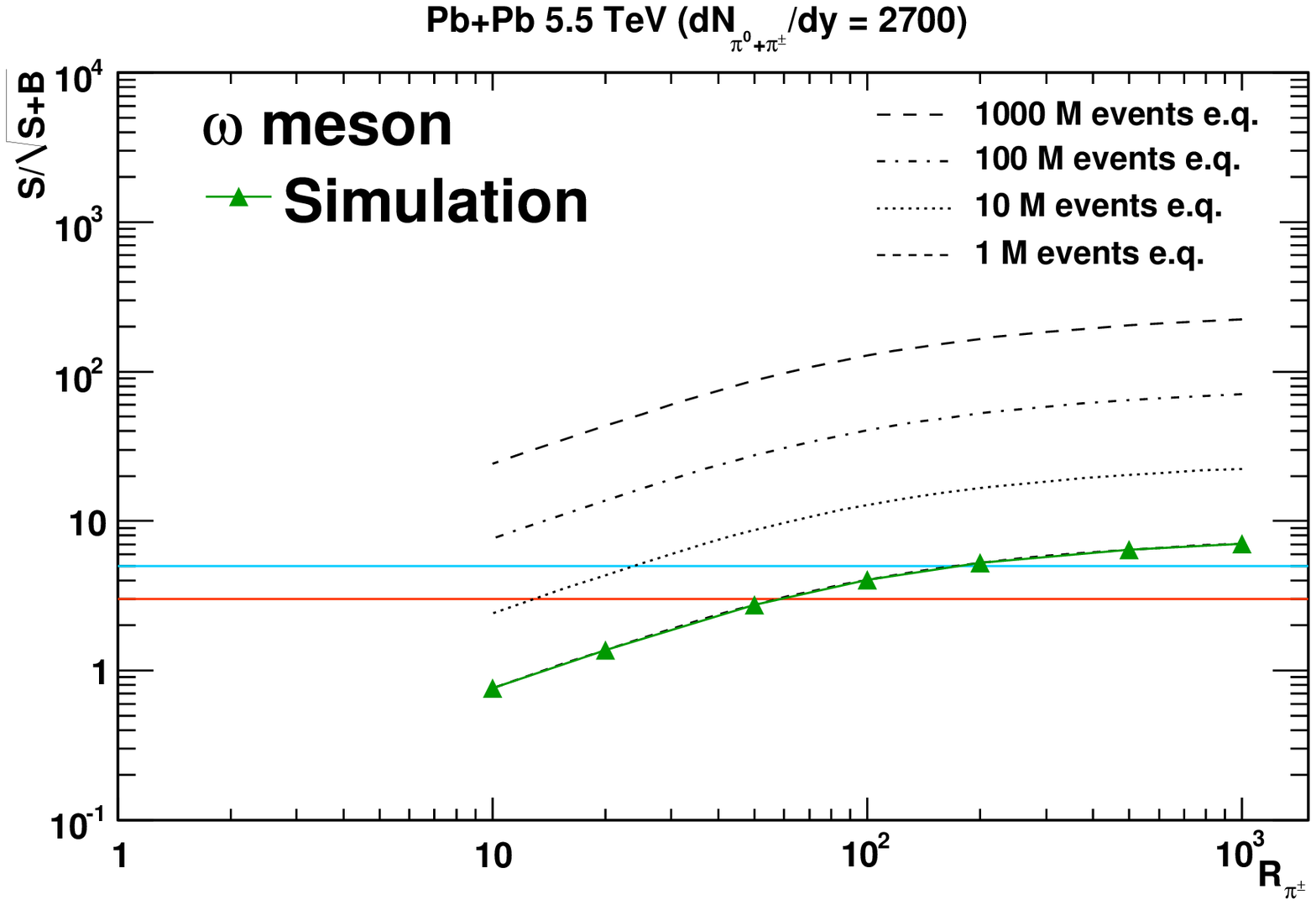}
	\end{center}
	\end{minipage}
\end{tabular}
\begin{tabular}{c}
	\begin{minipage}{0.5\hsize}
	\begin{center}
	\includegraphics[scale=0.39]{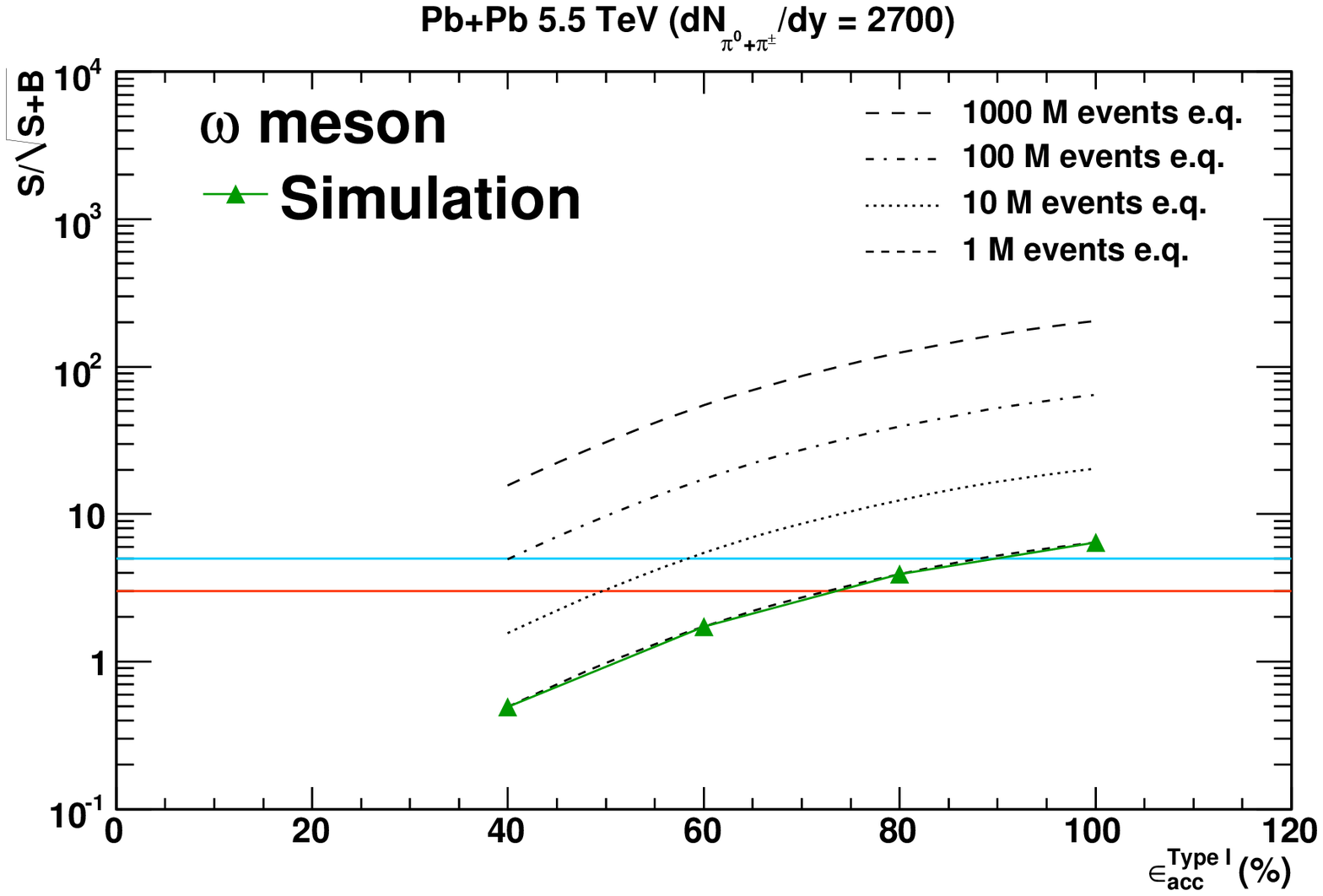}
	\end{center}
	\end{minipage}
	\begin{minipage}{0.5\hsize}
	\begin{center}
	\includegraphics[scale=0.39]{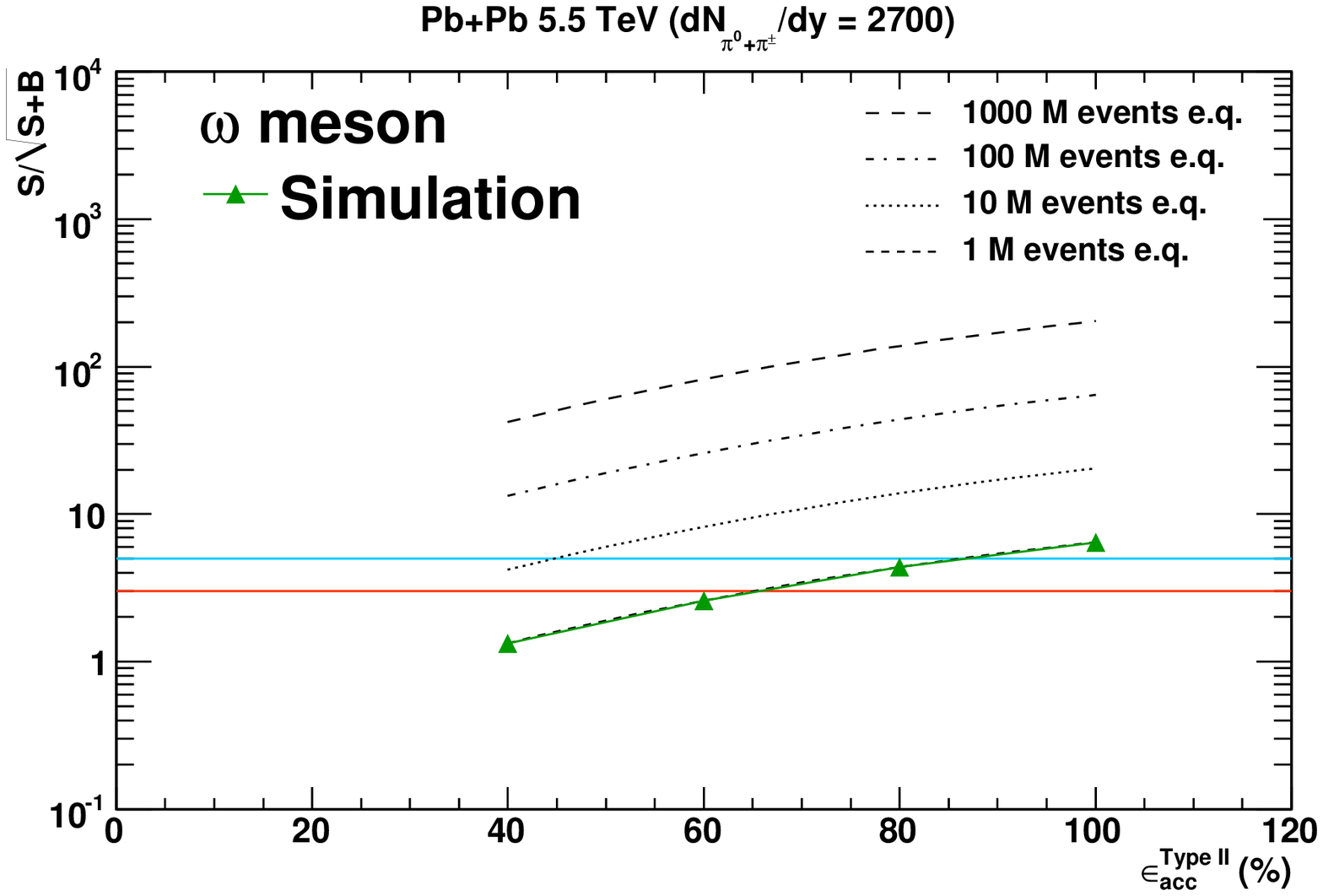}
	\end{center}
	\end{minipage}
\end{tabular}
\begin{tabular}{c}
	\begin{minipage}{0.5\hsize}
	\begin{center}
	\includegraphics[scale=0.39]{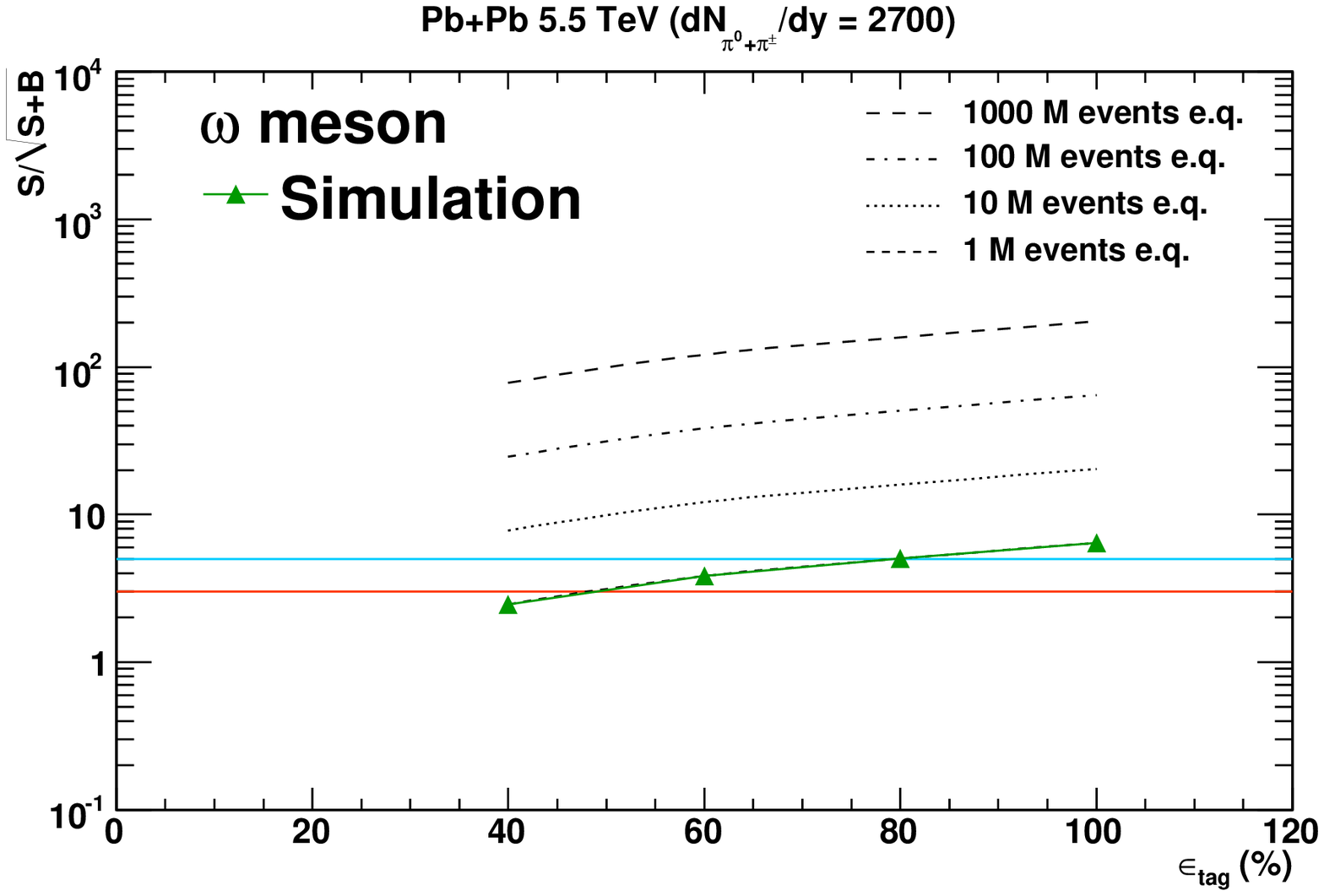}
	\end{center}
	\end{minipage}
	\begin{minipage}{0.5\hsize}
	\begin{center}
	\end{center}
	\end{minipage}
\end{tabular}
\caption{\label{ome_sg_mul2700} (color online) 
The statistical significance $S/\sqrt{S+B}$ of $\omega$ mesons as a function of the experimental parameters $P_{cnv}$, $R_{\pi^{\pm}}$, $\epsilon_{acc}$ and $\epsilon_{tag}$ in central Pb+Pb collisions at $\sqrt{s_{NN}}$ = 5.5 TeV ($dN_{\pi^{0} + \pi^{\pm}}/dy$ = 2700).
Only one parameter is changed by fixing the other parameters at the baseline values for each plot.
The results of the simulation are shown as the symbols and the empirical curves are superimposed on the data points as the solid curves.
The other dotted curves are the scaled curves with the square root of the expected number of events found in the highest centrality class.
Two horizontal lines indicate $S/\sqrt{S+B}$ = 3 and 5.
}
\end{center}
\end{figure*}

\begin{figure*}[!h]
\begin{center}
\begin{tabular}{c}
	\begin{minipage}{0.5\hsize}
	\begin{center}
	\includegraphics[scale=0.39]{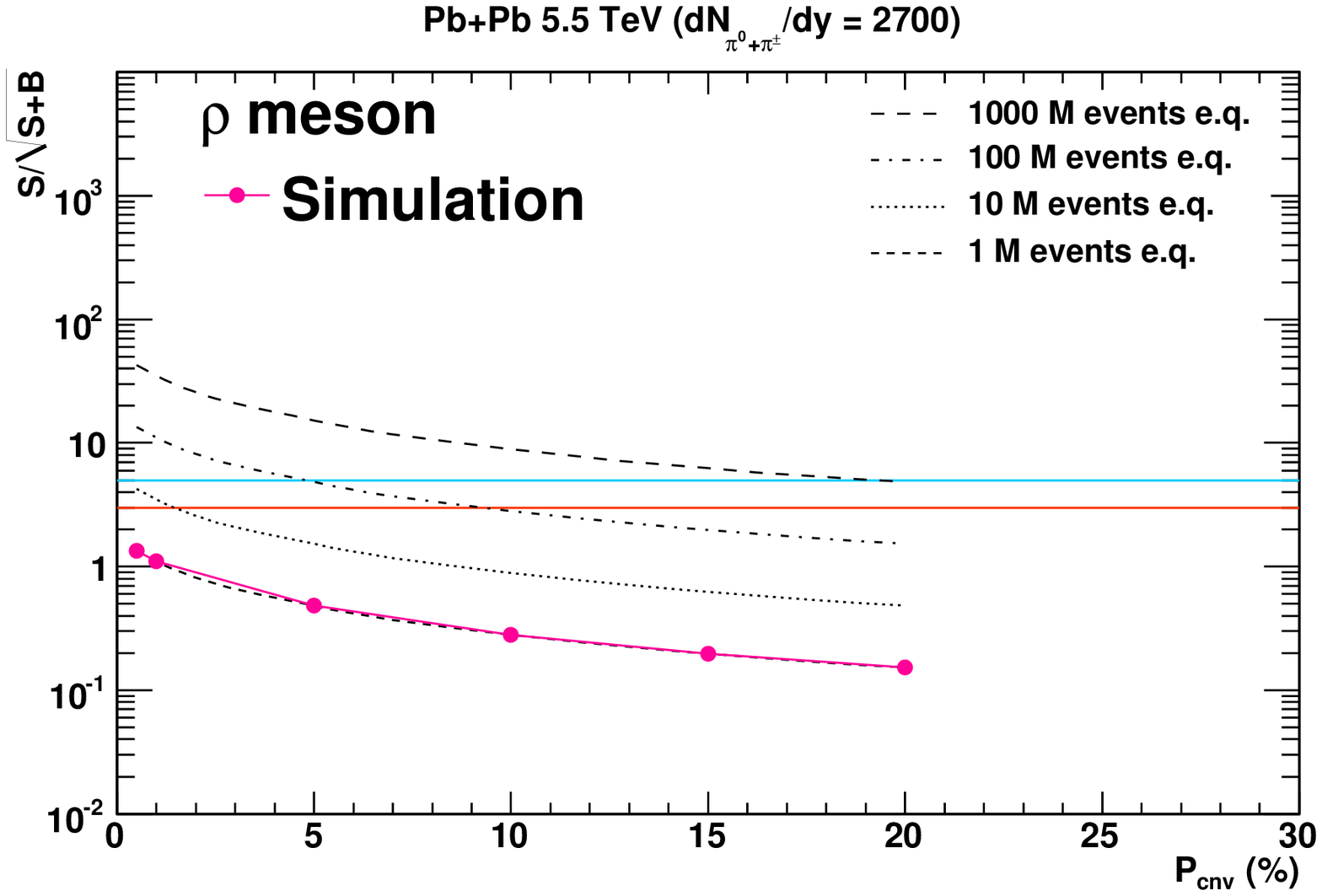}
	\end{center}
	\end{minipage}
	\begin{minipage}{0.5\hsize}
	\begin{center}
	\includegraphics[scale=0.39]{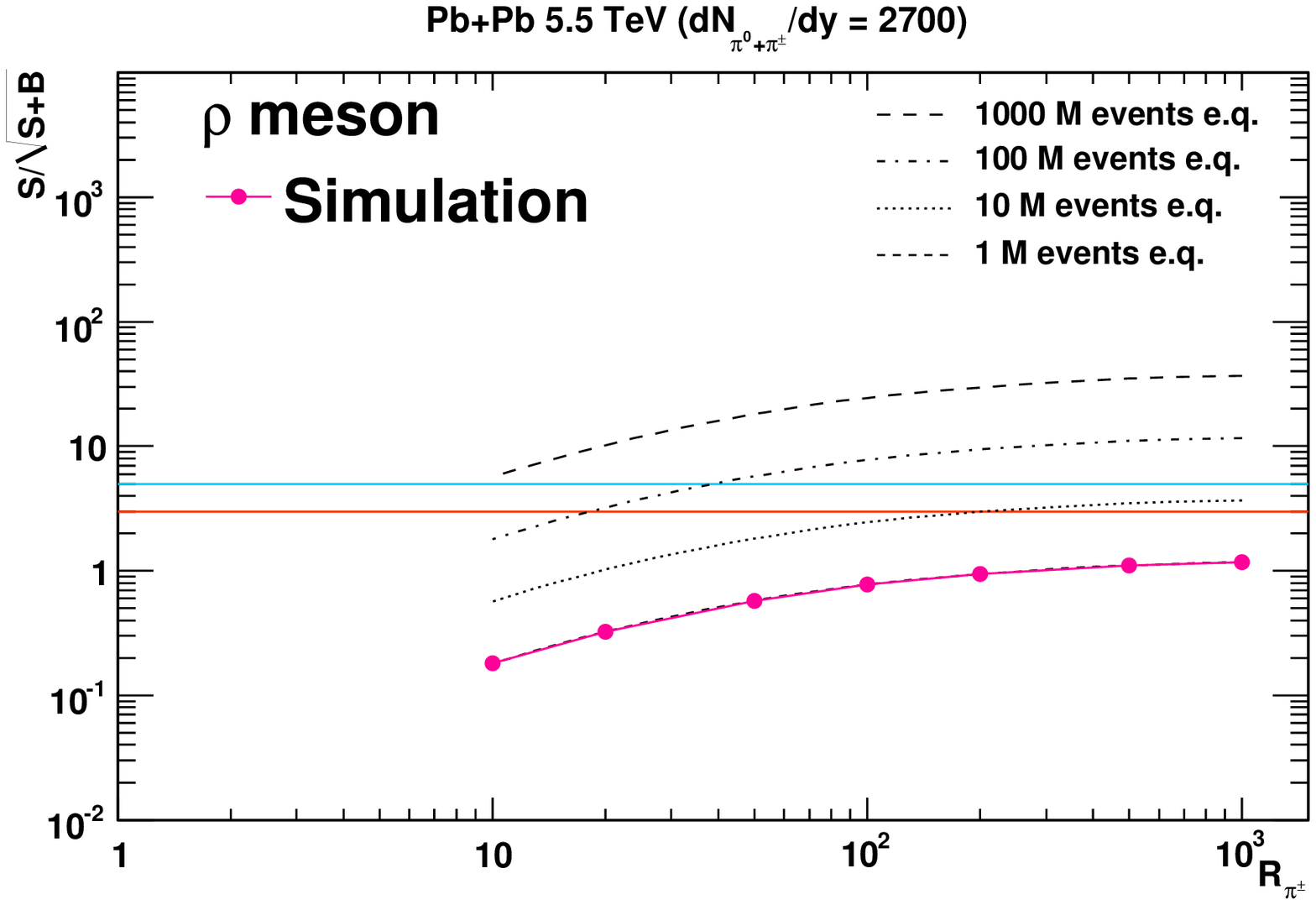}
	\end{center}
	\end{minipage}
\end{tabular}
\begin{tabular}{c}
	\begin{minipage}{0.5\hsize}
	\begin{center}
	\includegraphics[scale=0.39]{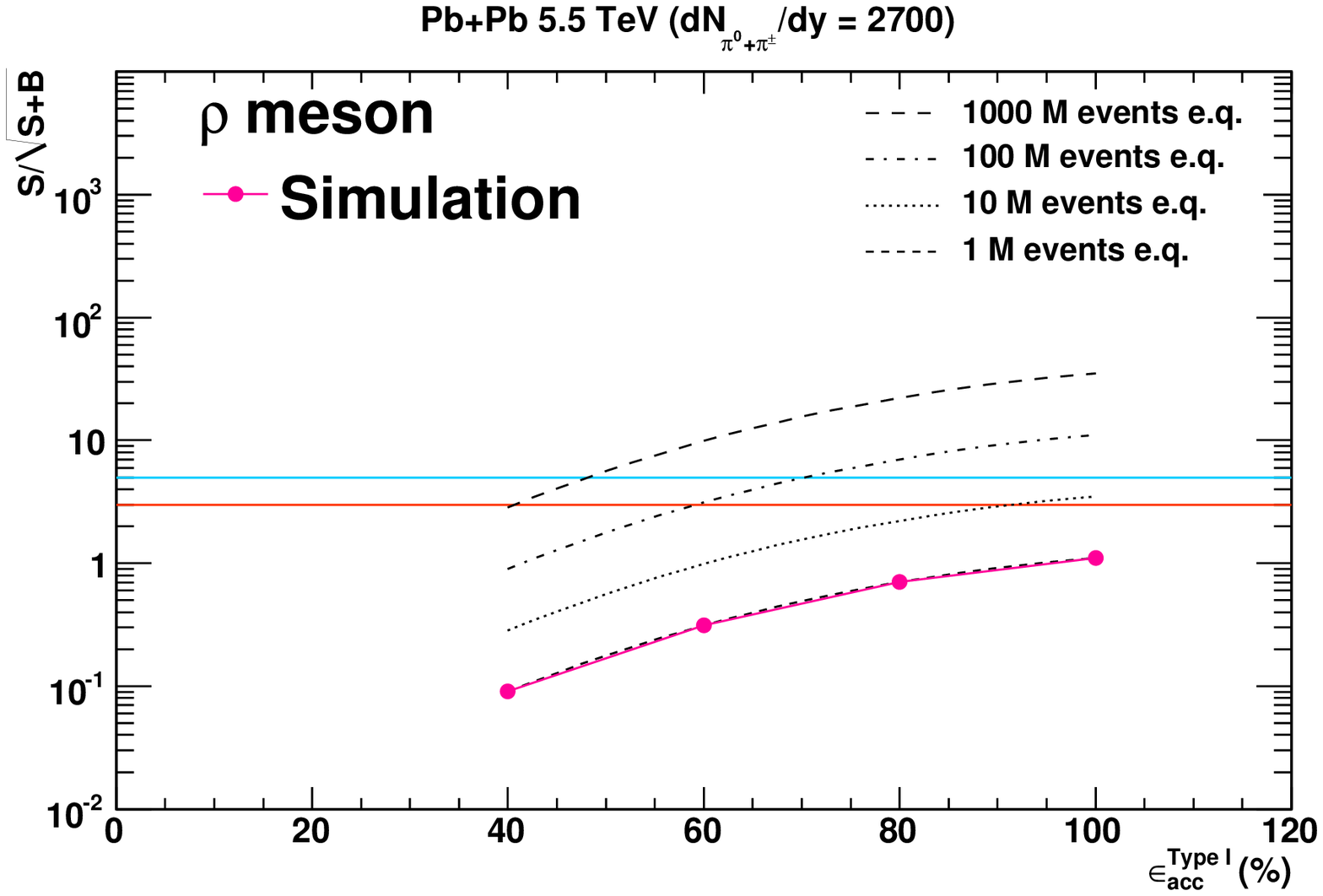}
	\end{center}
	\end{minipage}
	\begin{minipage}{0.5\hsize}
	\begin{center}
	\includegraphics[scale=0.39]{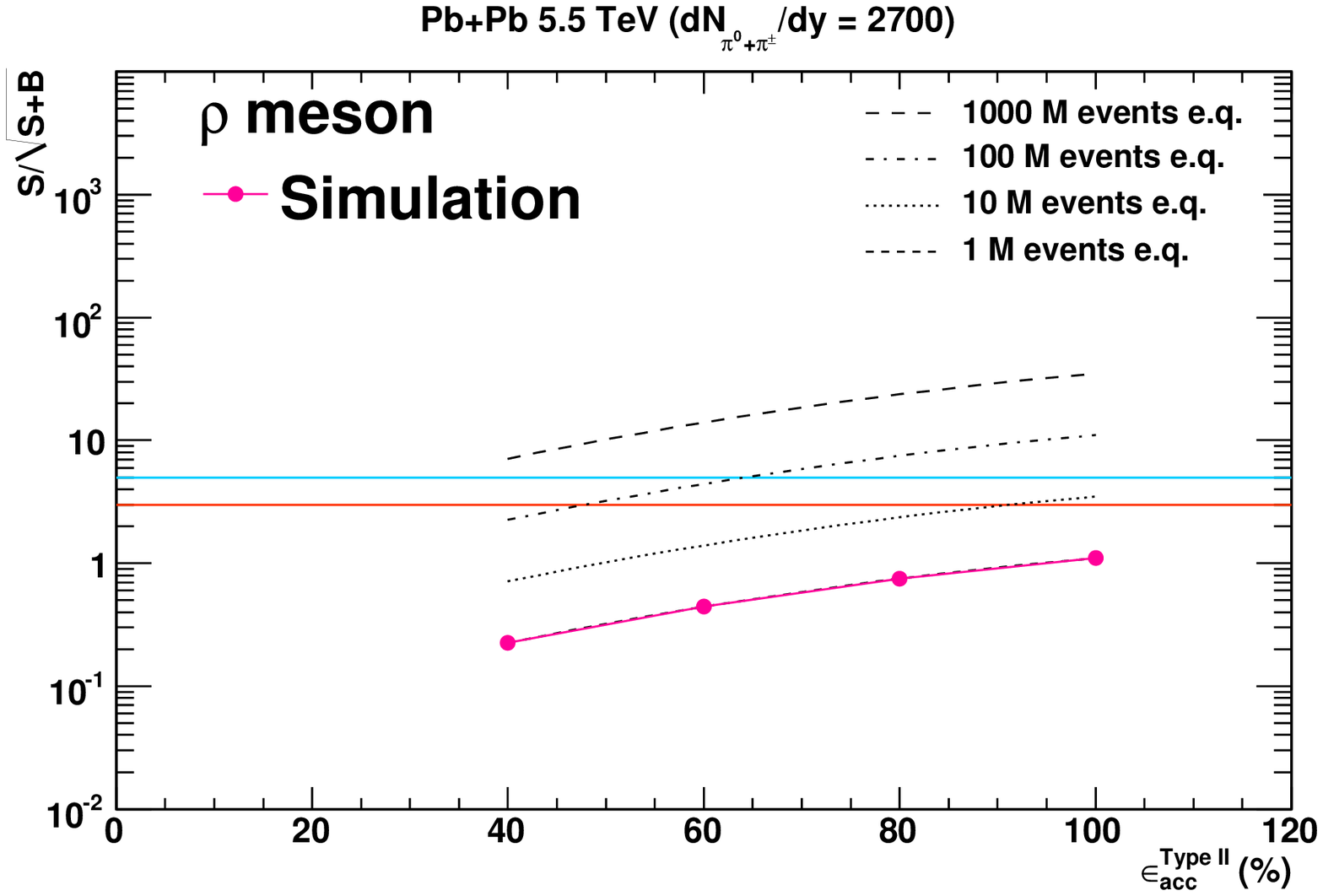}
	\end{center}
	\end{minipage}
\end{tabular}
\begin{tabular}{c}
	\begin{minipage}{0.5\hsize}
	\begin{center}
	\includegraphics[scale=0.39]{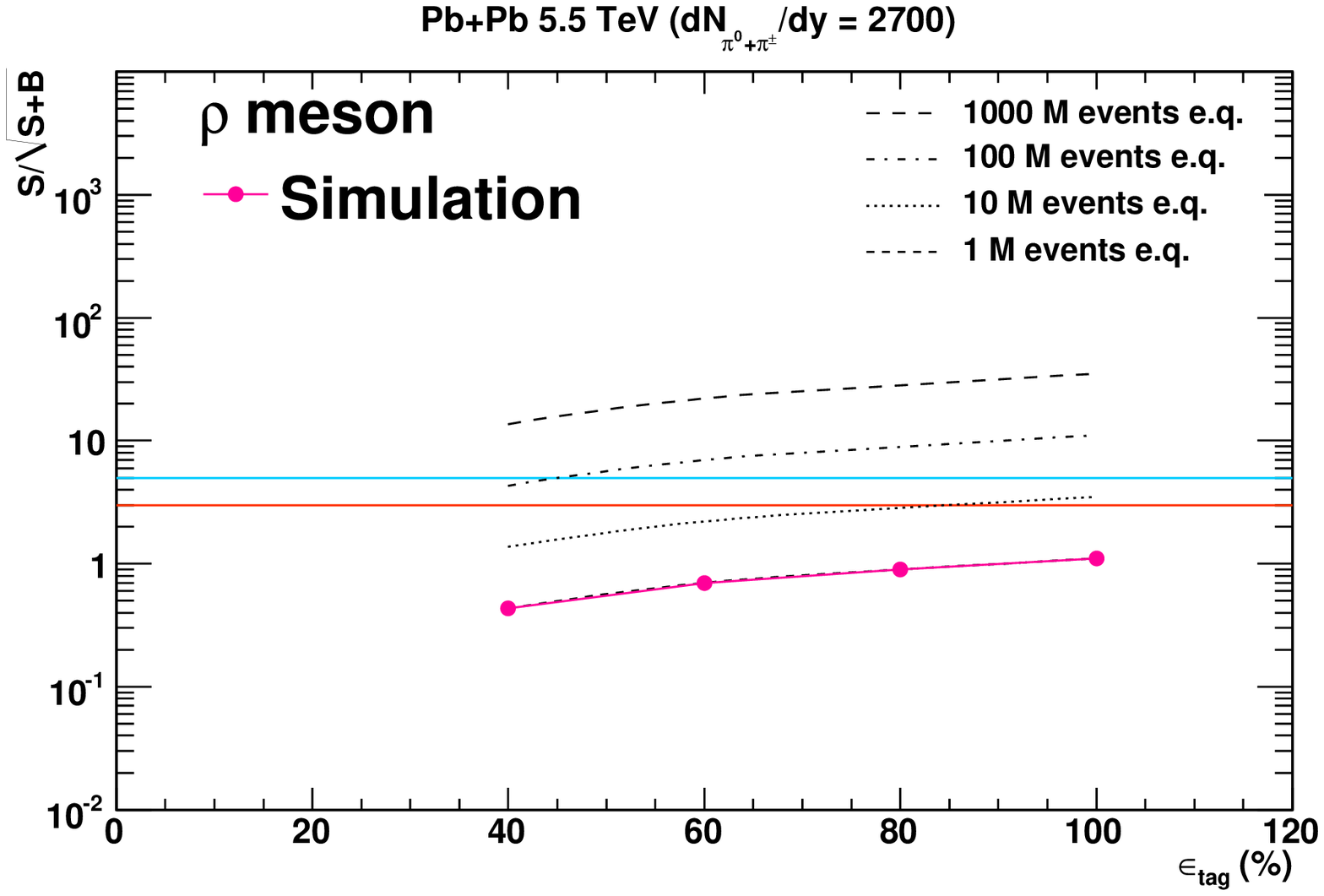}
	\end{center}
	\end{minipage}
	\begin{minipage}{0.5\hsize}
	\begin{center}
	\end{center}
	\end{minipage}
\end{tabular}
\caption{\label{rho_sg_mul2700} (color online) 
The statistical significance $S/\sqrt{S+B}$ of $\rho$ mesons as a function of the experimental parameters $P_{cnv}$, $R_{\pi^{\pm}}$, $\epsilon_{acc}$ and $\epsilon_{tag}$ in central Pb+Pb collisions at $\sqrt{s_{NN}}$ = 5.5 TeV ($dN_{\pi^{0} + \pi^{\pm}}/dy$ = 2700).
Only one parameter is changed by fixing the other parameters at the baseline values for each plot.
The results of the simulation are shown as the symbols and the empirical curves are superimposed on the data points as the solid curves.
The other dotted curves are the scaled curves with the square root of the expected number of events found in the highest centrality class.
Two horizontal lines indicate $S/\sqrt{S+B}$ = 3 and 5.
}
\end{center}
\end{figure*}

The statistical significance ${S/\sqrt{S+B}}$ depends on the square root of the number 
of events, in other words, depends on available luminosity in experiments.
Figure \ref{phi_sg_mul1000}-\ref{rho_sg_mul2700} show the statistical significance as a function of the experimental parameters in central Au+Au collisions 
at $\sqrt{s_{NN}}$ = 200 GeV ($dN_{\pi^{0} + \pi^{\pm}}/dy$ = 1000) and in central Pb+Pb collisions at $\sqrt{s_{NN}}$ = 
5.5 TeV  ($dN_{\pi^{0} + \pi^{\pm}}/dy$ = 2700) for $\phi$, $\omega$ and $\rho$ meson, separately.
The data points and the empirical curves are shown as fulfilled symbols and the solid curves in the figures.
The number of simulated events for central Au+Au collisions and central Pb+Pb collisions corresponds to 5M and 1M events, respectively.
The other dotted curves show the scaled curves with the square root of the expected number of events with the highest centrality selection.
The two horizontal lines indicate $S/\sqrt{S+B}$ = 3 and 5.
The $S/B$ is independent of the electron tagging efficiency $\epsilon_{tag}$,
whereas the $S/\sqrt{S+B}$ scales with the square root of the statistics.
Therefore we added the dependence on the $\epsilon_{tag}$ to the bottom figure for the discussion on the statistical significance.

Depending on the available statistics in the specific collision centrality and the detector conditions,
we can evaluate whether a detection system is able to measure the light vector mesons with a reasonable statistical significance or not. 

\clearpage
\section{Residual effects}
\label{sect4}
The signal-to-background ratios and the statistical significance of the light vector mesons are evaluated with the idealized detection system so far. 
In this section, we discuss the non-trivial aspects originating from the real data analysis and the correlations in the open charm production.
As the other issues beyond the scope of the numerical simulation, 
we mention the track reconstruction algorithm bias, the correlation between the electron identification and the rejection of charged hadrons,
and the fiducial effect on the acceptance to charged particles in the magnetic field. 
The studies of this section are performed by simulating 5 M events in Au+Au 200 GeV and Pb+Pb 5.5 TeV 
with the baseline parameters set: $P_{cnv}$ = 1 $\%$, $R_{\pi^{\pm}}$ = 500, $\epsilon_{acc}$ = 100 $\%$, 
$\epsilon_{tag}$ = 100 $\%$, $p_{T}^{th}$ = 0.1 GeV/c and $\sigma^{ref}_{p_{T}} = \sqrt{ \left(0.01 \cdot p_{T} \right)^{2} + \left( 0.0056 \right)^{2}}$ GeV/c. 

\begin{figure*}[!h]
\begin{center}
\begin{tabular}{c}
	\begin{minipage}{0.5\hsize}
	\begin{center}
	\includegraphics[scale=0.39]{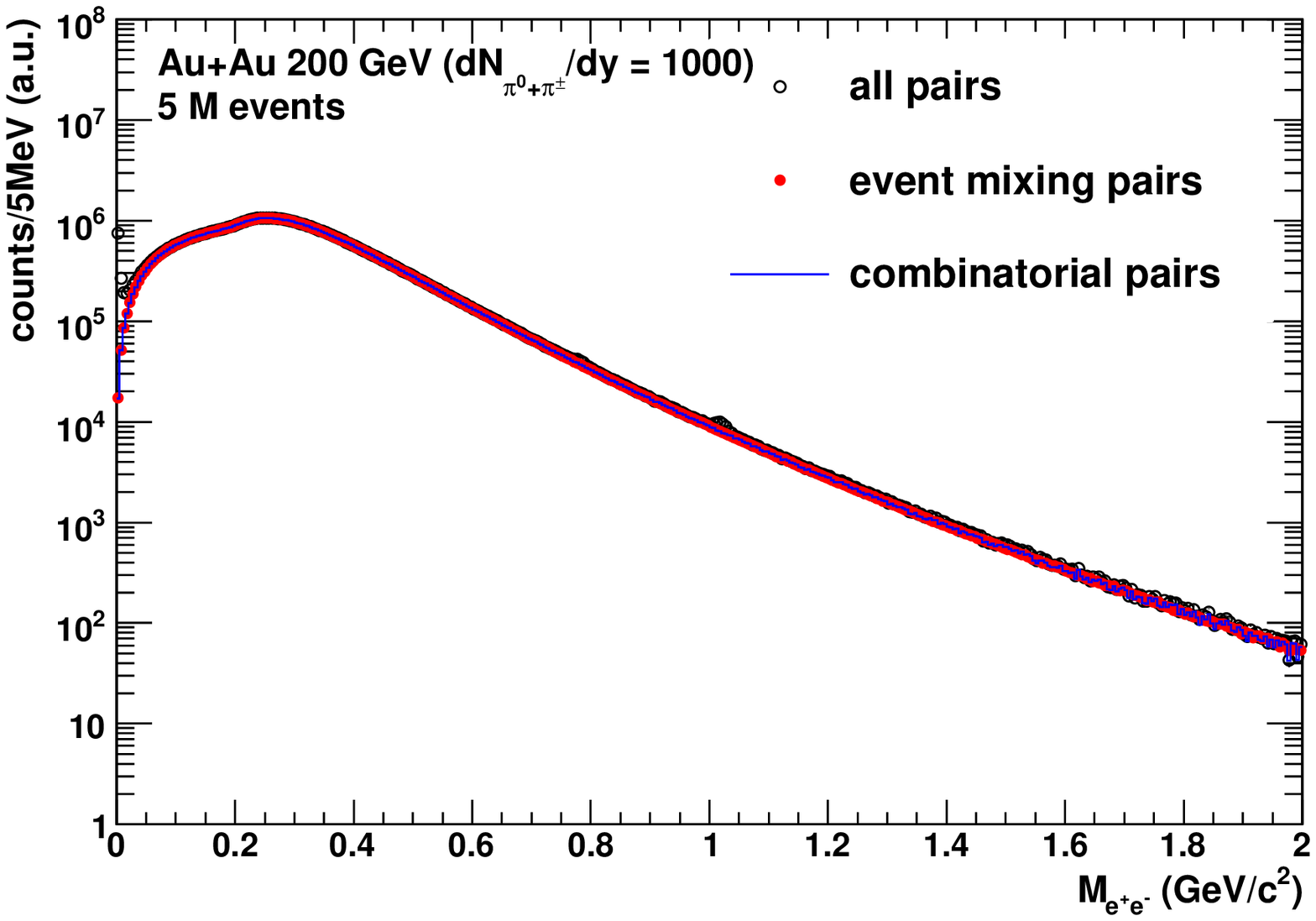}
	\end{center}
	\end{minipage}
	\begin{minipage}{0.5\hsize}
	\begin{center}
	\includegraphics[scale=0.39]{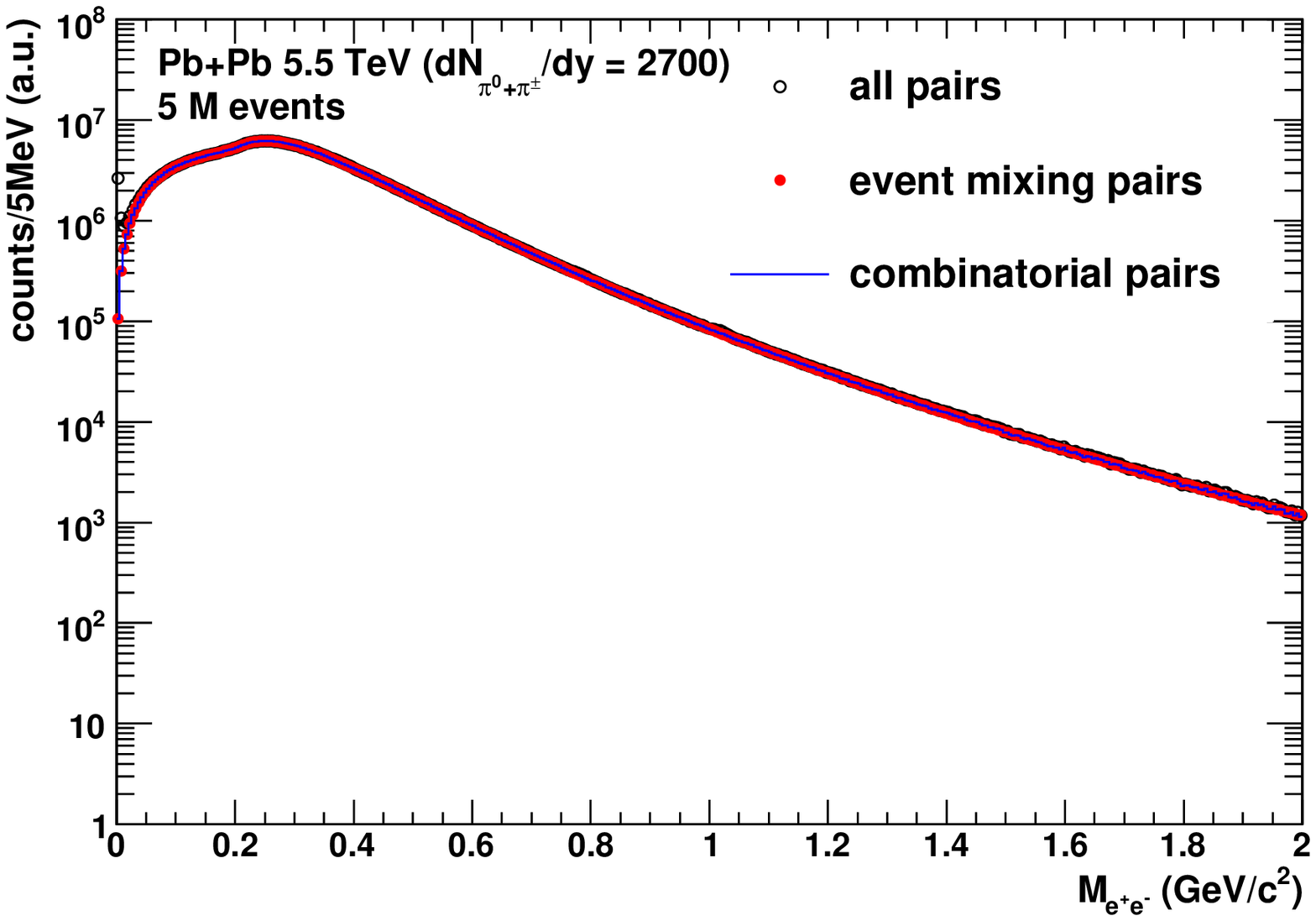}
	\end{center}
	\end{minipage}
\end{tabular}
\begin{tabular}{c}
	\begin{minipage}{0.5\hsize}
	\begin{center}
	\includegraphics[scale=0.39]{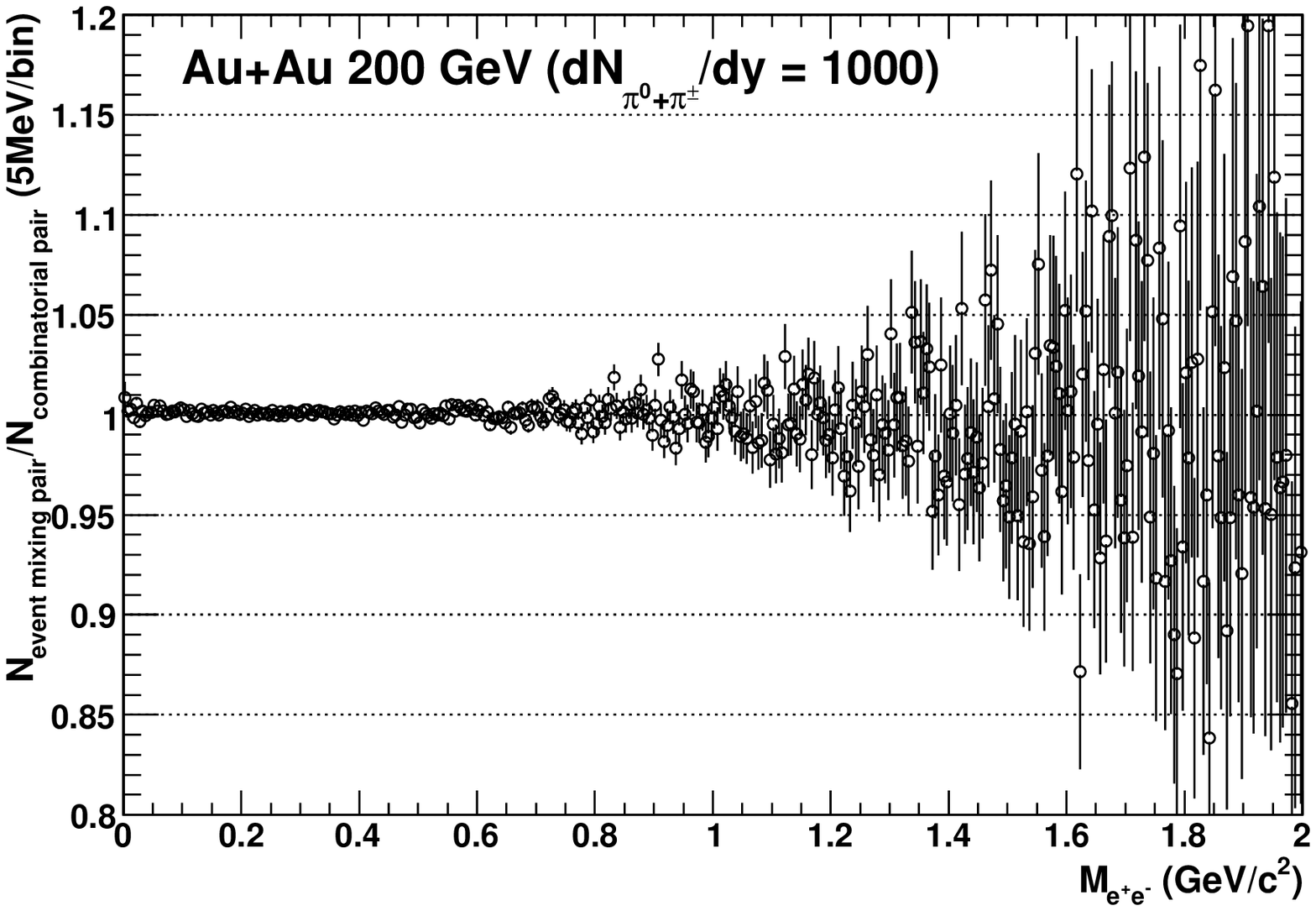}
	\end{center}
	\end{minipage}
	\begin{minipage}{0.5\hsize}
	\begin{center}
	\includegraphics[scale=0.39]{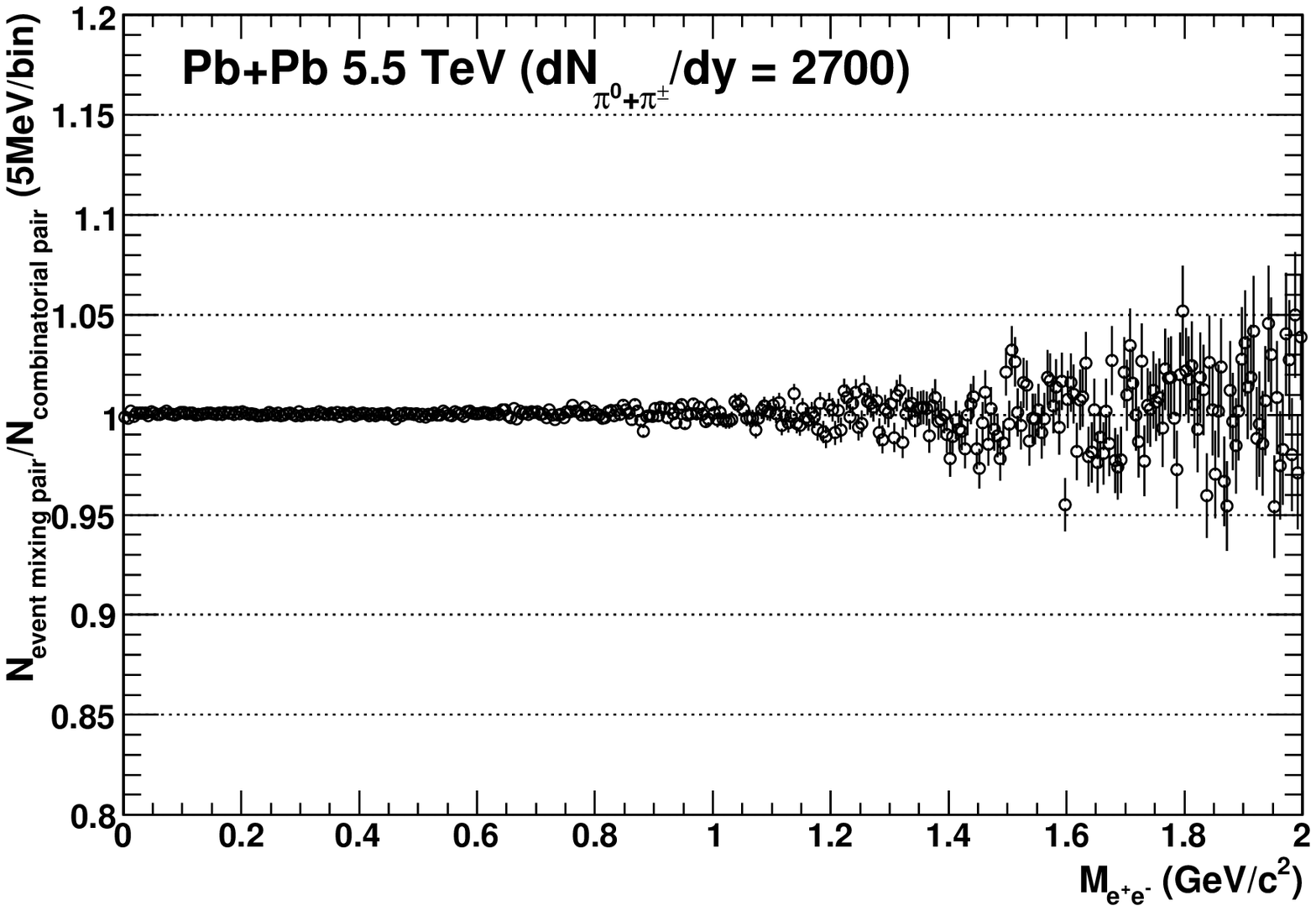}
	\end{center}
	\end{minipage}
\end{tabular}
\begin{tabular}{c}
	\begin{minipage}{0.5\hsize}
	\begin{center}
	\includegraphics[scale=0.39]{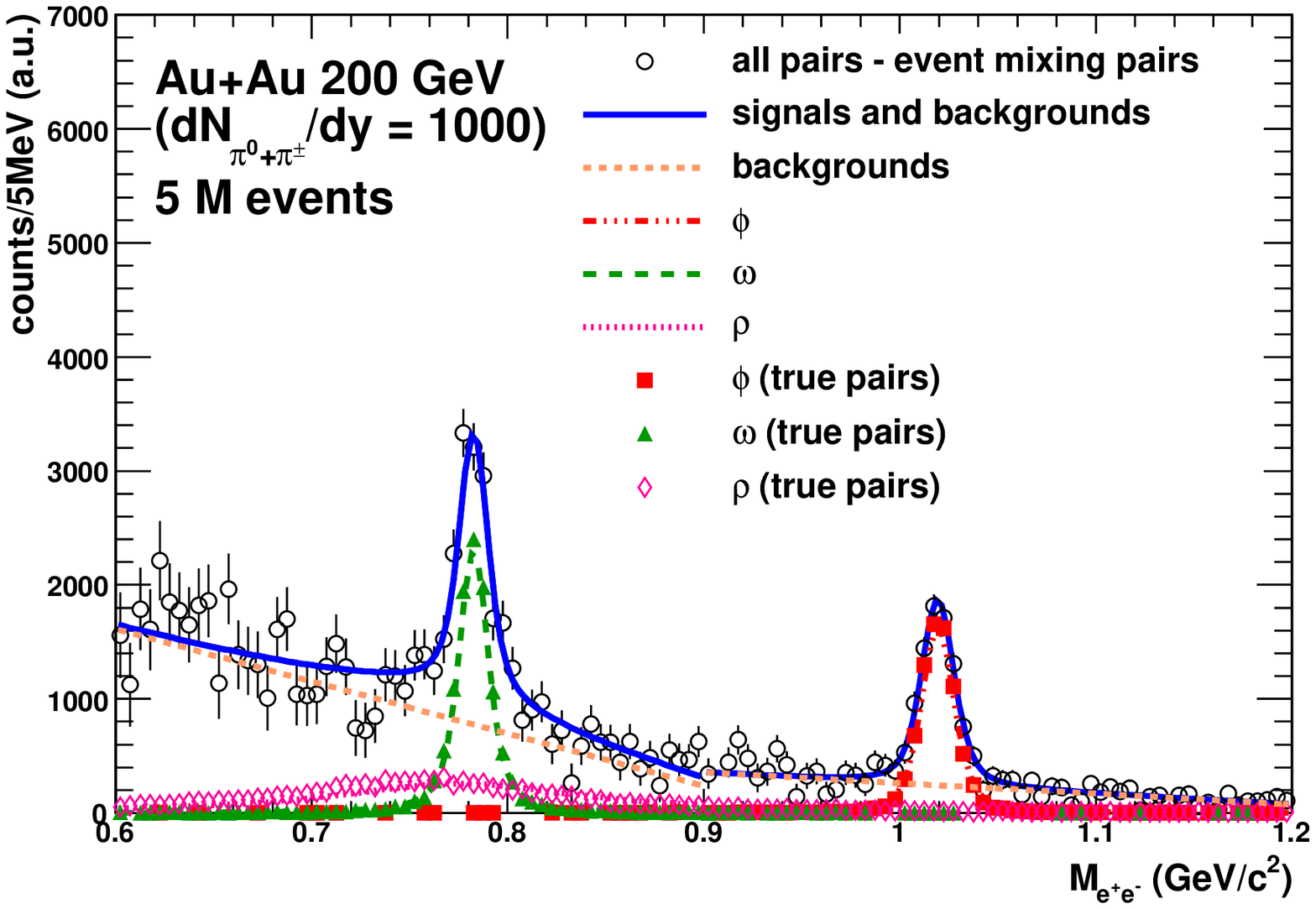}
	\end{center}
	\end{minipage}
	\begin{minipage}{0.5\hsize}
	\begin{center}
	\includegraphics[scale=0.39]{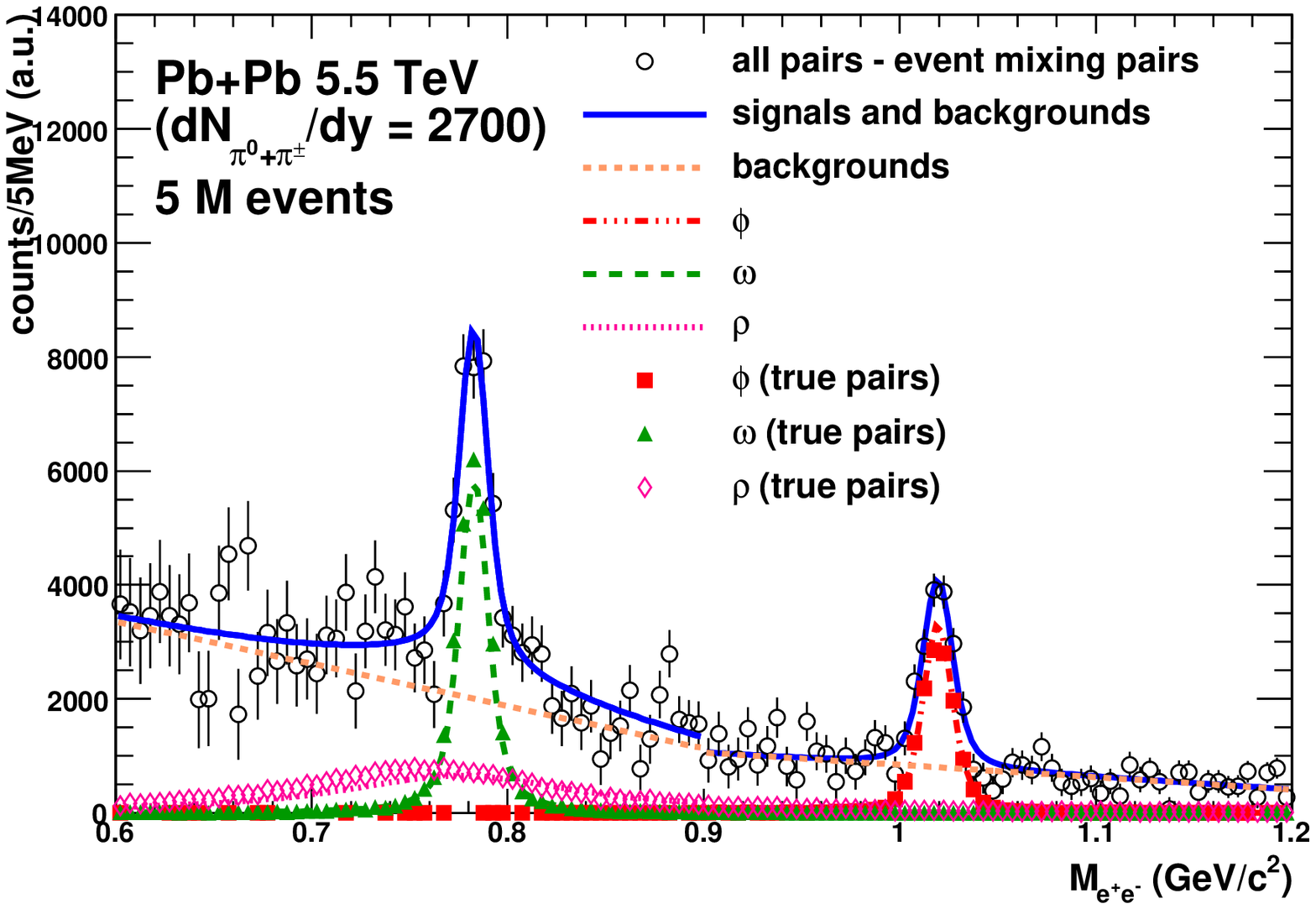}
	\end{center}
	\end{minipage}
\end{tabular}
\caption{\label{mass_real} (color online)
The invariant mass spectra in central Au+Au collisions at $\sqrt{s_{NN}}$ = 200 GeV ($dN_{\pi^{0} + \pi^{\pm}}/dy$ = 1000) and 
central Pb+Pb collisions at $\sqrt{s_{NN}}$ = 5.5 TeV ($dN_{\pi^{0} + \pi^{\pm}}/dy$ = 2700) with $P_{cnv}$ = 1 $\%$, $R_{\pi^{\pm}}$ = 500, $\epsilon_{acc}$ = 100 $\%$, $\epsilon_{tag}$ = 100 $\%$, $p_{T}^{th}$ = 0.1 GeV/c and $\sigma^{ref}_{p_{T}} = \sqrt{ \left(0.01 \cdot p_{T} \right)^{2} + \left( 0.0056 \right)^{2}}$ GeV/c.
The open symbols in top two figures show all reconstructed pairs. 
The closed circles show the event-mixing pairs after normalizing by $2\sqrt{N_{++}N_{--}}/N^{mix}_{+-}$.
The solid curves show the combinatorial pairs.
The ratio between the number of event-mixing pairs and that of combinatorial pairs is shown as a function of the invariant mass in the middle two figures. 
The bottom two figures show the invariant mass distributions after subtracting the combinatorial backgrounds.
The solid curves are the fitting results by the linear combination of the Breit-Wigner function convoluted with the Gauss function 
and a first-order polynomial function.
Each dotted curve is obtained by the fitting results and shows the components of the light vector mesons and the residual background sources. 
The squares, triangles and diamonds show the invariant mass of the true di-electron pairs decaying from the light vector mesons. 
}
\end{center}
\end{figure*}

The numerical simulation so far is performed under the assumption that we know the exact number of signals and backgrounds.
In the real data analysis, however, the source of any electron cannot be identified.
Therefore all electrons and positrons are combined into pairs and reconstructed into the invariant mass.
The mass distribution of pairs from one source ("true pairs") is extracted by subtracting that of pairs from different source ("combinatorial pairs") statistically.
The mass shape of combinatorial pairs is estimated by mixing an electron in an event and a positron in another event ("event mixing") \cite{event_mix1,event_mix2}.
The mass distribution of event-mixing pairs is normalized by $2\sqrt{N_{++}N_{--}}/N^{mix}_{+-}$, 
where $N_{++}$,  $N_{--}$ and  $N^{mix}_{+-}$ are the numbers of positron-positron pairs, electron-electron pairs and event-mixing electron-positron pairs, respectively.
This normalization factor is valid as long as electrons and positrons are produced as pairs and they have the same acceptance \cite{virtual_photon1}.
The mass distribution of true pairs includes the light vector mesons and the other sources.
The contributions from the light vector mesons and the background sources are separately estimated by 
the fits based on the linear combination between the Breit-Wigner function convoluted with the Gauss function and an empirical function.
We apply a series of procedures used in the real data analysis to the simulated data and evaluate 
how much the signal-to-background ratios change by applying these procedures.
The top plots of Fig.\ref{mass_real} show the comparison of the invariant mass spectra between the combinatorial pairs and the event-mixing pairs.
The comparison is performed in central Au+Au collisions at $\sqrt{s_{NN}}$ = 200 GeV ($dN_{\pi^{0} + \pi^{\pm}}/dy$ = 1000) and 
central Pb+Pb collisions at $\sqrt{s_{NN}}$ = 5.5 TeV ($dN_{\pi^{0} + \pi^{\pm}}/dy$ = 2700), respectively.
The ratio between the number of combinatorial pairs and that of event-mixing ones is close to unity within a few $\%$ of statistical fluctuations 
below the mass of 1.0 GeV/$c^{2}$ in both collision systems as shown in the middle plots of Fig.\ref{mass_real}.
Therefore the event-mixing pairs in the real data analysis can provide the reliable baseline representing the combinatorial pairs which is known only at the simulation study \footnote{
We note that the normalization factor of  $2 \sqrt{N_{++}N_{--}}/N^{mix}_{+-}$ overestimates the combinatorial backgrounds by 0.05-0.3 $\%$.
For instance, if a detection system has large amount of materials (i.e. $P_{cnv}$ is high) or has a poor capability of hadron rejection (i.e. $R_{\pi^{\pm}}$ is low), 
this estimation excessively subtracts the backgrounds.
}.
The bottom plots of Fig.\ref{mass_real} show the invariant mass distribution after subtracting the event-mixing distribution. 
The linear combination between the Breit-Wigner function convoluted with the Gauss function and a first-order polynomial function is used as the fitting function.
In the mass range of $\phi$ meson, we use
\begin{eqnarray}
\frac{dN_{e^{+}e^{-}}}{dM_{e^{+}e^{-}}} &=&
A \int F_{\phi} \left( M' \right) G_{gauss}\left(M_{e^{+}e^{-}}-M' \right)  dM'  \nonumber \\
& & + H_{bg} \left(  M_{e^{+}e^{-}} \right).  \label{eq:f1} 
\end{eqnarray}
In the mass range of $\omega/\rho$ meson, we use
\begin{eqnarray}
\frac{dN_{e^{+}e^{-}}}{dM_{e^{+}e^{-}}} &=& A\int \left\{ R F_{\omega} \left( M' \right) + \left( 1-R \right) F_{\rho} \left( M' \right) \right\} \nonumber \\
& & G_{gauss} \left( M_{e^{+}e^{-}}-M' \right) dM' 
+ H_{bg} \left(  M_{e^{+}e^{-}} \right),  \nonumber 
\end{eqnarray}
\begin{eqnarray}
R = \frac{N_{\omega} BR \left( \omega \rightarrow e^{+}e^{-} \right)}{N_{\omega} BR \left( \omega \rightarrow e^{+}e^{-} \right) + N_{\rho} BR \left( \rho \rightarrow e^{+}e^{-} \right) },  
\label{eq:f2}
\end{eqnarray}
where $N_{\omega}$ and $N_{\rho}$ are the inclusive yields of $\omega$ and $\rho$ meson, respectively.
The absolute values of the inclusive yields are fixed to the measured values listed in Table \ref{cross_section} of \ref{app2}.
$BR \left(  \omega \rightarrow e^{+}e^{-} \right)$ and $BR \left(  \rho \rightarrow e^{+}e^{-} \right)$ are the branching ratios to a di-electron for $\omega$ 
and $\rho$ meson, respectively.
$F_{\phi, \omega, \rho}\left( M' \right)$ in Eq.(\ref{eq:f1}) and (\ref{eq:f2}) indicate the Breit-Wigner function describing the intrinsic mass spectra of the light vector 
mesons and  $G_{gauss} \left( M_{e^{+}e^{-}} - M' \right)$ shows the Gauss function expressing the smearing effect caused by the transverse momentum resolution. 
The residual backgrounds are assumed to follow a first-order polynomial function $H_{bg}\left( M_{e^{+}e^{-}} \right)$.
These functions are expressed as
\begin{equation}
F_{\phi, \omega, \rho}\left( M' \right) = \frac{\Gamma_{\phi, \omega, \rho}/2\pi}{ \left( M'- M_{\phi, \omega, \rho} \right) ^{2} +
\left( \Gamma_{\phi, \omega, \rho}/2 \right) ^{2}},  \label{eq:f3} 
\end{equation}
\begin{equation}
G_{gauss}\left( M_{e^{+}e^{-}} - M' \right) = \frac{1}{\sqrt{2\pi}\sigma} e^{- \left( M_{e^{+}e^{-} }- M' \right) ^{2} / 2\sigma ^{2}},  \label{eq:f4}
\end{equation}
\begin{equation}
H_{bg}\left( M_{e^{+}e^{-}} \right) = B M_{e^{+}e^{-}}  + C , \label{eq:f5}
\end{equation}
where the mass center $M_{\phi,\omega, \rho}$ and the width $\Gamma_{\phi,\omega, \rho}$ of the light vector mesons 
are fixed to their intrinsic values \cite{PDG2010}, whereas the mass resolution $\sigma$ is a free parameter. 
$A$, $B$ and $C$ in the equations are normalization factors.
The fitting ranges are from 0.9 to 1.2 GeV/c$^{2}$ for $\phi$ meson and  from 0.6 to 0.9 GeV/c$^{2}$ for $\omega/\rho$ meson.
The number of the light vector mesons is counted by the integration of the convolution function over the signal mass region.
The definition of the signal mass region is mentioned in Section \ref{sect3}.  
The signal-to-background ratios estimated by fitting are 8.4 $\times$ 10$^{-2}$, 2.0 $\times$ 10$^{-2}$ and 4.1 $\times$ 10$^{-4}$ 
for $\phi$, $\omega$ and $\rho$ mesons, respectively, in central Au+Au collisions at $\sqrt{s_{NN}}$ = 200 GeV  ($dN_{\pi^{0} + \pi^{\pm}}/dy$ = 1000) 
with the baseline parameter set.
The differences of the signal-to-background ratios are 4.9 $\%$ ($\phi$), 7.4 $\%$ ($\omega$) and 8.8 $\%$ ($\rho$) compared to those by the simple counting of the simulated true pairs.
In central Pb+Pb collisions at $\sqrt{s_{NN}}$ = 5.5 TeV  ($dN_{\pi^{0} + \pi^{\pm}}/dy$ = 2700), 
The signal-to-background ratios are 1.7 $\times$ 10$^{-2}$, 6.7 $\times$ 10$^{-3}$ and 1.7 $\times$ 10$^{-4}$ for $\phi$, $\omega$ and $\rho$ mesons, respectively. 
They correspond to 3.2 $\%$ ($\phi$), 12.1 $\%$ ($\omega$) and 14.3 $\%$ ($\rho$) differences with respect to the case of the simple counting of the simulated true pairs. 
The differences depend on the experimental parameters.
It is unlikely for them to exceed 50 $\%$ at a realistic range of the experimental parameters \footnote{
The differences of the signal-to-background ratios between the estimation with the fit and the counting of the simulated true pairs 
originate from the background shape mainly depending on the subtraction of the combinatorial background. 
The differences are studied at the experimental parameter range of 1 $< P_{cnv} <$ 5 $\%$ or 100 $< R_{\pi^{\pm}} <$ 500  and result in $\sim$ 20 $\%$ for the three mesons in both collision systems.
}.

\begin{figure*}[!h]
\begin{center}
\begin{tabular}{c}
	\begin{minipage}{0.5\hsize}
	\begin{center}
	\includegraphics[scale=0.39]{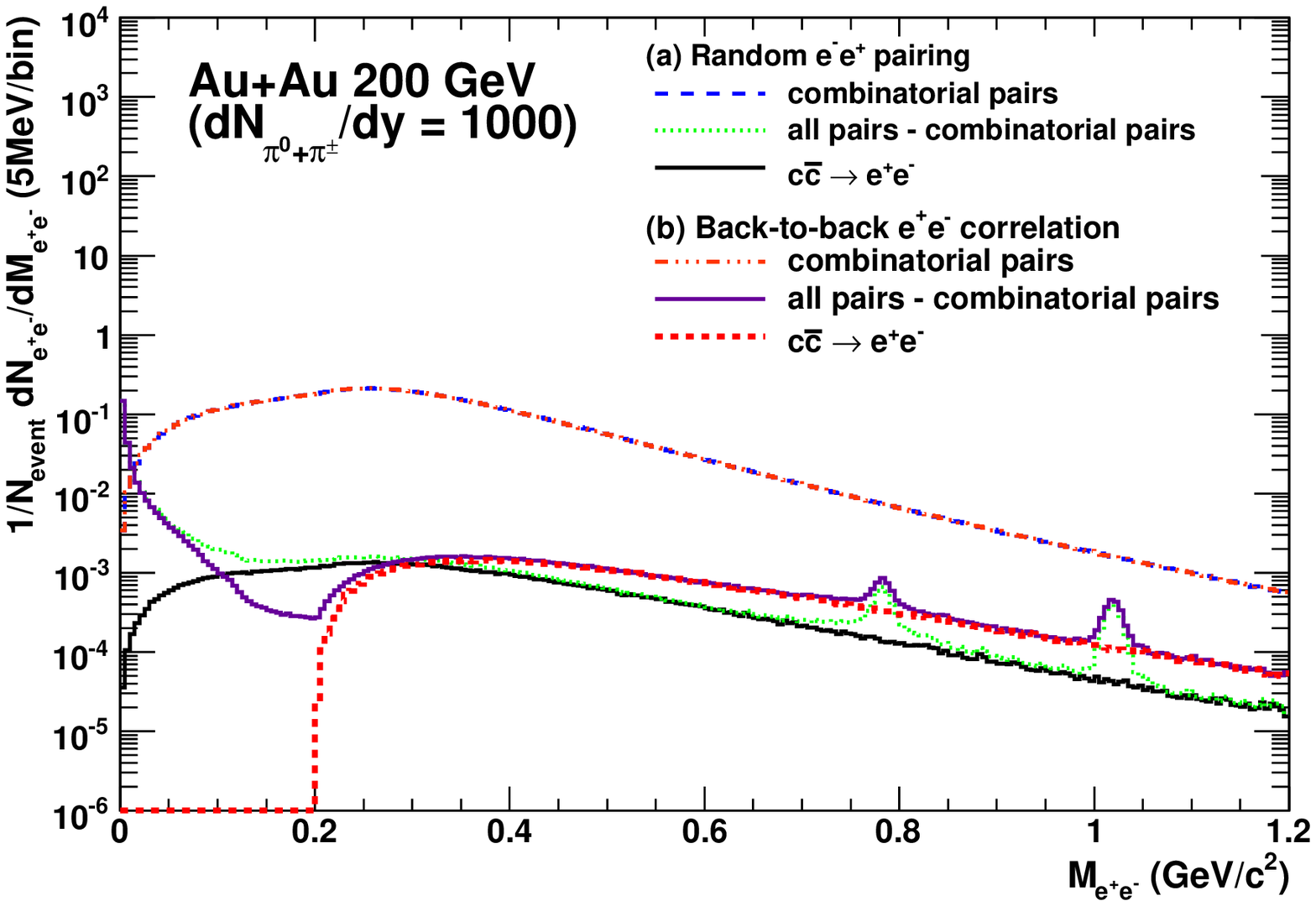}
	\end{center}
	\end{minipage}
	\begin{minipage}{0.5\hsize}
	\begin{center}
	\includegraphics[scale=0.39]{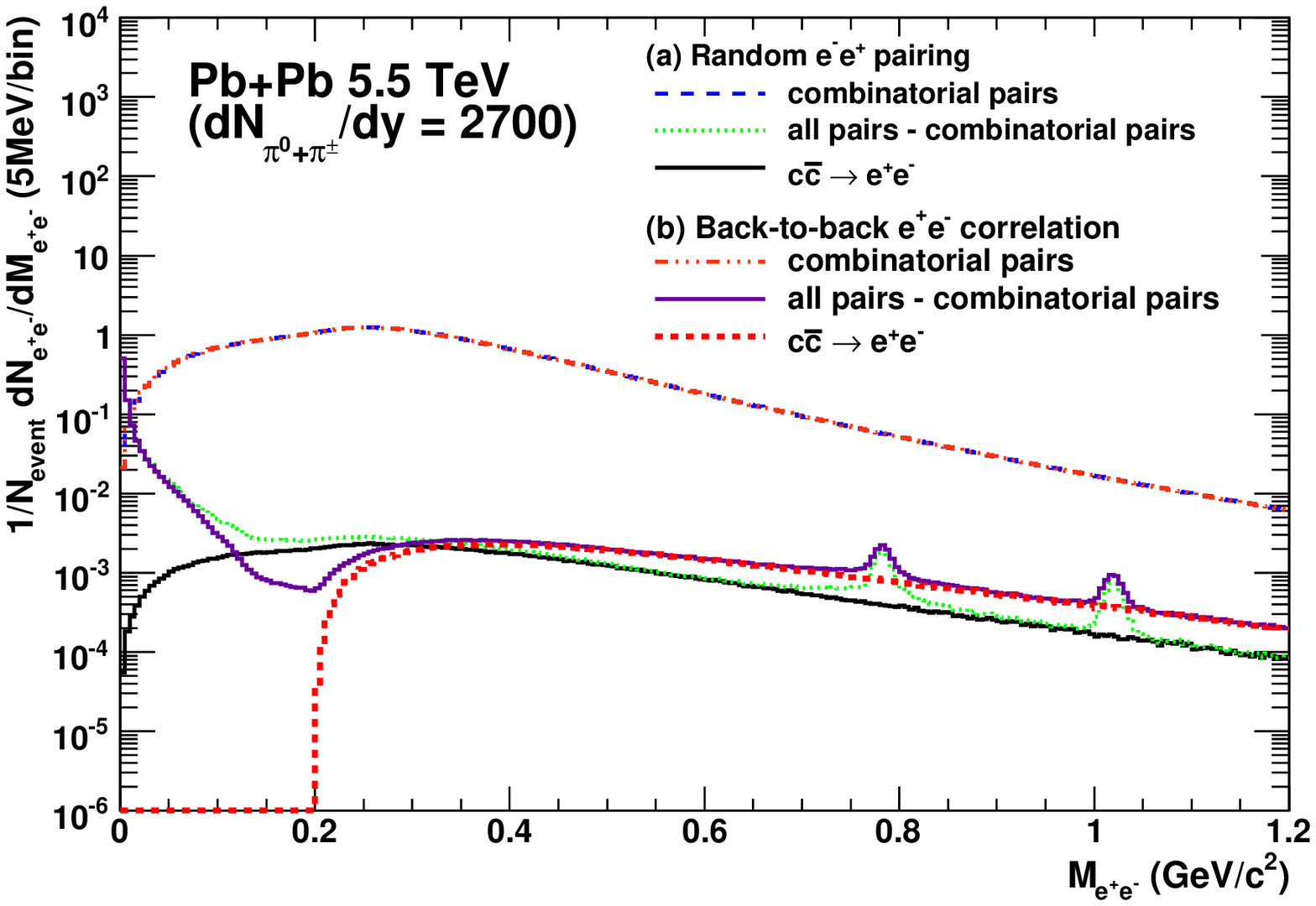}
	\end{center}
	\end{minipage}
\end{tabular}
\begin{tabular}{c}
	\begin{minipage}{0.5\hsize}
	\begin{center}
	\includegraphics[scale=0.39]{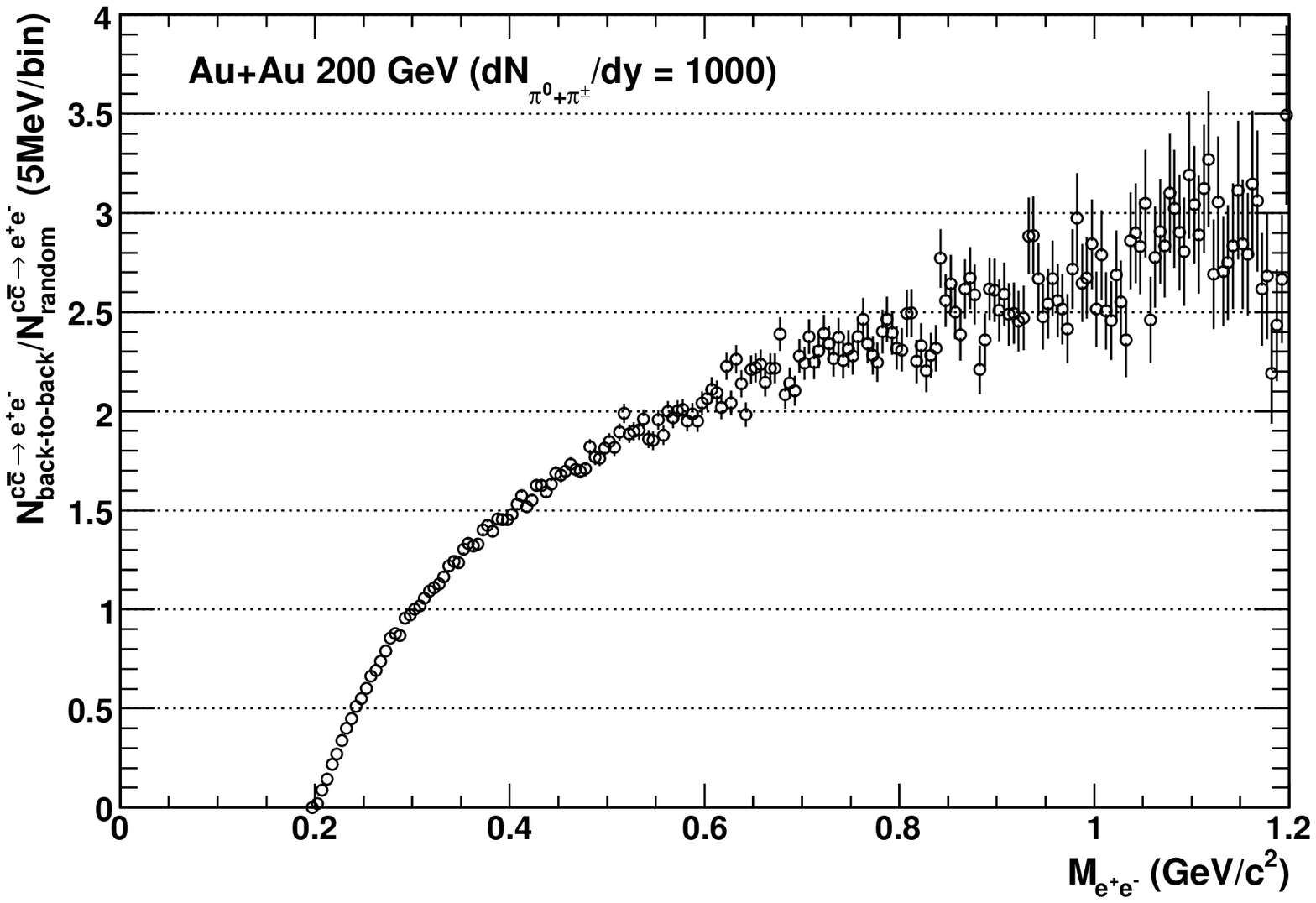}
	\end{center}
	\end{minipage}
	\begin{minipage}{0.5\hsize}
	\begin{center}
	\includegraphics[scale=0.39]{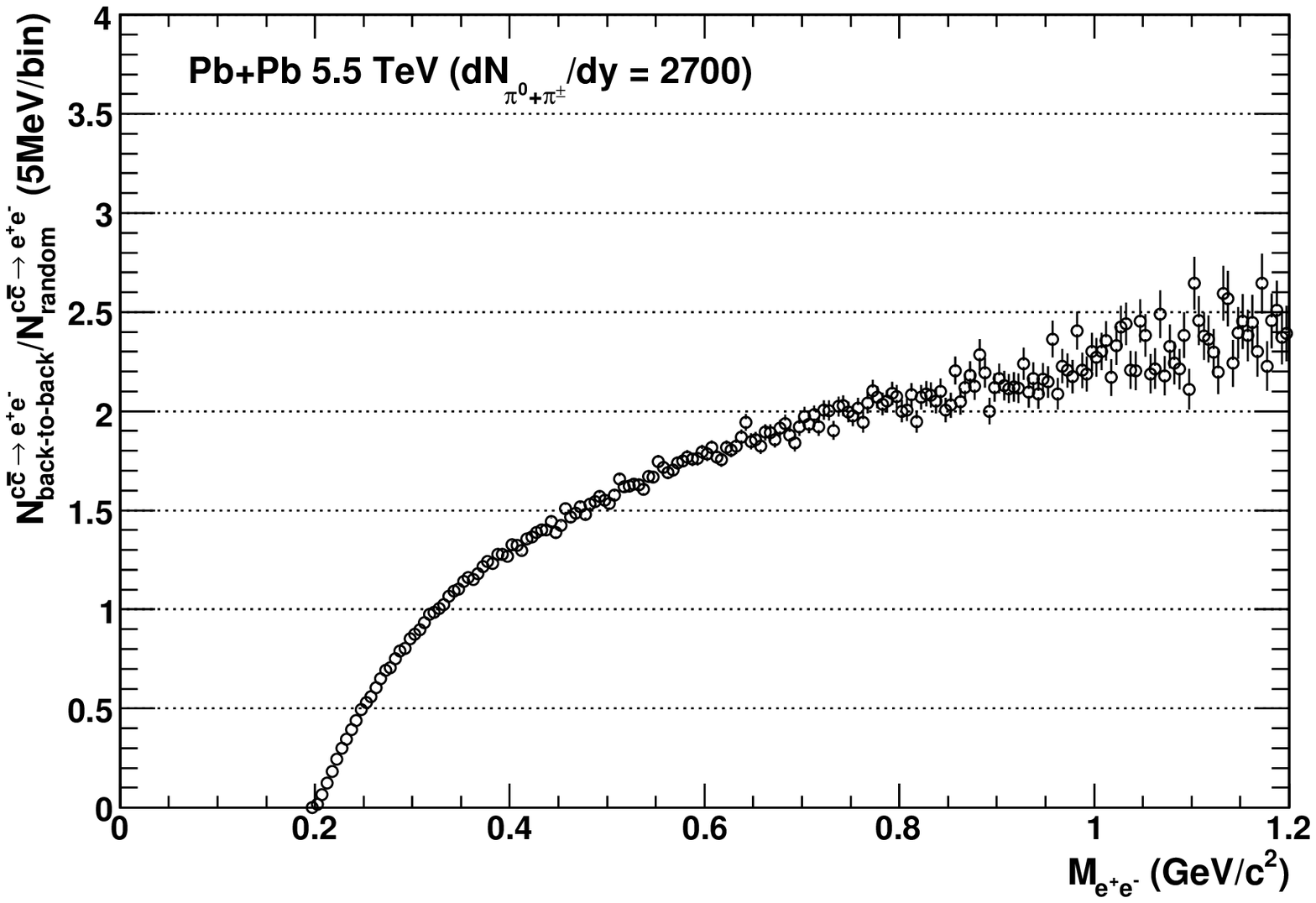}
	\end{center}
	\end{minipage}
\end{tabular}
\begin{tabular}{c}
	\begin{minipage}{0.5\hsize}
	\begin{center}
	\includegraphics[scale=0.39]{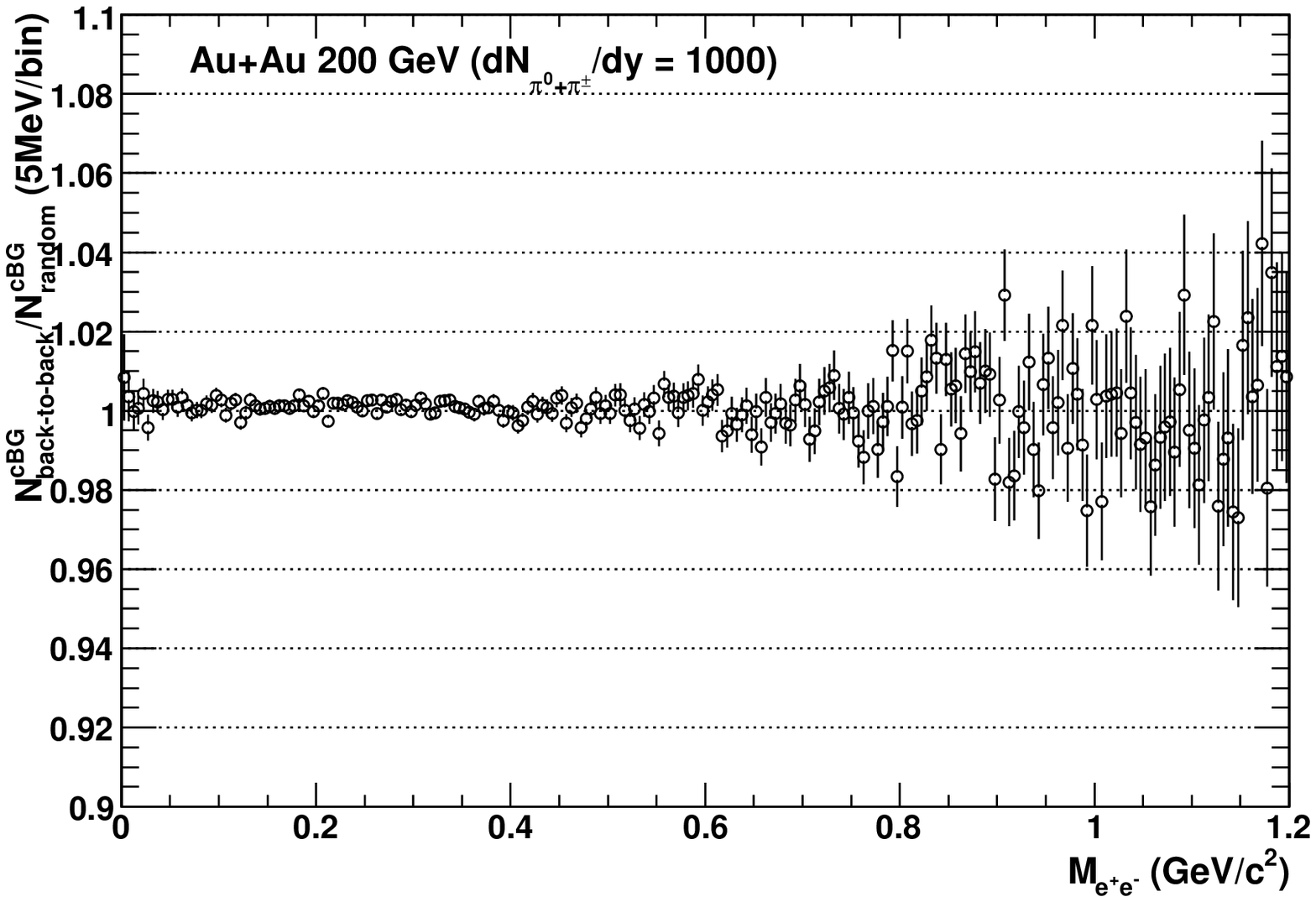}
	\end{center}
	\end{minipage}
	\begin{minipage}{0.5\hsize}
	\begin{center}
	\includegraphics[scale=0.39]{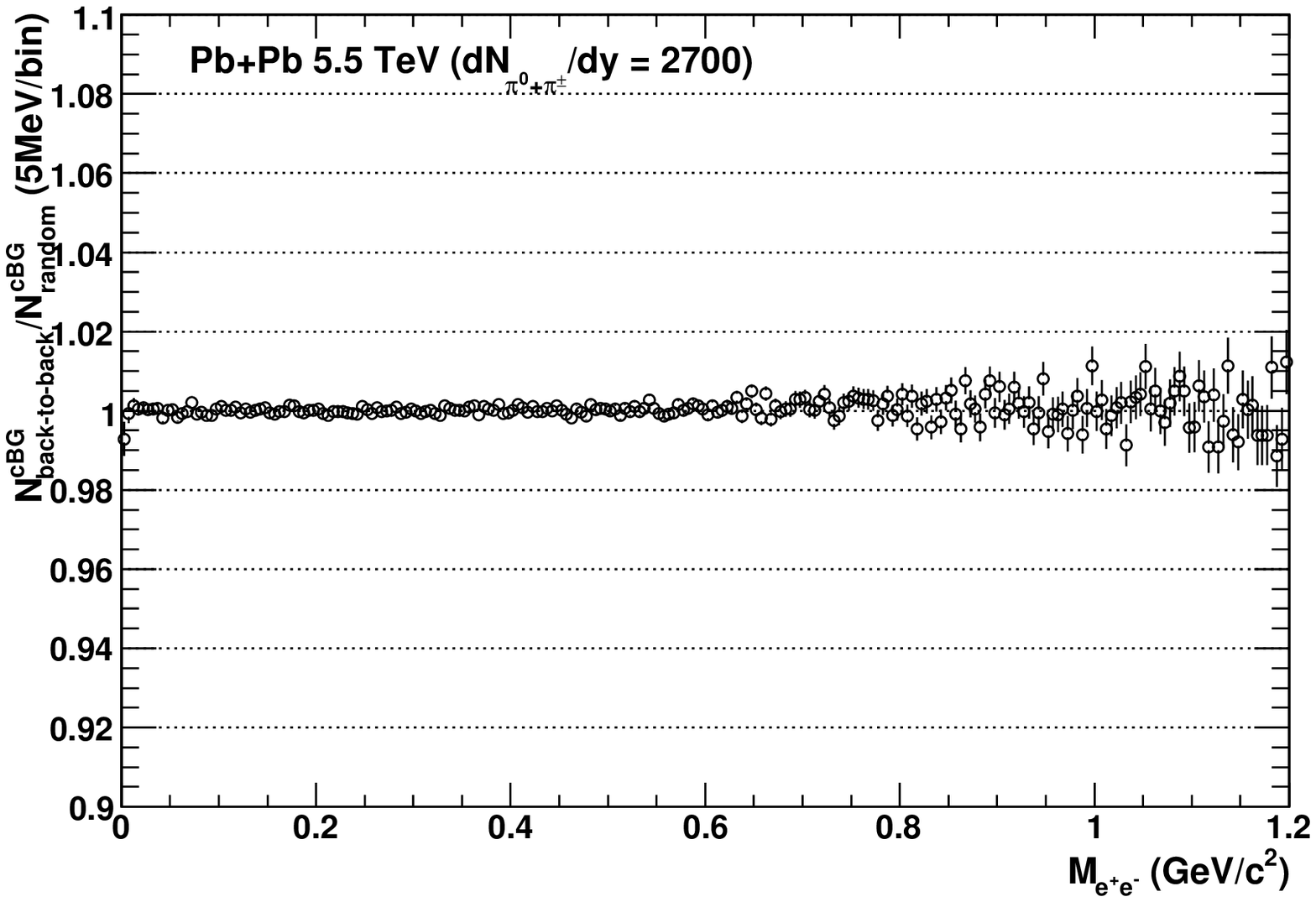}
	\end{center}
	\end{minipage}
\end{tabular}
\caption{\label{mass_cc} (color online)
The comparison of the invariant mass spectra for different correlations of di-electrons from charm quarks.
The panel (a) and (b) on the top figures show the invariant mass spectra in cases of random di-electron pairs and back-to-back correlated di-electron ones 
in azimuth, respectively.
The middle figures show the ratio between the number of $c\bar{c} \rightarrow e^{+}e^{-}$ in the random pairing case and in the back-to-back correlated one. 
The bottom figures show the ratio between the number of combinatorial pairs in the random pairing case and in the back-to-back correlated one.
}
\end{center}
\end{figure*}

Electrons and positrons from open charms are randomly generated and combined into pairs in this simulation.
These pairs are, in fact, azimuthally correlated at mid-rapidity, because they originate from the jets due to the large mass of charm quarks.
We assume the back-to-back $e^{+}e^{-}$ correlation in azimuth as the extreme case of the open charm production.
Realistic correlations would exist between the random pairing case and the back-to-back correlated case.
The top plots in Fig.\ref{mass_cc} show the invariant mass spectra reconstructed from all true pairs, combinatorial pairs and $c\bar{c} \rightarrow e^{+}e^{-}$, 
respectively, in central Au+Au collisions at $\sqrt{s_{NN}}$ = 200 GeV ($dN_{\pi^{0} + \pi^{\pm}}/dy$ = 1000) on the left and
in central Pb+Pb collisions at $\sqrt{s_{NN}}$ = 5.5 TeV ($dN_{\pi^{0} + \pi^{\pm}}/dy$ = 2700) on the right.  
The distributions of the random di-electron pairs and the back-to-back correlated ones in azimuth are superimposed in the same plot.
The middle plots show the ratio of the number of $c\bar{c} \rightarrow e^{+}e^{-}$ as a function of the invariant mass.
The denominator is the number of di-electrons with random pairing and the numerator is the number of di-electrons with the back-to-back correlation. 
This ratio varies by a factor of 1.5 to 3 around the mass range of the light vector mesons.
The ratio between the number of combinatorial pairs in the random pairing case and in the back-to-back correlated case is consistent within only a few $\%$ in both collision systems as shown in the bottom figures. 
Therefore the correlation of the $c\bar{c}$ production has little influence on the signal-to-background ratios of the light vector mesons. 

In addition to above issues, the other residual effects, which are beyond the scope of this study, are listed below. 
\begin{itemize}

\item Track reconstruction algorithms bias:
Track reconstruction algorithms can bias momentum measurement of charged particles.
For example, the algorithm based on the combinatorial Hough transform technique \cite{TrkReco1,TrkReco2,TrkReco3} 
reconstructs higher momentum than true one, especially for a charged particle producing from the off-axis point.
The photon-conversion electrons at the off-axis point contribute to the background shape in the relatively higher mass region.
In addition, especially under a high multiplicity environment, fake tracks are reconstructed by chance depending on the algorithms. 
These tracks can contribute as the additional backgrounds. 

\item The correlation between the tagging efficiency of electrons and the rejection factor of charged hadrons: 
The correlation between the electron tagging efficiency and the rejection factor of charged hadrons 
depends on the method of particle identification.
For instance, if particles are identified by $dE/dx$, the correlation has a trade-off relation. 
Another example is the degradation by the situation where a number of particles simultaneously pass through the detector.
If a hadron and an electron enter the same area of the electron identification device, either or both of them can be wrongly identified. 
In more general case, complicated correlations may appear, since particles are identified with a combination of multiple devices.
%
%

\item The fiducial effect in the magnetic field:
This simulation considers the detector acceptance under the assumption that di-electron kinematics is completely reconstructed.
In real experiments, charged particles are bent in the magnetic field and entered into the imperfect coverage of the detectors.
The fiducial effect becomes apparent at the edge of the acceptance.
Therefore the inefficiency of the electron detection should be taken into account as a function of  the magnetic field, 
the detector positions from the collision point and the detector configurations.

\end{itemize} 
Nevertheless, our simulation study would provide a useful guideline to evaluate the effect of the non-residual factors on measurability of the light vector mesons.

\section{Conclusion}
\label{sect5}
This paper provides a guideline to evaluate the minimum requirements
in order for a given idealized detection system to achieve 
reasonable statistical significance for the measurement of light vector mesons via di-electron decays 
in different temperature states, $dN_{\pi^{0} + \pi^{\pm}}/dy$ = 1000 and 2700
intended for the most central Au+Au collisions at $\sqrt{s_{NN}} = $ 200 GeV
and the most central Pb+Pb collisions at $\sqrt{s_{NN}} = $ 5.5 TeV, respectively.
The simulation codes used for this study are openly available \cite{simcode}. 

The results suggest that parameter ranges for the measurement of $\phi$ and $\omega$ mesons are selectable in designing a detector system 
depending on the number of events at the highest centrality class, even if the residual effects caused by the procedure of the real data analysis 
and the correlations in the open charm production are considered.
The statistical significance of $\rho$ mesons is less than that of  $\phi$ and $\omega$ mesons due to its broad mass shape and the limited mass resolution, 
even if sufficiently high luminosity is prospected.
However, the mass spectrum of $\rho$ mesons in the vacuum can be indirectly evaluated as long as $\phi$ and $\omega$ mesons spectra are accurately determined.
This would provide the baseline to understand the properties of the low-mass di-electron continuum. 

\clearpage
\section*{Acknowledgments}
\label{acknowledgments}
We thank Prof. T. Sugitate for his suggestions on this paper and supports especially for establishing the computing environment. 
We thank Prof. K. Shigaki and all the colleagues at Hiroshima university for many discussions.
The numerical simulation study has been performed on the parallel processing system at the Data Analysis Laboratory for High-Energy Physics at Hiroshima University.

\appendix
\section{Tsallis function and particle production}
\label{app1}
The Tsallis function \cite{tsallis} is widely used for explaining the properties of particle productions.
At midrapidity, the total energy of each particle is approximately represented by 
the transverse mass $m_{T} = \sqrt{p_T^{2} + m_0^{2}}$, where $m_{0}$ is the rest mass of a particle.
The Tsallis function is formulated as a function of  $m_{T}$ as follows,
\begin{eqnarray}
E\frac{d^3\sigma}{dp^{3}} = && \frac{1}{2\pi}\frac{d\sigma}{dy}\frac{\left( n-1 \right) \left( n-2 \right) }
{ \left( nT+m_{0} \left( n-1 \right) \right) \left( nT+m_{0} \right)} \nonumber \\
& & \left( \frac{nT+m_{T}}{nT+m_{0}} \right) ^{-n} , \label{eq:eq1} 
\end{eqnarray}
where $n=-1/ \left( 1-q \right)$, and $d\sigma/dy$ is the production cross section over all $p_{T}$ ranges at midrapidity.
Equation (\ref{eq:eq1}) simultaneously represents the power-law behavior at high $p_{T}$ and  the exponential behavior at low $p_{T}$  
by the two parameters $q$ and $T$. 

In the limit of $m_{0} \rightarrow 0$, Eq.(\ref{eq:eq1}) becomes \\
\begin{equation}
E\frac{d^3\sigma}{dp^{3}} = \frac{1}{2\pi}\frac{d\sigma}{dy}\frac{\left( n-1 \right) \left( n-2 \right)}{ \left( nT \right) ^2} \left( 1+\frac{m_{T}}{nT} \right)^{-n}. \label{eq:eq2} \\
\end{equation}
Equation (\ref{eq:eq2}) is similar to the pQCD expression by Hagedron \cite{hagedron}.
The parameter $n$ is directly related to the exponent $k$ of the pure power-law shape of pQCD calculations. 
The relation between $n$ and $k$ is expressed as
\begin{equation}
n = \frac{k m_{T}^2}{p_{T}^2 - k T m_{T}}. \label{eq:eq3} \\
\end{equation}

In the limit of  $m_{0} \rightarrow 0$ and $q \rightarrow 1$ (i.e. $n \rightarrow -\infty)$, Eq. (\ref{eq:eq1}) becomes 
\begin{equation}
E \frac{d^3\sigma}{dp^{3}} = Ae^{-m_{T}/T} ,\label{eq:eq4} \\
\end{equation}
where $A$ is a normalization factor.
Equation (\ref{eq:eq4}) is equivalent to the Boltzmann distribution.
The slope parameter $T$ characterizes the thermal production of particles.  

\section{The production cross sections and the inclusive yields over all $p_{T}$ ranges at midrapidity}
\label{app2}
\begin{table}[tbh]
\scalebox{0.9}{
\begin{tabular}{|c|c|c|c|c|c|c|} \hline
& \multicolumn{2}{|c|}{p+p 200 GeV} & \multicolumn{2}{|c|}{Au+Au 200 GeV}   & \multicolumn{2}{|c|}{p+p 7 TeV} \\ 
& \multicolumn{2}{|c|}{} & \multicolumn{2}{|c|}{(0-10$\%$)}   & \multicolumn{2}{|c|}{}  \\ \hline
Particle                &  $d\sigma/dy (mb) $ &   Ref.                      & $dN/dy$    &   Ref.                                           &  $d\sigma/dy (mb) $   &   Ref.                         \\ \hline
$\phi$                   & 0.41                           &  \cite{scale1}        &    5.8            &  \cite{phi_auau1}                      & 1.97         & \cite{phi7TeV}       \\
$\omega$            & 4.3                              &  \cite{scale1}        &  33.3            &  \cite{ome_auau1}                  & 17.7           &                                 \\
$\rho$                 & 7.4                               &  \cite{rho}             &  57.3            &                                                   & 30.4          &                                  \\
$\pi^{0}$             & 43.5                            &  \cite{npi1}           & 336.8           &  \cite{pi0_auau1,pi0_auau2}  &  178.7       & \cite{pi0eta7TeV}    \\
$\pi^{+}/\pi^{-}$ & 43.5                           &  \cite{ch1,ch2}      & 336.8           &  \cite{cpi_auau1}                      &  178.7      &                                  \\
$\eta$                  & 5.1                              &  \cite{eta1,eta2}    &  39.5           &  \cite{eta_auau1}                    & 15.7          & \cite{pi0eta7TeV}    \\
$K^{+}$/$K^{-}$   & 4.0                            &  \cite{ch1,ch2}      &  44.7             &  \cite{cpi_auau1}                     & 16.4          &                                   \\
$K^{0}_{s}$          & 4.0                             & \cite{scale1,k0}    &                       &                                                    &                   &                                     \\
$c \bar{c}$           & 0.18                          &  \cite{charm}         &   4.4              &  \cite{charm_auau1}                 & 1.48           & \cite{charm7TeV}   \\ \hline    
\end{tabular}
}
\caption{\label{cross_section} 
The production cross sections and the inclusive yields over all $p_{T}$ ranges at midrapidity for different collision systems.
The used data points to calculate the production cross sections and the inclusive yields are cited from the publications listed in the second column for each collision system.
The production cross section of single electrons is obtained by the Tsallis fit to the measured data points and converted into the $c\bar{c}$ cross section with the branching ratio of 9.5 $\%$ \cite{charm}.
The production cross sections of the other particles are obtained by fitting to the measured data points with the Tsallis function, or assuming the proper 
scaling for missing data points. The details are explained in Section \ref{sect2}. 
The errors of the production cross sections and the inclusive yields are expected to be from 10 to 30 $\%$ depending on particles.
These errors are neglected in the simulation.
}
\end{table}
\section{Masses, decay products and branching ratios}
\label{app3}
\begin{table}[th]
\begin{tabular}{|c|c|c|c|} \hline
                   Particle                         & Mass (GeV/$c^{2}$) & Decay products                      & Branching ratio              \\ \hline
                   $\phi$                          & $1.01946$               & $e^{+}e^{-}$                            & $2.954 \times 10^{-4}$   \\
                   $\omega$                     & $0.78265$               & $e^{+}e^{-}$                            & $7.28 \times 10^{-5}$    \\
                   $\rho$                           & $0.77549$               & $e^{+}e^{-}$                            & $4.72 \times 10^{-5}$    \\ 
                   $\pi^{0}$                    & $0.13498$               & $\gamma \gamma$               & $0.98823$                     \\
                                                          &                                  & $\gamma  e^{+}e^{-}$            & $0.01174$                     \\
                   $\eta$                            & $0.54785$               & $\gamma  \gamma$              & $0.03931$                    \\ 
                                                           &                                   & $\gamma  e^{+}e^{-}$           & $7.0 \times 10^{-3}$       \\     
                   $\pi^{+}$                        & $0.13957$               & N/A                                           & N/A                              \\
                   $\pi^{-}$                         & $0.13957$               & N/A                                           & N/A                               \\  
                   $K^{+}$                           & $0.49368$               & $e^{+} \pi^{0} \nu_{e}$          & $0.0507$                       \\
                   $K^{-}$                            & $0.49368$               & $e^{-} \pi^{0} \bar{\nu_{e}}$ & $0.0507$                       \\
                   $c \bar{c}$                      & N/A                           & $e^{+}e^{-} $                            & $0.095$               \\ \hline
\end{tabular}
\caption{\label{pdg_info} Masses, decay products and branching ratios of the light vector mesons and the other background particles.
Masses and branching ratios are cited from the particle data group \cite{PDG2010}. 
The branching ratio of $c \rightarrow e$ is assumed to be 9.5 $\%$ \cite{charm}.
}
\end{table}
\clearpage
\section{Dalitz decays of pseudo-scalar mesons}
\label{app4}
Pseudo-scalar mesons such as $\pi^0$ and $\eta$ mesons mainly decay into two photons.
The Dalitz decay corresponds to the case where photons become off-shell and subsequently decay into di-electrons. 
The relation between 2 $\gamma$ decay process ($P \rightarrow \gamma \gamma$) and the Dalitz decay process ($P \rightarrow \gamma e^+e^-$) 
is described by Kroll-Wada formula \cite{Kroll_Wada1, Kroll_Wada2} as follows,
\begin{eqnarray}
\frac{d\Gamma(P \rightarrow e^{+}e^{-} \gamma)}{dM_{e^{+}e^{-}} } & \propto & \sqrt{1-\frac{4m_{e}^{2}}{M_{e^{+}e^{-}}^{2}}} \left( 1+\frac{2m_{e}^{2}}{M_{e^{+}e^{-}}^{2}} \right) \frac{1}{M_{e^{+}e^{-}}} \nonumber \\
& & S \left( M_{e^{+}e^{-}} \right)  \Gamma(P \rightarrow \gamma\gamma), \nonumber \\ \nonumber \\
S\left( M_{e^{+}e^{-}} \right) & = &|F_P(Q^{2})|^{2} \left( 1-\frac{M_{e^{+}e^{-}}^{2}}{m_{P}^{2}} \right) ^{3},  \nonumber \\
F_{P} \left( Q^{2} \right) & =  &\left( 1-\frac{Q^{2}}{\Lambda_{P}^{2}} \right)^{-1}, 
\end{eqnarray}
where $M_{e^{+}e^{-}}$ is the invariant mass of di-electrons, $m_{e}$ is the rest mass of electron and $m_{P}$ is the rest mass of a parent meson. 
$F_{P} \left( Q^{2} \right) $ is the electro-magnetic transition form factor. $Q^{2}$ is equivalent to the square of the virtual photon mass (i.e. $Q=M_{e^{+}e^{-}}$). 
The measurements of the form factor by the experiments \cite{LeptonG, Cello} show $\Lambda _{P} \simeq M_{\rho}$, 
where $M_{\rho}$ is the rest mass of $\rho$ meson. 
The Kroll-Wada formula determines the branching ratio and the phase space of Dalitz decaying di-electron.


\begin{thebibliography}{99}
{\footnotesize
\bibitem{virtual_photon1} A. Adare \textit{et al.} (PHENIX Collaboration), Phys. Rev. C \textbf{81}, 034911 (2010).
\bibitem{virtual_photon2} A. Adare \textit{et al.} (PHENIX Collaboration), Phys. Rev. Lett. \textbf{104}, 132301 (2010).
\bibitem{direct_photon} M. Wilde \textit{et al.} (ALICE Collaboration), Nucl. Phys. A \textbf{904}, 573 (2013).
\bibitem{baryon_density1} B. I. Abelev \textit{et al.} (STAR Collaboration), Phys. Rev. C \textbf{79}, 034909 (2009).
\bibitem{baryon_density2} M. M. Aggarwal \textit{et al.} (STAR Collaboration), Phys. Rev. C \textbf{83}, 034910 (2011).
\bibitem{chiral1} R. D. Pisarski, Phys. Lett. B \textbf{110}, 155 (1982).
\bibitem{chiral2} M. Asakawa and C. M. Ko, Phys. Rev. C \textbf{48}, R526 (1993).
\bibitem{chiral3} M. Asakawa and C. M. Ko, Nucl. Phys. A \textbf{572}, 732 (1994).
\bibitem{chiral4} R. Rapp, Phys. Rev. C \textbf{63}, 054907 (2001).
\bibitem{chiral5} M. Harada and C. Sasaki, Phys. Rev. D \textbf{74}, 114006 (2006).
\bibitem{Lattice1} Y. Aoki, Z. Fodor, S. D. Katz and K. K. Szabo, Phys. Lett. B \textbf{643}, 46 (2006). 
\bibitem{Lattice2} M. Cheng \textit{et al.}, Phys. Rev. D \textbf{74}, 054507 (2006). 
\bibitem{Lattice3} A. Bazavov \textit{et al.}, Phys. Rev. D \textbf{85}, 054503 (2012). 
\bibitem{Lattice4} S. Borsanyi, S. Durr. Z. Fodor, C. Hoelbling, S. D. Katz, S. Krieg, D. Nogradi, K. K. Szabo, B. C. Toth and N. Trombitas, JHEP 1208,  (\textbf{2012}) 126. 
\bibitem{Lattice5} T. Umeda, S. Aoki, S. Ejiri, T. Hatsuda, K. Kanaya, Y. Maezawa and H. Ohno (WHOT-QCD Collaboration), PoS LATTICE \textbf{2012} (2012) 074. 
\bibitem{Lattice6} A. Bazavov \textit{et al}. (HotQCD Collaboration), Phys. Rev. D \textbf{86}, 094503 (2012). 
\bibitem{Lattice7} S. Borsanyi, Y. Delgado, S. Durr, Z. Fodor, S. D. Katz, S. Krieg, T. Lippert, D. Nogradi and K. K. Szabo, Phys. Lett. B \textbf{713}, 342 (2012). 
\bibitem{KEK_E325_1} M. Naruki \textit{et al.} (KEK-PS E325 Collaboration), Phys. Rev. Lett. \textbf{96}, 092301 (2006).
\bibitem{KEK_E325_2} R. Muto \textit{et al.} (KEK-PS E325 Collaboration), Phys. Rev. Lett. \textbf{98}, 042501 (2007).
\bibitem{JLab_1} M. H. Wood \textit{et al.} (CLAS Collaboration), Phys. Rev. C \textbf{78}, 015201 (2008).
\bibitem{TAPS_1} M. Nanova \textit{et al.} (CBELSA/TAPS Collaboration), Phys. Rev. C \textbf{82}, 035209 (2010).
\bibitem{continum1} J. Zhao \textit{et al.} (STAR Collaboration), J. Phys. G: Nucl. Part. Phys. \textbf{38}, 124134 (2011).
\bibitem{continum2} L. Adamczyk \textit{et al.} (STAR Collaboration), Phys. Rev. C \textbf{86}, 024906 (2012).
\bibitem{continum3} A. Adare \textit{et al.} (PHENIX Collaboration), Phys. Lett. B \textbf{670}, 313 (2009).
\bibitem{mul1}B. Alver \textit{et al.} (PHOBOS Collaboration), Phys. Rev. C \textbf{83}, 024913 (2011).
\bibitem{mul2} K. Adcox \textit{et al.} (PHENIX Collaboration), Phys. Rev. Lett. \textbf{86}, 3500 (2001).
\bibitem{mul3} K. Aamodt \textit{et al.} (ALICE Collaboration), Phys. Rev. Lett, \textbf{105}, 252301 (2010).
\bibitem{RICH} Y. Akiba \textit{et al.}, Nucl. Instrum. Methods Phys. Res., Sect. A \textbf{453}, 279 (2000). 
\bibitem{TPC1} M. Anderson \textit{et al.}, Nucl. Instrum. Methods Phys. Res., Sect. A \textbf{499}, 659 (2003). 
\bibitem{TPC2} P. Gros \textit{et al.} (ALICE Collaboration), Acta Phys. Pol. B \textbf{42}, 1401 (2011).
\bibitem{PHENIX_PID} M. Aizawa \textit{et al.}, Nucl. Instrum. Methods Phys. Res., Sect. A \textbf{499}, 508 (2003). 
\bibitem{HBD} W. Anderson \textit{et al.}, arXiv:physics.ins-det/1103.4277. 
\bibitem{ALICE_upgrade} B. Abelev \textit{et al.} (ALICE Collaboration), CERN Report No. CERN-LHCC-2012-012. LHCC-I-022 (2012). 
\bibitem{momreso_alice} K. Aamodt \textit{et al.} (ALICE Collaboration), Phys. Lett. B \textbf{696}, 30 (2011). 
\bibitem{rapidity1}K. Aamodt \textit{et al.} (ALICE Collaboration), Eur. Phys. J. C \textbf{68}, 89 (2010).
\bibitem{GSmodel} G. J. Gounaris and J. J. Sakurai, Phys. Rev. Lett. \textbf{21}, 244 (1968).
\bibitem{geant4} "Physics Reference Manual (Version: geant4 9.5.0)"
\bibitem{lepton_pair1} Y. S. Tsai, Rev. Mod. Phys. \textbf{46}, 815 (1974).
\bibitem{lepton_pair2} Y. S. Tsai, Rev. Mod. Phys. \textbf{49}, 421 (1977). 
\bibitem{Kroll_Wada1} N. M. Kroll and W. Wada, Phys. Rev. \textbf{98}, 1355 (1955).
\bibitem{Kroll_Wada2} L. G. Landsberg, Phys. Rep. \textbf{128}, 301 (1985). 
\bibitem{FONLL} M. Cacciari, P. Nason and R. Vogt, Phys. Rev. Lett. \textbf{95}, 122001 (2005).
\bibitem{bottom1} A. Adare \textit{et al.} (PHENIX Collaboration) . Phys. Rev. Lett. \textbf{103}, 082002 (2009).
\bibitem{bottom2} M. M. Aggarwal \textit{et al.} (STAR Collaboration), Phys. Rev. Lett. \textbf{105}, 202301 (2010).
\bibitem{bottom3} B. Abelev \textit{et al.} (ALICE Collaboration), Phys. Lett. B \textbf{721}, 13 (2013).
\bibitem{charm} A. Adare \textit{et al.} (PHENIX Collaboration), Phys. Rev. Lett. \textbf{97}, 252002 (2006).
\bibitem{EXODUS} the EXODUS Monte-Carlo-based event and decay generator.
\bibitem{GlauberModel} R. J. Glauber and G. Matthiae, Nucl. Phys. B \textbf{21}, 135 (1970).
\bibitem{GlauberMC} M. L. Miller, K. Reygers, S. J. Sanders and P. Steinberg, Annu. Rev. Nucl. Part. Sci. \textbf{57}, 205 (2007).
\bibitem{npi1} A. Adare \textit{et al.} (PHENIX Collaboration), Phys. Rev. D \textbf{76}, 051106(R) (2007).
\bibitem{ch1} A. Adare \textit{et al.} (PHENIX Collaboration), Phys. Rev. C \textbf{83}, 064903 (2011).
\bibitem{ch2} J.Adams \textit{et al.} (STAR Collaboration), Phys. Lett. B \textbf{616}, 8 (2005).
\bibitem{scale1} A. Adare \textit{et al.} (PHENIX Collaboration), Phys. Rev. D \textbf{83}, 052004 (2011).
\bibitem{k0}  B.I. Abelev \textit{et al.} (STAR Collaboration), Phys. Rev. C \textbf{75}, 064901 (2007).
\bibitem{eta1} A. Adare \textit{et al.} (PHENIX Collaboration), Phys. Rev. D \textbf{83}, 032001 (2011).
\bibitem{eta2} S. S. Adler \textit{et al.} (PHENIX Collaboration), Phys. Rev. C \textbf{75}, 024909 (2007).
\bibitem{rho} J.Adams. \textit{et al.} (STAR Collaboration), Phys. Rev. Lett. \textbf{92}, 092301 (2004).
\bibitem{pi0_auau1} A. Adare \textit{et al.} (PHENIX Collaboration), Phys. Rev. Lett. \textbf{101}, 232301 (2008). 
\bibitem{pi0_auau2} B. I. Abelev \textit{et al.} (STAR Collaboration), Phys. Rev. C \textbf{80}, 44905 (2009). 
\bibitem{cpi_auau1} S. S. Adler \textit{et al.} (PHENIX Collaboration), Phys. Rev. C \textbf{69}, 034909 (2004). 
\bibitem{eta_auau1} A. Adare \textit{et al.} (PHENIX Collaboration), Phys. Rev. C \textbf{82}, 011902(R) (2010). 
\bibitem{ome_auau1} A. Adare \textit{et al.} (PHENIX Collaboration), Phys. Rev. C \textbf{84}, 044902 (2011). 
\bibitem{phi_auau1} A. Adare \textit{et al.} (PHENIX Collaboration), Phys. Rev. C \textbf{83}, 024909 (2011). 
\bibitem{charm_auau1} A. Adare \textit{et al.} (PHENIX Collaboration), Phys. Rev. C \textbf{84}, 044905 (2011). 
\bibitem{pi0eta7TeV} B. Abelev \textit{et al.} (ALICE Collaboration), Phys. Lett. B \textbf{717}, 162 (2012). 
\bibitem{phi7TeV} B. Abelev \textit{et al.} (ALICE Collaboration), Eur. Phys. J. C \textbf{72}, 2183 (2012) . 
\bibitem{charm7TeV} B. Abelev \textit{et al.} (ALICE Collaboration), Phys. Rev. D \textbf{86}, 112007 (2012). 
\bibitem{PDG2010} K. Nakamura \textit{et al.} (Particle Data Group) J. Phys. G \textbf{37}, 075021 (2010).
\bibitem{ALICE_detector} K. Aamodt \textit{et al.} (ALICE Collaboration), JINST \textbf{3}, S08002 (2010).
\bibitem{ATLAS_detector} G. Aad \textit{et al.} (ATLAS Collaboration), JINST \textbf{3}, S08003 (2008).
\bibitem{event_mix1} G.I. Kopylov, Phys. Lett. B \textbf{50}, 472 (1974).
\bibitem{event_mix2} D. Drijard, H.G. Fischer and T. Nakada, Nucl. Instrum. Methods Phys. Res., Sect. A \textbf{225}, 367 (1984). 
\bibitem{TrkReco1} D. Ben-Tzvi and M. B. Sandler, Patt. Reco. Lett., \textbf{11}, 167 (1990).
\bibitem{TrkReco2} M. Ohlsson, C. Peterson and A. L. Yuille, Comp. Phys. Comm. \textbf{71}, 77 (1992).
\bibitem{TrkReco3} J. T. Mitchell \textit{et al.}, Nucl. Instrum. Methods Phys. Res., Sect. A \textbf{482}, 491 (2002).
\bibitem{simcode} Contact the corresponding author till the simulation codes for users are released on the web. 
\bibitem{tsallis} C. Tsallis, J. Stat. Phys. \textbf{52}, 479 (1988).
\bibitem{hagedron} R. Hagedorn, Riv. Nuovo Cimento Soc. Ital. Fis. \textbf{6N10}, 1 (1983).
\bibitem{LeptonG} R. I. Dzhelyadin \textit{et al.}, Phys. Lett. B \textbf{102}, 296 (1981).
\bibitem{Cello} H. J. Behrend \textit{et al.} (CELLO Collaboration), Z. Phys. C \textbf{49}, 401 (1991).
}
\end{thebibliography}
\end{document}